\documentclass[oneside,graybox,sectrefs]{svmult}

\usepackage{mathptmx}
\usepackage{helvet}
\usepackage{courier}
\usepackage{braket} 
\usepackage{algorithmicx}
\usepackage{algpseudocode} 
\usepackage{amsfonts}
\usepackage{amsmath}
\usepackage{simplewick}
\usepackage{type1cm}         
\usepackage{exercise}
\usepackage{makeidx}         
\usepackage{graphicx}        
\usepackage{multicol}        
\usepackage[bottom]{footmisc}
\usepackage[usenames,dvipsnames,x11names]{xcolor}
 \usepackage{listings}
 \usepackage{epic}
 \usepackage{eepic}
 \usepackage{a4wide}
 \usepackage{color}
 \usepackage{amsmath}
 \usepackage{amssymb}
 \usepackage[T1]{fontenc}
 \usepackage{cite} 
 \usepackage{shadow}
 \usepackage{hyperref}
 \usepackage{bezier}
 \usepackage{pstricks}
\usepackage{textcomp,type1ec,pdfpages}
\usepackage{bera}
\usepackage{slashed}
\usepackage{environ}
\usepackage{esvect}
\usepackage{xspace}
\usepackage{bm}

\definecolor{dkgreen}{rgb}{0,0.6,0}
\definecolor{gray}{rgb}{0.5,0.5,0.5}
\definecolor{mauve}{rgb}{0.58,0,0.82}

\definecolor{gray}{gray}{0.5}
\definecolor{green}{rgb}{0,0.5,0}

\lstnewenvironment{Python}[1]{
\lstset{
language=python,
basicstyle=\footnotesize\setstretch{1},
stringstyle=\color{red},
showstringspaces=false,
alsoletter={1234567890},
otherkeywords={\ , \}, \{},
keywordstyle=\color{blue},
emph={access,and,break,class,continue,def,del,elif ,else,%
except,exec,finally,for,from,global,if,import,in,is,%
lambda,not,or,pass,print,raise,return,try,while},
emphstyle=\color{black}\bfseries,
emph={[2]True, False, None, self},
emphstyle=[2]\color{red},
emph={[3]from, import, as},
emphstyle=[3]\color{blue},
upquote=true,
morecomment=[s]{"""}{"""},
commentstyle=\color{dkgreen}\slshape, 
emph={[4]1, 2, 3, 4, 5, 6, 7, 8, 9, 0},
emphstyle=[4]\color{blue},
framexleftmargin=1mm, framextopmargin=1mm, rulesepcolor=\color{blue},
breakatwhitespace=true,
tabsize=2
}}{}

\lstnewenvironment{C++}[1]{
\lstset{
language=c++,
basicstyle=\footnotesize\setstretch{1},
stringstyle=\color{red},
showstringspaces=false,
alsoletter={1234567890},
otherkeywords={\ , \}, \{},
keywordstyle=\color{blue},
emph={access,and,break,class,continue,def,del,elif ,else,%
except,exec,finally,for,from,global,if,import,in,is,%
lambda,not,or,pass,print,raise,return,try,while},
emphstyle=\color{black}\bfseries,
emph={[2]True, False, None, self},
emphstyle=[2]\color{red},
emph={[3]from, import, as},
emphstyle=[3]\color{blue},
upquote=true,
morecomment=[s]{"""}{"""},
commentstyle=\color{dkgreen}\slshape, 
emph={[4]1, 2, 3, 4, 5, 6, 7, 8, 9, 0},
emphstyle=[4]\color{blue},
framexleftmargin=1mm, framextopmargin=1mm, rulesepcolor=\color{blue},
breakatwhitespace=true,
tabsize=2
}}{}
\usepackage{calrsfs}
\DeclareMathAlphabet{\pazocal}{OMS}{zplm}{m}{n}
\newcommand {\bea}{\begin{eqnarray}}
\newcommand {\eea}{\end{eqnarray}}
\newcommand {\be}{\begin{equation}}
\newcommand {\ee}{\end{equation}}
\newcommand{\beq}{\begin{eqnarray}}
\newcommand{\eeq}{\end{eqnarray}}
\newcommand {\bc}{\begin{center}}
\newcommand {\ec}{\end{center}}
\def\lsim{\mathrel{\rlap{\lower4pt\hbox{\hskip1pt$\sim$}}
    \raise1pt\hbox{$<$}}}               
\def\gsim{\mathrel{\rlap{\lower4pt\hbox{\hskip1pt$\sim$}}
    \raise1pt\hbox{$>$}}}

\makeindex             

\newcommand{\element}[3]{\bra{#1}#2\ket{#3}}
\newcommand{\normord}[1]{\left\{#1\right\}}
\newcommand*{\kpr}[1]{\left\{#1\right\}}
\newcommand*{\fpr}[1]{\left[#1\right]}
\newcommand*{\for}[3]{\langle#1|#2|#3\rangle} 

\begin{document}

\title{Computational Nuclear Physics and Post Hartree-Fock Methods}
\author{Justin G.~Lietz, Samuel Novario, Gustav R.~Jansen, Gaute
  Hagen, and Morten Hjorth-Jensen} 
\institute{Justin G.~Lietz \at
  Department of Physics and Astronomy and National Superconducting
  Cyclotron Laboratory, Michigan State University, East Lansing,
  Michigan, USA, \email{lietz@nscl.msu.edu}, \and Samuel Novario \at
  Department of Physics and Astronomy and National Superconducting
  Cyclotron Laboratory, Michigan State University, East Lansing,
  Michigan, USA, \email{novarios@nscl.msu.edu}, \and Gustav R.~Jansen
  \at National Center for Computational Sciences and Physics Division, Oak Ridge National Laboratory, Oak Ridge,
  Tennessee, USA, \email{jansen@ornl.gov},
  \and Gaute Hagen \at Physics Division, Oak Ridge National Laboratory, Oak Ridge, Tennessee, USA and Department of Physics and
  Astronomy, University of Tennessee, Knoxville, Tennessee, USA,
  \email{hageng@ornl.gov}, \and Morten Hjorth-Jensen \at Department of
  Physics and Astronomy and National Superconducting Cyclotron
  Laboratory, Michigan State University, East Lansing, Michigan, USA
  and Department of Physics, University of Oslo, Oslo, Norway, \email{hjensen@msu.edu}}

\maketitle 
\abstract{We present a computational approach to infinite
  nuclear matter employing Hartree-Fock theory, many-body perturbation
  theory and coupled cluster theory. These lectures are closely linked
  with those of chapters 9, 10 and 11 and serve as input for the
  correlation functions employed in Monte Carlo calculations in
  chapter 9, the in-medium similarity renormalization group theory of
  dense fermionic systems of chapter 10 and the Green's function
  approach in chapter 11.  We provide extensive code examples and
  benchmark calculations, allowing thereby an eventual reader to start
  writing her/his own codes. We start with an object-oriented serial
  code and end with discussions on strategies for porting the
  code to present and planned high-performance computing facilities. }


\section{Introduction}\label{sec:chap8intro}

Studies of dense baryonic matter are of central importance to our
basic understanding of the stability of nuclear matter, spanning from
matter at high densities and temperatures to matter as found within
dense astronomical objects like neutron stars.  An object like a
neutron star offers an intriguing interplay between nuclear processes
and astrophysical observables, spanning many orders of magnitude in
density and several possible compositions of matter, from the crust of
the star to a possible quark matter phase in its interior, see for
example Refs.~\cite{shapiro,prakash2001,Lattimer2007,steiner2010,steiner2012,lattimer2012,weber1999,hh2000}
for discussions.  A central issue in studies of infinite nuclear
matter is the determination of the equation of state (EoS), which can
in turn be used to determine properties like the mass range, the
mass-radius relationship, the thickness of the crust and the rate by
which a neutron star cools down over time. The EoS is also an
important ingredient in studies of the energy release in supernova
explosions.

The determination and our understanding of the EoS for dense nuclear
matter is intimately linked with our capability to solve the nuclear
many-body problem. In particular, to be able to provide precise
constraints on the role of correlations beyond the mean field is
crucial for improved and controlled calculations of the EoS.  
In recent years, there has been a considerable
algorithmic development of first principle (or {\em ab initio})
methods for solving the nuclear many-body problem. Linked with a
similar progress in the derivation of nuclear forces based on
effective field theory (EFT), see chapters four, five and six of the
present text and Refs.~\cite{Weinberg:1990rz,Weinberg:1991um,Ordonez:1992xp,Ordonez:1993tn,vanKolck:1994yi,vankolck1999,machleidt2011,epelbaum2009},
we are now in a situation where reliable results can be provided at
different levels of approximation.  The nuclear Hamiltonians which are
now used routinely in nuclear structure and nuclear matter
calculations, include both nucleon-nucleon ($NN$) and three-nucleon forces (3NFs) derived from EFT, see for
example
Refs.~\cite{Hagen:2015yea,Ekstrom:2015rta,hagen2016,coraggio2014,sammarruca2015,roth2012,binder2013,hergert2016,navratil2016,steiner2012,carbone2013,steiner2012}.
Parallel to the development of nuclear forces from EFT, which employ
symmetries of quantum chromodynamics, there are recent and promising
approaches to derive the EoS using forces constrained from lattice
quantum chromodynamics calculations \cite{tetsuo2013}, see chapters 2
and 3 of the present text.

Theoretical studies of nuclear matter and the pertinent EoS span back
to the very early days of nuclear many-body physics. These early
developments are nicely summarized in for example the review of Day
\cite{day1967} from 1967. These early state-of-the-art calculations
were performed using what is known as Brueckner-Bethe-Goldstone theory
\cite{brueckner1954,brueckner1955}, see for example
Refs.~\cite{hh2000,baldo2012,baldo2012a,coraggio2013,coraggio2014,sammarruca2015} for recent reviews and
developments.  In these calculations, mainly particle-particle
correlations were summed to infinite order.  Other correlations were
often included in a perturbative way. A systematic inclusion of other
correlations in a non-perturbative way are nowadays accounted for in
many-body methods like coupled cluster theory
\cite{bartlett2007,shavittbartlett2009,baardsen2013,hagenmatter,Hagen:2015yea,hagen2016,binder2013,jansen2016} (this chapter), various Monte Carlo methods
\cite{carlson2003,gandolfi2009,gandolfi2009b,gezerlis2010,gandolfi2012,lovato2012,gezerlis2013,carlson2016} (chapter 9), Green's function approaches
\cite{baldo2012a,carbone2013,ch11_Soma2014s2n,dickhoff2004} (chapter 11) and similarity renormalization group methods \cite{morris2015,hergert2016} (chapter 10), just to mention a few of the actual
many-body methods which are used for nuclear matter studies. Many of these methods
are discussed in detail in this and the following chapters.

The aim of this part of the present lecture notes (comprising this chapter and the
three subsequent ones) is to provide the necessary ingredients for
performing studies of neutron star matter (or matter in
$\beta$-equilibrium) and symmetric nuclear matter.  We will mainly
focus on pure neutron matter, but the framework and formalism can
easily be extended to other dense and homogeneous fermionic systems
such as the electron gas in two and three dimensions. The introductory
material we present here forms also the basis for the next three
chapters, starting with the definition of the single-particle basis
and our Hamiltonians as well as Hartree-Fock theory. For infinite
matter, due to the translational invariance of the Hamiltonian, the
single-particle basis, in terms of plane waves, is unchanged under
Hartree-Fock calculations.

Neutron star matter at densities of 0.1 fm$^{-3}$ and greater, is
often assumed to consist mainly neutrons, protons, electrons and
muons in beta equilibrium. However, other baryons like various
hyperons may exist, as well as possible mesonic condensates and
transitions to quark degrees of freedom at higher densities
\cite{hh2000,Lattimer2007,prakash1996,steiner2010}.  In this chapter we limit ourselves to matter composed
of neutrons only.  Furthermore, we will also consider matter at
temperatures much lower than the typical Fermi energies.

In the next section we present some of the basic quantities that enter
the different many-body methods discussed in this and the three
subsequent chapters. All these methods start with some single-particle
basis states, normally obtained via the solution of mean-field
approaches like Hartree-Fock theory. Contributions from correlations
beyond such a mean-field basis to selected observables, are then
obtained via a plethora of many-body methods. These methods represent
different mathematical algorithms used to solve either
Schr\"{o}dinger's or Dirac's equations for many interacting
fermions. After the definitions of our basis states, we derive the
Hartree-Fock equations in the subsequent section and move on with
many-body perturbation theory, full configuration interaction theory
and coupled cluster theory.  Monte Carlo methods, Green's function
theory approaches and Similarity Renormlization group approaches are
discussed in the subsequent three chapters.

The strengths and weaknesses of these methods are discussed throughout
these chapters, with applications to either a simple pairing model
and/or pure neutron matter. Our focus will be on pure neutron matter,
starting with a simple model for the interaction between
nucleons. This allows us to focus on pedagogical and algorithmic
aspects of the various many-body methods, avoiding thereby the full
complexity of nuclear forces.  If properly written however, the codes
can easily be extended to include models of the nuclear interactions
based on effective field theory (see chapters four, five and six of
the present text) and other baryon species than just neutrons. In our
conclusions we point back to models for nuclear forces and their links
to the underlying theory of the strong interaction discussed in the
first chapters of this book, bridging thereby the gap between the
theory of nuclear Hamiltonians and many-body methods.

\section{Single-particle basis, Hamiltonians and models for the nuclear force}\label{sec:chap8forces}

\subsection{Introduction to nuclear matter and Hamiltonians}

Although our focus here and in the coming chapters is on neutron
matter only, our formalism lends itself easily to studies of nuclear
matter with a given proton fraction and electrons. In this section we
outline some of the background details, with a focus on the
calculational basis and the representation of a nuclear Hamiltonian.

Neutron star matter is not composed of only neutrons. Rather, matter
is composed of various baryons and leptons in chemical and charge
equilibrium.  The equilibrium conditions are governed by the weak
processes (normally referred to as the processes for
$\beta$-equilibrium)
\begin{equation*} 
      b_1 \rightarrow b_2 + l +\bar{\nu}_l \hspace{1cm} b_2 +l
      \rightarrow b_1 +\nu_l,
\end{equation*}
where $b_1$ and $b_2$ refer to different types of baryons, for example
a neutron and a proton.  The symbol $l$ is either an electron or a
muon and $\bar{\nu}_l $ and $\nu_l$ their respective anti-neutrinos
and neutrinos. Leptons like muons appear at a density close to nuclear
matter saturation density $\rho_0$, the latter being
\[
     \rho_0 \approx 0.16 \pm 0.02 \hspace{0.1cm} \mathrm{fm}^{-3},
\]
with a corresponding energy per baryon ${\cal E}_0$ for symmetric
nuclear matter at saturation density of
\[
     {\cal E}_0 = B/A=-15.6\pm 0.2 \hspace{0.1cm} \mathrm{MeV}.
\]
The energy per baryon is the central quantity in the present
studies. From the energy per baryon, we can define the pressure $P$
which counteracts the gravitional forces and hinders the gravitational
collapse of a neutron star. The pressure is defined through the
relation
\[
    P=\rho^2\frac{\partial {\cal E}}{\partial \rho}= \rho\frac{\partial
      \varepsilon}{\partial \rho}-\varepsilon,
\]
where $\varepsilon$ is the energy density and $\rho$ the density.  Similarly, the chemical
potential for particle species $i$ is given by
\[
     \mu_i = \left(\frac{\partial \varepsilon}{\partial \rho_i}\right).
\]
In calculations of properties of neutron star matter in
$\beta$-equilibrium, we need to calculate the energy per baryon ${\cal E}$ 
for several proton fractions $x_p$. The proton fraction
corresponds to the ratio of protons as compared to the total nucleon
number ($Z/A$). It is defined as
\[
    x_p = \frac{\rho_p}{\rho},
\]
where $\rho=\rho_p+\rho_n$ is the total baryonic density if neutrons and
protons are the only baryons present (the subscripts used here,
$n,p,e,\mu$, refer to neutrons, protons, electrons and
muons, respectively).
If this is the case, the total
Fermi momentum $k_F$ and the Fermi momenta $k_{Fp}$, $k_{Fn}$ for
protons and neutrons, respectively, are related to the total nucleon density $n$ by
\[
     \rho = \frac{2}{3\pi^2} k_F^3 = x_p \rho + (1-x_p) \rho= \frac{1}{3\pi^2} k_{Fp}^3 + \frac{1}{3\pi^2}k_{Fn}^3.
\]
The energy per baryon will thus be labelled as ${\cal E}(\rho,x_p)$. The
quantity ${\cal E}(\rho,0)$ refers then to the energy per baryon for pure
neutron matter while ${\cal E}(\rho,\frac{1}{2})$ is the
corresponding value for symmetric nuclear matter. 

An important ingredient in the discussion of the EoS and the criteria
for matter in $\beta$-equilibrium is the so-called symmetry energy
${\cal S} (\rho)$, defined as the difference in energy for symmetric
nuclear matter and pure neutron matter
\[
      {\cal S} (\rho) = {\cal E} (\rho,x_p=0) - {\cal E} (\rho,x_p=1/2 ).
\]
If we expand the energy per baryon in the case of nucleonic degrees of
freedom only in the proton concentration $x_p$ about the value of the
energy for SNM ($x_p=\frac{1}{2}$), we obtain,
\[
     {\cal E} (\rho,x_p)={\cal E} (n,x_p=\frac{1}{2})+
     \frac{1}{2}\frac{d^2 {\cal E}}{dx_p^2}(\rho)\left(x_p-1/2\right)^2+\dots ,
\]
where the term $d^2 {\cal E}/dx_p^2$ is to be associated with the
symmetry energy ${\cal S} (\rho)$ in the empirical mass formula. If we
assume that higher order derivatives in the above expansion are small,
then through the conditions for $\beta$-equilbrium (see for example Ref.~\cite{hh2000}) we can define the proton fraction by the
symmetry energy as
\[
    \hbar c\left(3\pi^2\rho x_p\right)^{1/3} = 4{\cal S}
    (\rho)\left(1-2x_p\right),
\]
where the electron chemical potential is given by $\mu_e = \hbar c
k_F$, i.e.\ ultrarelativistic electrons are assumed.  Thus, the
symmetry energy is of paramount importance for studies of neutron star
matter in $\beta$-equilibrium.  One can extract information about the
value of the symmetry energy at saturation density $\rho_0$ from
systematic studies of the masses of atomic nuclei. However, these
results are limited to densities around $\rho_0$ and for proton fractions
close to $\frac{1}{2}$.  Typical values for ${\cal S} (\rho)$ at
$\rho_0$ are in the range $27-38$ MeV \cite{horowitz2014}.  For densities greater than $\rho_0$
it is more difficult to get a reliable information on the symmetry
energy, and thereby the related proton fraction.

With this background, we are now ready to define our basic inputs and
approximations to the various many-body theories discussed in this
chapter and the three subsequent ones.  We will assume that the
interacting part of the Hamiltonian can be approximated by a two-body
interaction.  This means that our Hamiltonian is written as the sum of
a one-body part and a two-body part
\begin{equation*}
    \hat{H} = \hat{H}_0 + \hat{H}_I = \sum_{i=1}^A \hat{h}_0(x_i) +
    \sum_{i < j}^A \hat{v}(r_{ij}),
\label{Hnuclei}
\end{equation*}
with
\begin{equation*}
  \hat{H}_0=\sum_{i=1}^A \hat{h}_0(x_i), 
\label{hinuclei}
\end{equation*}
being the so-called unperturbed part of the Hamiltonian defined by the one-body operator
\[
\hat{h}_0(x_i)=\hat{t}(x_i) + \hat{u}_{\mathrm{ext}}(x_i),
\]
where $\hat{t}$ represents the kinetic energy and $x_i$ represents
both spatial and spin degrees of freedom.  For many-body calculations
of finite nuclei, the external potential $u_{\mathrm{ext}}(x_i)$ is
normally approximated by a harmonic oscillator or Woods-Saxon
potential. For atoms, the external potential is defined by the Coulomb
interaction an electron feels from the atomic nucleus. However, other
potentials are fully possible, such as one derived from the
self-consistent solution of the Hartree-Fock equations to be discussed
below. Since we will work with infinite matter and plane wave basis states, the one-body operator is
simply given by the kinetic energy operator.
Finally, the term $\hat{H}_I$ represents the residual two-body interaction 
\[
  \hat{H}_I= \sum_{i < j}^A \hat{v}(r_{ij}).
\]

Our Hamiltonian is invariant under the permutation (interchange) of
two particles.  Since we deal with fermions however, the total wave
function is anti-symmetric and we assume that we can approximate the exact eigenfunction
for say the ground state with a Slater determinant
\begin{equation*}
   \Phi_0(x_1, x_2,\dots ,x_A,\alpha,\beta,\dots,
   \sigma)=\frac{1}{\sqrt{A!}}  \left| \begin{array}{ccccc}
     \psi_{\alpha}(x_1)& \psi_{\alpha}(x_2)& \dots & \dots &
     \psi_{\alpha}(x_A)\\ \psi_{\beta}(x_1)&\psi_{\beta}(x_2)& \dots &
     \dots & \psi_{\beta}(x_A)\\ \dots & \dots & \dots & \dots & \dots
     \\ \dots & \dots & \dots & \dots & \dots
     \\ \psi_{\sigma}(x_1)&\psi_{\sigma}(x_2)& \dots & \dots &
     \psi_{\sigma}(x_A)\end{array} \right|,
\end{equation*}
where $x_i$ stand for the coordinates and spin values of particle
$i$ and $\alpha,\beta,\dots, \gamma$ are quantum numbers needed to
describe remaining quantum numbers.

The single-particle function $\psi_{\alpha}(x_i)$ are eigenfunctions
of the one-body Hamiltonian $h_0$, that is
\[
\hat{h}_0(x_i) \psi_{\alpha}(x_i)=\left(\hat{t}(x_i) +
\hat{u}_{\mathrm{ext}}(x_i)\right)\psi_{\alpha}(x_i)=\varepsilon_{\alpha}\psi_{\alpha}(x_i).
\]
The energies $\varepsilon_{\alpha}$ are the so-called non-interacting
single-particle energies, or unperturbed energies.  The total energy
is in this case the sum over all single-particle energies, if no
two-body or more complicated many-body interactions are present.

The properties of the determinant lead to a straightforward
implementation of the Pauli principle since no two particles can be at
the same place (two columns being the same in the above determinant)
and no two particles can be in the same state (two rows being the
same).  As a practical matter, however, Slater determinants beyond
$N=4$ quickly become unwieldy. Thus we turn to the occupation
  representation or second quantization to simplify
calculations. For a good introduction to second quantization see for
example Refs.~\cite{blaizot,Ch11_Dickhoff2008,Ch11_Mattuck1992,shavittbartlett2009}.

We start with a set of orthonormal single-particle states $\{
\psi_{\alpha}(x) \}$.  To each single-particle state
$\psi_{\alpha}(x)$ we associate a creation operator
$a^\dagger_{\alpha}$ and an annihilation operator $a_{\alpha}$.  When
acting on the vacuum state $| 0 \rangle$, the creation operator
$a^\dagger_{\alpha}$ causes a particle to occupy the single-particle
state $\psi_{\alpha}(x)$
\[
\psi_{\alpha}(x) \rightarrow a^\dagger_{\alpha} |0 \rangle.
\]
But with multiple creation operators we can occupy multiple states
\[
\psi_{\alpha}(x) \psi_{\beta}(x^\prime) \psi_{\delta}(x^{\prime
  \prime}) \rightarrow a^\dagger_{\alpha} a^\dagger_{\beta}
a^\dagger_{\delta} |0 \rangle.
\]

Now we impose anti-symmetry by having the fermion operators satisfy
the anti-commutation relations
\[
a^\dagger_{\alpha} a^\dagger_{\beta} + a^\dagger_{\beta}
a^\dagger_{\alpha} = \left\{ a^\dagger_{\alpha}
,a^\dagger_{\beta}\right\}= 0,
\]
with the consequence that
\[
a^\dagger_{\alpha} a^\dagger_{\beta} = - a^\dagger_{\beta}
a^\dagger_{\alpha}.
\]
Because of this property, we obtain $a^\dagger_{\alpha}
a^\dagger_{\alpha} = 0$, enforcing the Pauli exclusion principle.
Thus we can represent a Slater determinant using creation operators as
\[
a^\dagger_{\alpha} a^\dagger_{\beta} a^\dagger_{\delta} \ldots |0\rangle,
\]
where each index $\alpha,\beta,\delta, \ldots$ has to be unique.

We will now find it convenient to define a Fermi (F) level and introduce a
new reference vacuum. The Fermi level is normally  defined in terms of all
occupied single-particle states below a certain
single-particle energy.  With the definition of a Fermi level, we can
in turn define our ansatz for the ground state, represented by a
Slater determinant $\Phi_0$.  We will throughout the rest of this
text use creation and annihilation operators to represent quantum
mechanical operators and states.  It means that our compact
representation for a given Slater determinant in Fock space is
\[
  \Phi_{0}=|i_1 \dots i_A\rangle= a_{i_1}^{\dagger} \dots
  a_{i_A}^{\dagger} |0\rangle
\]
where $\vert 0\rangle$ is the true vacuum and we have defined the
creation and annihilation operators as
    \[
        a_p^\dagger|0\rangle = |p\rangle, \quad a_p |q\rangle =
        \delta_{pq}|0\rangle
    \]
with the anti-commutation relations
\[
  \delta_{pq} = \left\{a_p, a_q^\dagger \right\},
\]
and
\[
\left\{a_p^\dagger, a_q \right\} = \left\{a_p, a_q \right\} =
\left\{a_p^\dagger, a_q^\dagger \right\}=0.
\]

We can rewrite the ansatz for the ground state as
\[
\vert\Phi_0\rangle = \prod_{i\le F}a_{i}^{\dagger} |0\rangle,
\]
where we have introduced the shorthand labels for states below the
Fermi level $F$ as $i,j,\ldots \leq F$. For single-particle states
above the Fermi level we reserve the labels $a,b,\ldots > F$, while
the labels $p,q, \ldots$ represent any possible single-particle state.

Since our focus is on infinite systems, the one-body part of the
Hamiltonian is given by the kinetic energy operator only.  In second
quantization it is defined as
\[
\hat{H}_0=\hat{T} = \sum_{pq} \langle p|\hat{t}|q\rangle a_p^\dagger
a_q,
\]
where the matrix elements $\langle p|\hat{t}|q\rangle$ represent the expectation value of the kinetic energy operator (see the discussion below as well).  The two-body interaction reads
\[
\hat{H}_I=\hat{V} = \frac{1}{4} \sum_{pqrs} \langle
pq|\hat{v}|rs\rangle_{AS} a_p^\dagger a_q^\dagger a_s a_r,
\]
where we have defined the anti-symmetrized matrix elements
\[
\langle pq|\hat{v}|rs\rangle_{AS} = \langle pq|\hat{v}|rs\rangle -
\langle pq|\hat{v}|sr\rangle.
\]

We can also define a three-body operator
\[
\hat{V}_3 = \frac{1}{36} \sum_{pqrstu} \langle
pqr|\hat{v}_3|stu\rangle_{AS} a_p^\dagger a_q^\dagger a_r^\dagger a_u
a_t a_s,
\]
with the anti-symmetrized matrix element
\begin{align*}
            \langle pqr|\hat{v}_3|stu\rangle_{AS} = \langle
            pqr|\hat{v}_3|stu\rangle + \langle
            pqr|\hat{v}_3|tus\rangle + \langle
            pqr|\hat{v}_3|ust\rangle- \langle pqr|\hat{v}_3|sut\rangle
            - \langle pqr|\hat{v}_3|tsu\rangle - \langle
            pqr|\hat{v}_3|uts\rangle.
\end{align*}
In this and the forthcoming chapters we will limit ourselves to
two-body interactions at most.  Throughout this chapter and the subsequent three we will drop the subscript $AS$ and use only anti-symmetrized matrix elements.

Using the ansatz for the ground state $\vert \Phi_0\rangle$ as new reference
vacuum state, we need to redefine the anticommutation relations to
\[
\left\{a_p^\dagger, a_q \right\}= \delta_{pq}, \hspace{0.1cm} p, q \leq F,
\]
and
\[
\left\{a_p, a_q^\dagger \right\} = \delta_{pq}, \hspace{0.1cm} p, q > F.
\]
It is easy to see that
\[
        a_i|\Phi_0\rangle = |\Phi_i\rangle\ne 0, \hspace{0.5cm}
        a_a^\dagger|\Phi_0\rangle = |\Phi^a\rangle\ne 0,
\]
and
\[
a_i^\dagger|\Phi_0\rangle = 0 \hspace{0.5cm} a_a|\Phi_0\rangle = 0.
\]
With the new reference vacuum state the Hamiltonian can be rewritten
as, see problem \ref{problem:prob8.1},
\[
\hat{H}=E_{\mathrm{Ref}}+\hat{H}_N,
\]
with the reference energy defined as the expectation value of the
Hamiltonian using the reference state $\Phi_0$
\[
E_{\mathrm{Ref}}=\langle \Phi_0 \vert \hat{H} \vert \Phi_0\rangle =
\sum_{i\le F} \langle i|\hat{h}_0|i\rangle + \frac{1}{2} \sum_{ij\le
  F}\langle ij|\hat{v}|ij\rangle,
\]
and the new normal-ordered Hamiltonian is defined as
\begin{equation}\label{eq:Hnormalorder}
\hat{H}_N = \sum_{pq} \langle p|\hat{h}_0|q\rangle \left\{a^\dagger_p
a_q\right\}+\frac{1}{4} \sum_{pqrs} \langle pq|\hat{v}|rs\rangle
\left\{a^\dagger_p a^\dagger_q a_s a_r\right\}+\sum_{pq,i\le F}
\langle pi|\hat{v}|qi\rangle \left\{a^\dagger_p a_q\right\},
\end{equation}
where the curly brackets represent normal-ordering with respect to the
new reference vacuum state.  The normal-ordered Hamiltonian can be
rewritten in terms of a new one-body operator and a two-body operator
as
\[
\hat{H}_N=\hat{F}_N+\hat{V}_N,
\]
with
\begin{equation}\label{eq:hfn}
\hat{F}_N=\sum_{pq} \langle p|\hat{f}|q\rangle \left\{a^\dagger_pa_q\right\},
\end{equation}
where
\[
\langle p|\hat{f}|q\rangle= \langle p|\hat{h}_0|q\rangle +\sum_{i\le F}
\langle pi|\hat{v}|qi\rangle.
\]
The last term on the right hand side represents a medium modification
to the single-particle Hamiltonian due to the two-body interaction.
Finally, the two-body interaction is given by
\begin{equation*}
\hat{V}_N = \frac{1}{4} \sum_{pqrs} \langle pq|\hat{v}|rs\rangle
\left\{a^\dagger_p a^\dagger_q a_s a_r\right\}.
\end{equation*}	     

\subsection{Single-particle basis for infinite matter}

Infinite nuclear or neutron matter is a homogeneous system and the
one-particle wave functions are given by plane wave functions
normalized to a volume $\Omega$ for a box with length $L$ (the limit
$L\rightarrow \infty$ is to be taken after we have computed various
expectation values)
\[
\psi_{\mathbf{k}\sigma}(\mathbf{r})=
\frac{1}{\sqrt{\Omega}}\exp{(i\mathbf{kr})}\xi_{\sigma}
\]
where $\mathbf{k}$ is the wave number and $\xi_{\sigma}$ is the spin
function for either spin up or down nucleons
\[ 
\xi_{\sigma=+1/2}=\left(\begin{array}{c} 1
  \\ 0 \end{array}\right) \hspace{0.5cm}
\xi_{\sigma=-1/2}=\left(\begin{array}{c} 0 \\ 1 \end{array}\right).
\]
As an interesting aside, the recent works of Binder {\em et al} \cite{binder2016} and McElvain and Haxton
\cite{haxton2016} offer new perspectives on the construction of effective Hamiltonians and choices of basis functions.

We focus first on the kinetic energy operator.  We assume that we have
periodic boundary conditions which limit the allowed wave numbers to
\[
k_i=\frac{2\pi n_i}{L}\hspace{0.5cm} i=x,y,z \hspace{0.5cm} n_i=0,\pm
1,\pm 2, \dots
\]
The operator for the kinetic energy can be written as
\[
\hat{T}=\sum_{\mathbf{p}\sigma_p}\frac{\hbar^2k_P^2}{2m}a_{\mathbf{p}\sigma_p}^{\dagger}a_{\mathbf{p}\sigma_p}.
\]
When using periodic boundary conditions, the discrete-momentum
single-particle basis functions (excluding spin and/or isospin degrees
of freedom) result in the following single-particle energy
\begin{align*}
  \varepsilon_{n_{x}, n_{y}, n_{z}} = \frac{\hbar^{2}}{2m} \left(
  \frac{2\pi }{L}\right)^{2} \left( n_{x}^{2} + n_{y}^{2} +
  n_{z}^{2}\right)=\frac{\hbar^2}{2m}\left(k_{n_x}^2+k_{n_y}^2+k_{n_z}^2\right),
\end{align*} 
for a three-dimensional system with
\[
k_{n_i}=\frac{2\pi n_i}{L}, \hspace{0.2cm} n_i = 0, \pm 1, \pm 2,
\dots,
\]
We will select the single-particle basis such that both the occupied
and unoccupied single-particle states have a closed-shell
structure. This means that all single-particle states corresponding to
energies below a chosen cutoff are included in the basis. We study
only the unpolarized spin phase, in which all orbitals are occupied
with one spin-up and one spin-down fermion (neutrons and protons in
our case).  With the kinetic energy rewritten in terms of the
discretized momenta we can set up a list  and obtain (assuming
identical particles one and including spin up and spin down solutions)
for energies less than or equal to $n_{x}^{2}+n_{y}^{2}+n_{z}^{2}\le
3$, as shown in for example Table \ref{tab:table1}
\begin{table}
\begin{center}
\caption{Total number of particle filling $N_{\uparrow \downarrow }$
  for various $n_{x}^{2}+n_{y}^{2}+n_{z}^{2}$ values for one spin-1/2
  fermion species.  Borrowing from nuclear shell-model terminology,
  filled shells corresponds to all single-particle states for one
  $n_{x}^{2}+n_{y}^{2}+n_{z}^{2}$ value being occupied.  For matter
  with both protons and neutrons, the filling degree increased with a
  factor of $2$.} \label{tab:table1}
\begin{tabular}{ccccc}
\hline \multicolumn{1}{c}{ $n_{x}^{2}+n_{y}^{2}+n_{z}^{2}$ } &
\multicolumn{1}{c}{ $n_{x}$ } & \multicolumn{1}{c}{ $n_{y}$ } &
\multicolumn{1}{c}{ $n_{z}$ } & \multicolumn{1}{c}{ $N_{\uparrow
    \downarrow }$ } \\ \hline 0 & 0 & 0 & 0 & 2 \\ \hline 1 & -1 & 0 &
0 & \\ 1 & 1 & 0 & 0 & \\ 1 & 0 & -1 & 0 & \\ 1 & 0 & 1 & 0 & \\ 1 & 0
& 0 & -1 & \\ 1 & 0 & 0 & 1 & 14 \\ \hline 2 & -1 & -1 & 0 & \\ 2 & -1
& 1 & 0 & \\ 2 & 1 & -1 & 0 & \\ 2 & 1 & 1 & 0 & \\ 2 & -1 & 0 & -1 &
\\ 2 & -1 & 0 & 1 & \\ 2 & 1 & 0 & -1 & \\ 2 & 1 & 0 & 1 & \\ 2 & 0 &
-1 & -1 & \\ 2 & 0 & -1 & 1 & \\ 2 & 0 & 1 & -1 & \\ 2 & 0 & 1 & 1 &
38 \\ \hline 3 & -1 & -1 & -1 & \\ 3 & -1 & -1 & 1 & \\ 3 & -1 & 1 &
-1 & \\ 3 & -1 & 1 & 1 & \\ 3 & 1 & -1 & -1 & \\ 3 & 1 & -1 & 1 & \\ 3
& 1 & 1 & -1 & \\ 3 & 1 & 1 & 1 & 54 \\ \hline
\end{tabular}
\end{center}
\end{table}

Continuing in this way we get for $n_{x}^{2}+n_{y}^{2}+n_{z}^{2}=4$ a
total of 12 additional states, resulting in $66$ as a new magic
number. For the lowest six energy values the degeneracy in energy
gives us $2$, $14$, $38$, $54$, $66$ and $114$ as magic numbers. These
numbers will then define our Fermi level when we compute the energy in
a Cartesian basis. When performing calculations based on many-body
perturbation theory, coupled cluster theory or other many-body
methods, we need then to add states above the Fermi level in order to
sum over single-particle states which are not occupied.

If we wish to study infinite nuclear matter with both protons and
neutrons, the above magic numbers become $4, 28, 76, 108, 132, 228,
\dots$.

Every number of particles for filled shells defines also the number of
particles to be used in a given calculation. The number of particles
can in turn be used to define the density $\rho$ (or the Fermi momentum)
of the system via
\[
\rho = g \frac{k_F^3}{6\pi^2},
\]
where $k_F$ is the Fermi momentum and the degeneracy $g$, which is two
for one type of spin-$1/2$ particles and four for symmetric nuclear
matter.  With the density defined and having fixed the number of
particles $A$ and the Fermi momentum $k_F$, we can define the length
$L$ of the box used with periodic boundary contributions via the
relation
\[
  V= L^3= \frac{A}{\rho}.
\]
With $L$ we can to define the spacing between various
$k$-values given by
\[
  \Delta k = \frac{2\pi}{L}.
\]
Here, $A$ is the number of nucleons. If we deal with the electron
gas only, this needs to be replaced by the number of electrons $N$.
Exercise \ref{problem:spbasissetup} deals with the set up of a program
that establishes the single-particle basis for nuclear matter
calculations with input a given number of nucleons and a user
specificied density or Fermi momentum. 

\subsection{Two-body interaction\label{sec:interaction}}

As mentioned above, we will employ a plane wave basis for our
calculations of infinite matter properties. With a cartesian basis it
means that we can calculate directly the various matrix elements. 
However, a cartesian basis
represents an approximation to the thermodynamical limit. In order to
compare the stability of our basis with results from the
thermodynamical limit, it is convenient to rewrite the nucleon-nucleon
interaction in terms of a partial wave expansion. This will allow us
to compute the Hartree-Fock energy of the ground state in the
thermodynamical limit (with the caveat that we need to limit the
number of partial waves). In order to find the expressions for the
Hartree-Fock energy in a partial wave basis, we will find it
convenient to rewrite our two-body force in terms of the relative and
center-of-mass motion momenta.

The direct matrix element, with single-particle three-dimensional
momenta $\mathbf{k}_p$, spin $\sigma_p$ and isospin $\tau_p$, is
defined as
\[
\langle \mathbf{k}_p\sigma_p\tau_p \mathbf{k}_q\sigma_q\tau_q \vert
\hat{v}\vert \mathbf{k}_r\sigma_r\tau_r \mathbf{k}_s\sigma_s\tau_s
\rangle,
\]
or in a more compact form as $\langle \mathbf{p}\mathbf{q}\vert
\hat{v} \vert \mathbf{r}\mathbf{s} \rangle$ where the boldfaced
letters $\mathbf{p}$ etc represent the relevant quantum numbers, here
momentum, spin and isospin. Introducing the relative momentum
\[
\mathbf{k} = \frac{1}{2}\left(\mathbf{k}_p-\mathbf{k}_q\right),
\]
and the center-of-mass momentum
\[
\mathbf{K} = \mathbf{k}_p+\mathbf{k}_q,
\]
we have
\[
\langle \mathbf{k}_p\sigma_p\tau_p \mathbf{k}_q\sigma_q\tau_q \vert
\hat{v}\vert \mathbf{k}_r\sigma_r\tau_r \mathbf{k}_s\sigma_s\tau_s
\rangle=\langle \mathbf{k}\mathbf{K}\sigma_p\tau_p \sigma_q\tau_q
\vert \hat{v}\vert \mathbf{k}'\mathbf{K}'\sigma_r\tau_r \sigma_s\tau_s
\rangle.
\]
The nucleon-nucleon interaction conserves the total momentum and
charge, implying that the above uncoupled matrix element reads
\[
\langle \mathbf{k}\mathbf{K}\sigma_p\tau_p \sigma_q\tau_q \vert
\hat{v}\vert \mathbf{k}'\mathbf{K}'\sigma_r\tau_r \sigma_s\tau_s
\rangle=\delta_{T_z,T_z'}\delta(\mathbf{K}-\mathbf{K}')\langle
\mathbf{k}T_zS_z=(\sigma_a+\sigma_b) \vert \hat{v}\vert
\mathbf{k}'T_zS_z'=(\sigma_c+\sigma_d) \rangle,
\]
where we have defined the isospin projections $T_z=\tau_p+\tau_q$ and
$T_z'=\tau_r+\tau_s$.  Defining
$\hat{v}=\hat{v}(\mathbf{k},\mathbf{k}' )$, we can rewrite the
previous equation in a more compact form as
\[
\delta_{T_z,T_z'}\delta(\mathbf{K}-\mathbf{K}')\langle
\mathbf{k}T_zS_z=(\sigma_p+\sigma_q) \vert \hat{v}\vert
\mathbf{k}'T_zS_z'=(\sigma_r+\sigma_s)
\rangle=\delta_{T_z,T_z'}\delta(\mathbf{K}-\mathbf{K}')\langle
T_zS_z\vert\hat{v}(\mathbf{k},\mathbf{k}' ) \vert T_zS_z' \rangle.
\]
These matrix elements can in turn be rewritten in terms of the total
two-body quantum numbers for the spin $S$ of two spin-1/2 fermions as
\[
\langle \mathbf{k}T_zS_z \vert \hat{v}(\mathbf{k},\mathbf{k}' )\vert
\mathbf{k}'T_zS_z' \rangle=\sum_{SS'}\langle
\frac{1}{2}\sigma_p\frac{1}{2}\sigma_q\vert SS_z\rangle \langle
\frac{1}{2}\sigma_r\frac{1}{2}\sigma_s\vert S'S_z'\rangle \langle
\mathbf{k}T_zSS_z\vert \hat{v}(\mathbf{k},\mathbf{k}' )\vert
\mathbf{k}T_zS'S_z' \rangle
\]
The coefficients $\langle \frac{1}{2}\sigma_p\frac{1}{2}\sigma_q\vert
SS_z\rangle$ are so-called Clebsch-Gordan recoupling coefficients.  We
will assume that our interactions conserve charge. We will refer to $T_z=0$ as the $pn$ (proton-neutron)
channel, $T_z=-1$ as the $pp$ (proton-proton) channel and $T_z=1$ as
the $nn$ (neutron-neutron) channel.

The nucleon-nucleon force is often derived and analyzed theoretically
in terms of a partial wave expansion. A state with linear momentum
$\mathbf{k}$ can be written as
\[
\vert \mathbf{k} \rangle =
\sum_{l=0}^{\infty}\sum_{l_l=-l}^{L}\imath^lY_{l}^{m_l}(\hat{k}\vert
klm_l\rangle.
\]

In terms of the relative and center-of-mass momenta $\mathbf{k}$ and
$\mathbf{K}$, the potential in momentum space is related to the
nonlocal operator $V(\mathbf{r},\mathbf{r}')$ by
\begin{equation*}
      \langle \mathbf{k'K'}\vert \hat{v} \vert \mathbf{k'K} \rangle=
      \int d\mathbf{r}d \mathbf{r'} e^{-\imath
        \mathbf{k'r'}}V(\mathbf{r'},\mathbf{r}) e^{\imath \mathbf{kr}}
      \delta(\mathbf{K},\mathbf{K'}).
\end{equation*}
We will assume that the interaction is spherically symmetric and use
the partial wave expansion of the plane waves in terms of spherical
harmonics.  This means that we can separate the radial part of the
wave function from its angular dependence. The wave function of the
relative motion is described in terms of plane waves as
\begin{equation*}
       e^{\imath \mathbf{kr}} = \langle\mathbf{r}\vert
       \mathbf{k}\rangle = 4\pi \sum_{lm} \imath ^{l} j_{l} (kr)
       Y_{lm}^{*}(\mathbf{\hat{k}}) Y_{lm}(\mathbf{\hat{r}}),
\end{equation*}
where $j_l$ is a spherical Bessel function and $Y_{lm}$ the spherical
harmonic.  This partial wave basis is useful for defining the operator
for the nucleon-nucleon interaction, which is symmetric with respect
to rotations, parity and isospin transformations. These symmetries
imply that the interaction is diagonal with respect to the quantum
numbers of total angular momentum $J$, spin $S$ and isospin $T$. Using
the above plane wave expansion, and coupling to final $J$, $S$ and $T$
we get
\begin{equation*}
      \langle \mathbf{k'}\vert V \vert \mathbf{k}\rangle = (4\pi)^2
      \sum_{JM}\sum_{lm}\sum_{l'm'} \imath ^{l+l'}
      Y_{lm}^{*}(\mathbf{\hat{k}}) Y_{l'm'}(\mathbf{\hat{k}'}) {\cal
        C}_{m'M_SM}^{l'SJ}{\cal C}_{mM_SM}^{lSJ} \langle k'l'STJM
      \vert V \vert klSTJM \rangle,
\end{equation*}
where we have defined
\begin{equation*}
    \langle k'l'STJM\vert V \vert klSTJM\rangle = \int
    j_{l'}(k'r')\langle l'STJM\vert V(r',r)\vert lSTJM \rangle j_l(kr)
    {r'}^2 dr' r^2 dr.
\end{equation*}
We have omitted the momentum of the center-of-mass motion $\mathbf{K}$
and the corresponding orbital momentum $L$, since the interaction is
diagonal in these variables. 

The interaction we will use for these calculations is a semirealistic
nucleon-nucleon potential known as the Minnesota potential
\cite{minnesota} which has the form, $V_{\alpha}\left(
r\right)=V_{\alpha}\exp{(-\alpha r^{2})}$. The spin and isospin
dependence of the Minnesota potential is given by,
\begin{equation*}
V\left( r\right)=\frac{1}{2}\left( V_{R}+\frac{1}{2}\left(
1+P_{12}^{\sigma}\right) V_{T}+\frac{1}{2}\left(
1-P_{12}^{\sigma}\right) V_{S}\right)\left(
1-P_{12}^{\sigma}P_{12}^{\tau}\right),
\end{equation*}
where $P_{12}^{\sigma}=\frac{1}{2}\left(
1+\sigma_{1}\cdot\sigma_{2}\right)$ and
$P_{12}^{\tau}=\frac{1}{2}\left( 1+\tau_{1}\cdot\tau_{2}\right)$ are
the spin and isospin exchange operators, respectively. A Fourier
transform to momentum space of the radial part $V_{\alpha}\left(
r\right)$ is rather simple, see problem \ref{problem:fourier}, since
the radial depends only on the magnitude of the relative distance and
thereby the relative momentum
$\vec{q}=\frac{1}{2}\left(\vec{k}_{p}-\vec{k}_{q}-\vec{k}_{r}+\vec{k}_{s}\right)$. Omitting
spin and isospin dependencies, the momentum space version of the
interaction reads
\begin{equation*}
\langle \mathbf{k}_p \mathbf{k}_q \vert V_{\alpha}\vert
\mathbf{k}_r\mathbf{k}_s\rangle=\frac{V_{\alpha}}{L^{3}}\left(\frac{\pi}{\alpha}\right)^{3/2}\exp{(\frac{-q^{2}}{4\alpha})}\delta_{\vec{k}_{p}+\vec{k}_{q},\vec{k}_{r}+\vec{k}_{s}}
\end{equation*}
The various parameters defining the interaction model used in this
work are listed in Table \ref{tab:minnesotatab}.
\begin{table}
\caption{Parameters used to define the Minnesota interaction model
  \cite{minnesota}.}\label{tab:minnesotatab}
\begin{center}
  \begin{tabular}{| l | l | l |}
    \hline $\alpha$ & $V_{\alpha}$ & $\kappa_{\alpha}$ \\ \hline $R$ &
    200 $\mathrm{MeV}$ & 1.487 $\mathrm{fm}^{-2}$ \\ \hline $T$ & 178
    $\mathrm{MeV}$ & 0.639 $\mathrm{fm}^{-2}$ \\ \hline $S$ & 91.85
    $\mathrm{MeV}$ & 0.465 $\mathrm{fm}^{-2}$ \\ \hline
  \end{tabular}
\end{center}
\end{table}

\subsection{Models from Effective field theory for the two- and three-nucleon interactions}\label{subsec:forcemodels}

During the past two decades it has been demonstrated that chiral
effective field theory represents a powerful tool to deal with
hadronic interactions at low energy in a systematic and
model-independent way (see
Refs.~\cite{weinberg1990,ordonez1996,vankolck1994,vankolck1999,machleidt2011,epelbaum2009,ekstrom2013,Ekstrom:2015rta,ekstromPRX}).
Effective field theories (EFTs) are defined in terms of effective Lagrangians which are given by
an infinite series of terms with increasing number of derivatives
and/or nucleon fields, with the dependence of each term on the pion
field prescribed by the rules of broken chiral symmetry.  Applying
this Lagrangian to a particular process, an unlimited number of
Feynman graphs can be drawn. Therefore, a scheme is needed that makes
the theory manageable and calculable.  This scheme which tells us how
to distinguish between large (important) and small (unimportant)
contributions is chiral perturbation theory (ChPT).  Chiral perturbation theory  allows for
an expansion in terms of $(Q/\Lambda_\chi)^\nu$, where $Q$ is generic
for an external momentum (nucleon three-momentum or pion
four-momentum) or a pion mass, and $\Lambda_\chi \sim 1$ GeV is the
chiral symmetry breaking scale.  Determining the power $\nu$ has
become known as power counting.

\begin{figure}[t]\centering
\scalebox{0.14}{\includegraphics{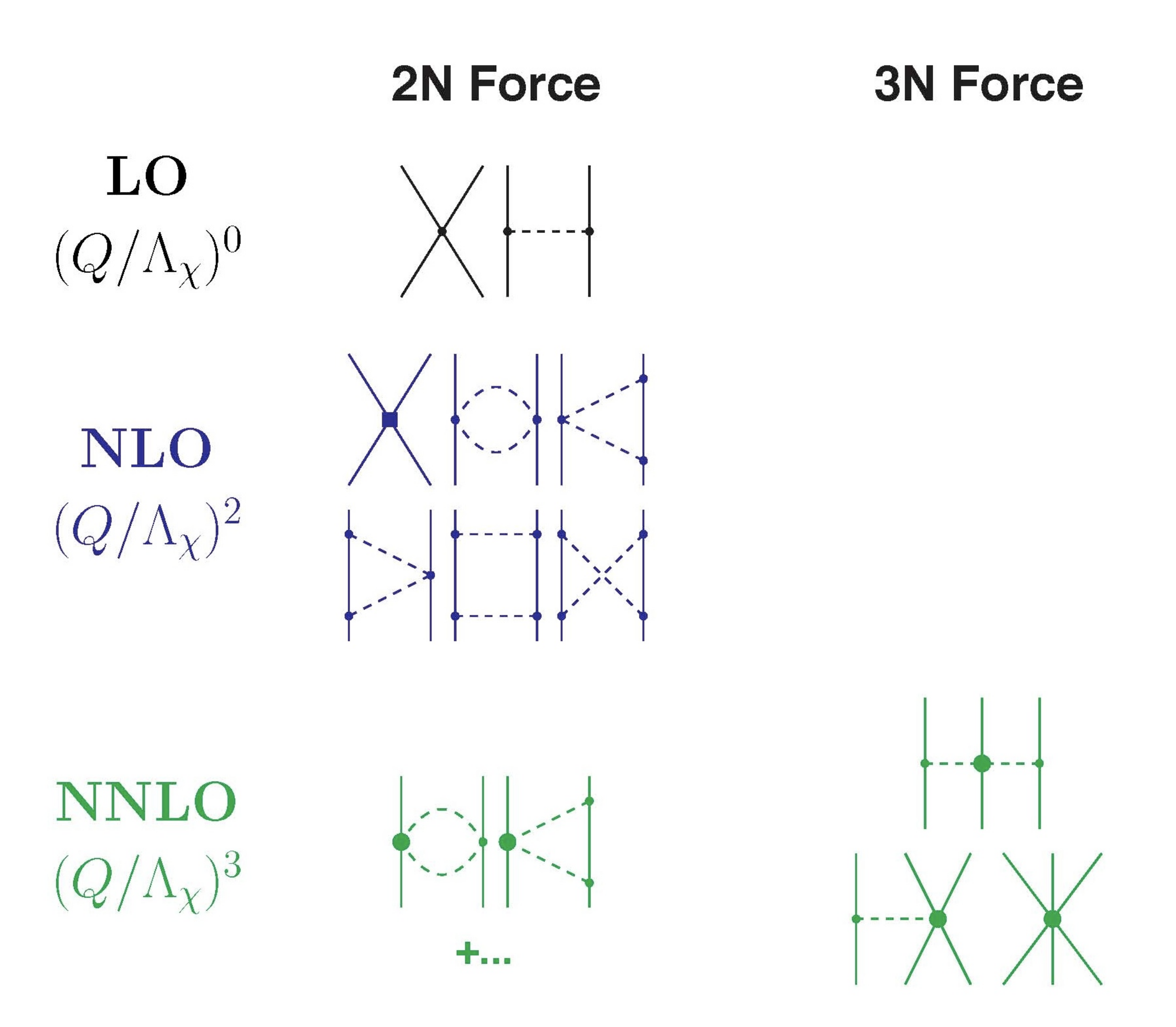}}
\caption{Nuclear forces in ChPT up to NNLO. Solid lines represent
  nucleons and dashed lines pions.  Small dots, large solid dots, and
  solid squares denote vertices of index $\Delta_i= \, $ 0, 1, and 2,
  respectively.}
\label{fig_diagNNLO}
\end{figure}

Nuclear potentials are defined as sets of irreducible graphs up to a
given order.  The power $\nu$ of a few-nucleon diagram involving $A$
nucleons is given in terms of naive dimensional analysis by:
\begin{equation} 
\nu = -2 +2A - 2C + 2L + \sum_i \Delta_i \, ,
\label{eq_nu} 
\end{equation}
with
\begin{equation*}
\Delta_i \equiv d_i + \frac{n_i}{2} - 2 \, ,
\end{equation*}
where $A$ labels the number of nucleons, $C$ denotes the number of separately connected pieces and $L$
the number of loops in the diagram; $d_i$ is the number of derivatives
or pion-mass insertions and $n_i$ the number of nucleon fields
(nucleon legs) involved in vertex $i$; the sum runs over all vertices
contained in the diagram under consideration.  Note that $\Delta_i
\geq 0$ for all interactions allowed by chiral symmetry.  In this work we will focus on the 
simple Minnesota model discussed above. It is however possible, see also the exercises, to
include two- and three-nucleon forces at order
NNLO, as indicated in Fig.~\ref{fig_diagNNLO}.  

Below we revisit briefly the formalism and results presented in
Refs.~\cite{ekstromPRX}. For further details on chiral effective
field theory and nuclear interactions, see for example
Refs.~\cite{machleidt2011,epelbaum2009,ekstrom2013,Ekstrom:2015rta,ekstromPRX}
For an irreducible $NN$ diagram (``two-nucleon potential'', $A=2$,
$C=1$), Eq.~(\ref{eq_nu}) collapses to
\begin{equation*} 
\nu = 2L + \sum_i \Delta_i \, .
\end{equation*}
Thus, in terms of naive dimensional analysis or ``Weinberg counting''
\cite{weinberg1990}, the various orders of the
irreducible graphs which define the chiral $NN$ potential are given by
(see Fig.~\ref{fig_diagNNLO})
\begin{eqnarray*}
V_{\rm LO} & = & V_{\rm ct}^{(0)} + V_{1\pi}^{(0)}
\label{eq_VLO}
\\ V_{\rm NLO} & = & V_{\rm LO} + V_{\rm ct}^{(2)} + V_{1\pi}^{(2)} +
V_{2\pi}^{(2)}
\label{eq_VNLO*}
\\ V_{\rm NNLO} & = & V_{\rm NLO} + V_{1\pi}^{(3)} + V_{2\pi}^{(3)}
\label{eq_VNNLO*}
\end{eqnarray*}
where the superscript denotes the order $\nu$ of the low-momentum
expansion, LO stands for leading order, NLO for next-to-leading order
and NNLO stands for next-to-next-to leading order.  Contact potentials
carry the subscript ``ct'' and pion-exchange potentials can be
identified by an obvious subscript.

The charge-independent one-pion-exchange (1PE) potential reads
\begin{equation}
V_{1\pi} ({\vec k}~', \vec k) = -
\frac{g_A^2}{4f_\pi^2} \: {\vec \tau}_1 \cdot {\vec \tau}_2 \: \frac{
  \vec \sigma_1 \cdot \vec q \,\, \vec \sigma_2 \cdot \vec q} {q^2 +
  m_\pi^2} \,,
\label{eq:eq_1PEci}
\end{equation}
where ${\vec k}~'$ and $\vec k$ represent the final and initial
nucleon momenta in the center-of-mass system (CMS) and $\vec q \equiv
{\vec k}~' - \vec k$ is the momentum transfer; $\vec \sigma_{1,2}$ and
$\vec \tau_{1,2}$ are the spin and isospin operators of nucleon 1 and
2; $g_A$, $f_\pi$, and $m_\pi$ denote axial-vector coupling constant,
the pion decay constant, and the pion mass, respectively.  Since
higher order corrections contribute only to mass and coupling constant
renormalizations and since, on shell, there are no relativistic
corrections, the on-shell 1PE has the form of Eq.~(\ref{eq:eq_1PEci})
to all orders.

It is well known that for high-precision $NN$ potentials, charge
dependence is important.  To take into account the charge
dependence of the 1PE contribution we define a pion-mass dependent
1PE by
\[
V_{1\pi} (m_\pi) \equiv - \,
\frac{g_A^2}{4f_\pi^2} \, \frac{ \vec \sigma_1 \cdot \vec q \,\, \vec
  \sigma_2 \cdot \vec q} {q^2 + m_\pi^2}.
\]
The 1PE for proton-proton ($pp$) and neutron-neutron ($nn$) terms
are then given by
\[
V_{1\pi}^{(pp)} ({\vec k}~', \vec k) = V_{1\pi}^{(nn)} ({\vec k}~',
\vec k) = V_{1\pi} (m_{\pi^0}) \,,
\]
while for the neutron-proton ($np$) part  we have
\[
V_{1\pi}^{(np)} ({\vec k}~', \vec k) = -V_{1\pi} (m_{\pi^0}) +
(-1)^{T+1}\, 2\, V_{1\pi} (m_{\pi^\pm}) \,,
\]
where $T$ denotes the isospin of the two-nucleon system.  The pion masses are defined  as
$m_{\pi^0}=134.9766$ MeV and $m_{\pi^\pm}=139.5702$ MeV.  For the
leading-order, next-to-leading order and NNLO, we refer the reader to
Refs.~\cite{machleidt2011,ekstromPRX}.  The final interaction at
order NNLO is multiplied with the following factors
\cite{machleidt2011},
\begin{equation*}
\widehat{V}({\vec k}~',{\vec k}) \equiv \frac{1}{(2\pi)^3}
\sqrt{\frac{M_N}{E_{p'}}}\: {V}({\vec p}~',{\vec p})\:
\sqrt{\frac{M_N}{E_{p}}}
\end{equation*}
with $E_p=\sqrt{M_N^2+p^2}$ and where the factor $1/(2\pi)^3$ is just
added for convenience.  The potential $\widehat{V}$ satisfies the
nonrelativistic Lippmann-Schwinger (LS) equation, see Ref.~\cite{machleidt2011} for discussions,
\begin{equation*}
 \widehat{T}({\vec k}~',{\vec k})= \widehat{V}({\vec k}~',{\vec k})+
 \int d^3p''\: \widehat{V}({\vec k}~',{\vec k}~'')\: \frac{M_N} {{
     k}^{2}-{k''}^{2}+i\epsilon}\: \widehat{T}({\vec k}~'',{\vec k})
 \, .
\label{eq_LS}
\end{equation*}
In $pp$ scattering, we use $M_N=M_p=938.2720$ MeV, and in $nn$
scattering, $M_N=M_n=939.5653$ MeV.  Moreover, the on-shell momentum
is simply
\begin{equation*}
p^2 = \frac12 M_N T_{\rm lab} \,,
\end{equation*}
where $T_{\rm lab}$ denotes the kinetic energy of the incident nucleon
in the laboratory system (``Lab.\ Energy'').  For $np$ scattering, we
have the relations
\begin{eqnarray*}
M_N &=& \frac{2M_pM_n}{M_p+M_n} = 938.9182 \mbox{ MeV, and} \\ p^2 & =
& \frac{M_p^2 T_{\rm lab} (T_{\rm lab} + 2M_n)} {(M_p + M_n)^2 +
  2T_{\rm lab} M_p} \,,
\end{eqnarray*}
which are based upon relativistic kinematics.

Iteration of $\widehat V$ in the Lippman-Schwinger equation, Eq.~(\ref{eq_LS}),
requires cutting $\widehat V$ off for high momenta to avoid
infinities.  This is consistent with the fact that ChPT is a
low-momentum expansion which is valid only for momenta $Q \ll
\Lambda_\chi \approx 1$ GeV.  Therefore, the potential $\widehat V$ is
multiplied with the regulator function $f(k',k)$,
\begin{equation*}
{\widehat V}(\vec{ k}~',{\vec k}) \longmapsto {\widehat V}(\vec{
  k}~',{\vec k}) \, f(k',k)
\end{equation*}
with
\begin{equation*}
f(p',p) = \exp[-(p'/\Lambda)^{2n}-(p/\Lambda)^{2n}],
\label{eq:eq_f}
\end{equation*}
as a possible example.

Up to NNLO in chiral perturbation theory there are, in addition to the
two-body interaction diagrams discussed above, also a few three-body
interaction diagrams, see Fig.~\ref{fig_diagNNLO}. In chiral
perturbation theory, the orders are generated systematically, and at a
given chiral order the number of Feynman diagrams is
finite. Consistency requires that a calculation includes all diagrams
which are present at the chosen order. 
There are in total five contact terms that determine the strength of
the NNLO three-nucleon force (3NF); $c_1,c_3,$ and $c_4$ are
associated with the three-body two-pion-exchange (2PE) diagram, $c_D$
and $c_E$ determine the strength of the one-pion-exchange plus contact
(1PE) diagram and the pure contact (CNT) diagram,
respectively. The work of Refs.~\cite{epelbaum2002,epelbaum2009} 
gives an extensive discussions of these terms and we refer the reader to these references for further details.

\section{Hartree-Fock theory}\label{sec:chap8hf}

Hartree-Fock (HF) theory is an algorithm for finding an approximative
expression for the ground state of a given Hamiltonian. The basic
ingredients contain a single-particle basis $\{\psi_{\alpha}\}$
defined by the solution of the following eigenvalue problem
\[ 
\hat{h}^{\mathrm{HF}}\psi_{\alpha} =\varepsilon_{\alpha}\psi_{\alpha},
\]
with the Hartree-Fock Hamiltonian defined as
\[
\hat{h}^{\mathrm{HF}}=\hat{t}+\hat{u}_{\mathrm{ext}}+\hat{u}^{\mathrm{HF}}.
\]

The term $\hat{u}^{\mathrm{HF}}$ is a single-particle potential to be
determined by the HF algorithm. The HF algorithm means to select
$\hat{u}^{\mathrm{HF}}$ in order to have
\[ \langle \hat{H} \rangle = E^{\mathrm{HF}}= \langle \Phi_0^{HF} | \hat{H}|\Phi_0^{HF} \rangle,
\]
as a local minimum with a Slater determinant
$\Phi_0^{HF}$ being the ansatz for the ground state.  The variational
principle ensures that $E^{\mathrm{HF}} \ge E_0$, with $E_0$ representing the exact
ground state energy.

We will show that the Hartree-Fock Hamiltonian $\hat{h}^{\mathrm{HF}}$
equals our definition of the operator $\hat{f}$ discussed in
connection with the new definition of the normal-ordered Hamiltonian,
that is we have, for a specific matrix element
\[
\langle p |\hat{h}^{\mathrm{HF}}| q \rangle =\langle p |\hat{f}| q
\rangle=\langle p|\hat{t}+\hat{u}_{\mathrm{ext}}|q \rangle +\sum_{i\le
  F} \langle pi | \hat{V} | qi\rangle,
\]
meaning that
\[
\langle p|\hat{u}^{\mathrm{HF}}|q\rangle = \sum_{i\le F} \langle pi |
\hat{V} | qi\rangle.
\]
The so-called Hartree-Fock potential $\hat{u}^{\mathrm{HF}}$ adds an
explicit medium dependence due to the summation over all
single-particle states below the Fermi level $F$. It brings also in an
explicit dependence on the two-body interaction (in nuclear physics we
can also have complicated three- or higher-body forces). The two-body
interaction, with its contribution from the other bystanding fermions,
creates an effective mean field in which a given fermion moves, in
addition to the external potential $\hat{u}_{\mathrm{ext}}$ which
confines the motion of the fermion. For systems like nuclei or
infinite nuclear matter, there is no external confining
potential. Nuclei and nuclear matter are examples of self-bound
systems, where the binding arises due to the intrinsic nature of the
strong force. For nuclear systems thus, there would be no external
one-body potential in the Hartree-Fock Hamiltonian.

Another possibility is to expand the single-particle functions in a
known basis and vary the coefficients, that is, the new
single-particle wave function is written as a linear expansion in
terms of a fixed chosen orthogonal basis (for example the well-known
harmonic oscillator functions or the hydrogen-like functions etc).  We
define our new Hartree-Fock single-particle basis by performing a
unitary transformation on our previous basis (labelled with greek
indices) as
\begin{equation}
\psi_p^{HF} = \sum_{\lambda}
C_{p\lambda}\phi_{\lambda}. \label{eq:newbasis}
\end{equation}
In this case we vary the coefficients $C_{p\lambda}$. If the basis has
infinitely many solutions, we need to truncate the above sum.  We
assume that the basis $\phi_{\lambda}$ is orthogonal. A unitary
transformation keeps the orthogonality, as discussed in problem
\ref{problem:unitarity} below.

It is normal to choose a single-particle basis defined as the
eigenfunctions of parts of the full Hamiltonian. The typical situation
consists of the solutions of the one-body part of the Hamiltonian,
that is we have
\[
\hat{h}_0\phi_{\lambda}=\epsilon_{\lambda}\phi_{\lambda}.
\]
For infinite nuclear matter $\hat{h}_0$ is given by the kinetic energy
operator and the states are given by plane wave functions. Due to the
translational invariance of the two-body interaction, the Hartree-Fock
single-particle eigenstates are also given by the same functions. For
infinite matter thus, it is only the single-particle energies that
change when we solve the Hartree-Fock equations.

The single-particle wave functions $\phi_{\lambda}({\bf r})$, defined
by the quantum numbers $\lambda$ and ${\bf r}$ are defined as the
overlap
\[
   \phi_{\lambda}({\bf r}) = \langle {\bf r} | \lambda \rangle .
\]
In our discussions we will use our definitions of single-particle
states above and below the Fermi ($F$).

We use greek letters to refer to our original single-particle
basis. The expectation value for the energy with the ansatz $\Phi_0$
for the ground state reads (see problem \ref{problem:referenceE}, with
application to infinite nuclear matter)
\[
  E[\Phi_0] = \sum_{\mu\le F} \langle \mu | h | \mu \rangle +
  \frac{1}{2}\sum_{{\mu},{\nu}\le F} \langle
  \mu\nu|\hat{v}|\mu\nu\rangle.
\]
Now we are interested in defining a new basis defined in terms of a
chosen basis as defined in Eq.~(\ref{eq:newbasis}). We define the energy functional as
\begin{equation}
  E[\Phi^{HF}] = \sum_{i\le F} \langle i | h | i \rangle +
  \frac{1}{2}\sum_{ij\le F}\langle
  ij|\hat{v}|ij\rangle, \label{FunctionalEPhi2}
\end{equation}
where $\Phi^{HF}$ is the new Slater determinant defined by the new
basis of Eq.~(\ref{eq:newbasis}).

Using Eq.~(\ref{eq:newbasis}) we can rewrite
Eq.~(\ref{FunctionalEPhi2}) as
\begin{equation}
  E[\Psi] = \sum_{i\le F} \sum_{\alpha\beta}
  C^*_{i\alpha}C_{i\beta}\langle \alpha | h | \beta \rangle +
  \frac{1}{2}\sum_{ij\le F}\sum_{{\alpha\beta\gamma\delta}}
  C^*_{i\alpha}C^*_{j\beta}C_{i\gamma}C_{j\delta}\langle
  \alpha\beta|\hat{v}|\gamma\delta\rangle. \label{FunctionalEPhi3}
\end{equation}

In order to find the variational minimum of the above functional, we introduce a set of
Lagrange multipliers, noting that since $\langle i | j \rangle =
\delta_{i,j}$ and $\langle \alpha | \beta \rangle =
\delta_{\alpha,\beta}$, the coefficients $C_{i\gamma}$ obey the
relation
\[
 \langle i | j \rangle=\delta_{i,j}=\sum_{\alpha\beta}
 C^*_{i\alpha}C_{i\beta}\langle \alpha | \beta \rangle= \sum_{\alpha}
 C^*_{i\alpha}C_{i\alpha},
\]
which allows us to define a functional to be minimized that reads
\begin{equation}
  F[\Phi^{HF}]=E[\Phi^{HF}] - \sum_{i\le F}\epsilon_i\sum_{\alpha}
  C^*_{i\alpha}C_{i\alpha}.
\end{equation}

Minimizing with respect to $C^*_{i\alpha}$ (the equations for
$C^*_{i\alpha}$ and $C_{i\alpha}$ can be written as two independent
equations) we obtain
\[
\frac{d}{dC^*_{i\alpha}}\left[ E[\Phi^{HF}] -
  \sum_{j}\epsilon_j\sum_{\alpha} C^*_{j\alpha}C_{j\alpha}\right]=0,
\]
which yields for every single-particle state $i$ and index $\alpha$
(recalling that the coefficients $C_{i\alpha}$ are matrix elements of
a unitary matrix, or orthogonal for a real symmetric matrix) the
following Hartree-Fock equations
\[
\sum_{\beta} C_{i\beta}\langle \alpha | h | \beta \rangle+
\sum_{j\le F}\sum_{\beta\gamma\delta}
C^*_{j\beta}C_{j\delta}C_{i\gamma}\langle
\alpha\beta|\hat{v}|\gamma\delta\rangle=\epsilon_i^{HF}C_{i\alpha}.
\]

We can rewrite this equation as (changing dummy variables)
\[
\sum_{\beta} \left\{\langle \alpha | h | \beta \rangle+
\sum_{j\le F}\sum_{\gamma\delta} C^*_{j\gamma}C_{j\delta}\langle
\alpha\gamma|\hat{v}|\beta\delta\rangle\right\}C_{i\beta}=\epsilon_i^{HF}C_{i\alpha}.
\]
Note that the sums over greek indices run over the number of basis set
functions (in principle an infinite number).

Defining
\[
h_{\alpha\beta}^{HF}=\langle \alpha | h | \beta \rangle+
\sum_{j\le F}\sum_{\gamma\delta} C^*_{j\gamma}C_{j\delta}\langle
\alpha\gamma|\hat{v}|\beta\delta\rangle,
\]
we can rewrite the new equations as
\begin{equation}
\sum_{\gamma}h_{\alpha\beta}^{HF}C_{i\beta}=\epsilon_i^{HF}C_{i\alpha}. \label{eq:newhf}
\end{equation}
The latter is nothing but a standard eigenvalue problem.  Our
Hartree-Fock matrix is thus
\[
\hat{h}_{\alpha\beta}^{HF}=\langle \alpha | \hat{h}_0 | \beta \rangle+
\sum_{j\le F}\sum_{\gamma\delta} C^*_{j\gamma}C_{j\delta}\langle
\alpha\gamma|\hat{v}|\beta\delta\rangle.
\]
The Hartree-Fock equations are solved in an iterative way starting
with a guess for the coefficients $C_{j\gamma}=\delta_{j,\gamma}$ and
solving the equations by diagonalization till the new single-particle
energies $\epsilon_i^{\mathrm{HF}}$ do not change anymore by a
user defined small  quantity.

Normally we assume that the single-particle basis $|\beta\rangle$
forms an eigenbasis for the operator $\hat{h}_0$, meaning that the
Hartree-Fock matrix becomes
\[
\hat{h}_{\alpha\beta}^{HF}=\epsilon_{\alpha}\delta_{\alpha,\beta}+
\sum_{j\le F}\sum_{\gamma\delta} C^*_{j\gamma}C_{j\delta}\langle
\alpha\gamma|\hat{v}|\beta\delta\rangle.
\]

\subsection{Hartree-Fock algorithm with simple Python code}

The equations are often rewritten in terms of a so-called density matrix,
which is defined as 
\begin{equation}
\rho_{\gamma\delta}=\sum_{i\le F}\langle\gamma|i\rangle\langle i|\delta\rangle = \sum_{i=1}^{N}C_{i\gamma}C^*_{i\delta}.
\end{equation}
It means that we can rewrite the Hartree-Fock Hamiltonian as
\[
\hat{h}_{\alpha\beta}^{HF}=\epsilon_{\alpha}\delta_{\alpha,\beta}+
\sum_{\gamma\delta} \rho_{\gamma\delta}\langle \alpha\gamma|V|\beta\delta\rangle.
\]
It is convenient to use the density matrix since we can precalculate in every iteration the product of the eigenvector components $C$. 

The Hartree-Fock equations are, in their simplest form, solved in an
iterative way, starting with a guess for the coefficients
$C_{i\alpha}$. We label the coefficients as $C_{i\alpha}^{(n)}$, where
the superscript $n$ stands for iteration $n$.  To set up the algorithm
we can proceed as follows.

\begin{svgraybox}
\begin{enumerate}
\item We start with a guess
  $C_{i\alpha}^{(0)}=\delta_{i,\alpha}$. Alternatively, we could have
  used random starting values as long as the vectors are
  normalized. Another possibility is to give states below the Fermi
  level a larger weight. We construct then the density matrix and the 
Hartree-Fock Hamiltonian. 
\item The Hartree-Fock matrix simplifies then to
\[
\hat{h}_{\alpha\beta}^{HF}(0)=\epsilon_{\alpha}\delta_{\alpha,\beta}+
\sum_{\gamma\delta} \rho_{\gamma\delta}^{(0)}\langle \alpha\gamma|V|\beta\delta\rangle.
\]
Solving the Hartree-Fock eigenvalue problem yields then new eigenvectors $C_{i\alpha}^{(1)}$ and eigenvalues
$\epsilon_i^{\mathrm{HF}}(1)$. 
\item With the new eigenvectors we can set up a new Hartree-Fock potential 
\[
\sum_{\gamma\delta} \rho_{\gamma\delta}^{(1)}\langle \alpha\gamma|V|\beta\delta\rangle.
\]
The diagonalization with the new Hartree-Fock potential yields new eigenvectors and eigenvalues.
\item This process is continued till a user defined test is
  satisfied. As an example, we can require that
\[
\frac{\sum_{p} |\epsilon_i^{(n)}-\epsilon_i^{(n-1)}|}{m} \le \lambda,
\]
where $\lambda$ is a small number defined by the user ($\lambda \sim
10^{-8}$ or smaller) and $p$ runs over all calculated single-particle
energies and $m$ is the number of single-particle states.

\end{enumerate}
\end{svgraybox}
The following simple Python program implements the above algorithm using the density matrix formalism outlined above.
We have omitted the functions that set up the single-particle basis and the anti-symmetrized two-body interaction matrix elements.
These have to be provided, see \url{https://github.com/ManyBodyPhysics/LectureNotesPhysics/tree/master/Programs/Chapter8-programs/python/hfnuclei.py}
for full code and matrix elements.

https://github.com/ManyBodyPhysics/LectureNotesPhysics/tree/master/Programs
\begin{lstlisting}
# We skip here functions that set up the one- and two-body parts of the Hamiltonian 
# These functions need to be defined by the user. The two-body interaction below is
# calculated by calling the function TwoBodyInteraction(alpha,gamma,beta,delta)
# Similalry, the one-body part is computed by the function singleparticleH(alpha)
# We have omitted specific quantum number tests as well (isopsin conservation, 
# momentum conservation etc)
import numpy as np 
from decimal import Decimal

if __name__ == '__main__':
	
	""" Star HF-iterations, preparing variables and density matrix """

        """ Coefficients for setting up density matrix, assuming only one along the diagonals """
	C = np.eye(spOrbitals) # HF coefficients
        DensityMatrix = np.zeros([spOrbitals,spOrbitals])
        for gamma in range(spOrbitals):
            for delta in range(spOrbitals):
                sum = 0.0
                for i in range(Nparticles):
                    sum += C[gamma][i]*C[delta][i]
                DensityMatrix[gamma][delta] = Decimal(sum)
        maxHFiter = 100
        epsilon =  1.0e-5 
        difference = 1.0
	hf_count = 0
	oldenergies = np.zeros(spOrbitals)
	newenergies = np.zeros(spOrbitals)
	while hf_count < maxHFiter and difference > epsilon:
   	        HFmatrix = np.zeros([spOrbitals,spOrbitals])		
		for alpha in range(spOrbitals):
			for beta in range(spOrbitals):
                            """  Setting up the Fock matrix using the density matrix and antisymmetrized two-body interaction  """
     		            sumFockTerm = 0.0
                            for gamma in range(spOrbitals):
                                for delta in range(spOrbitals):
                                    sumFockTerm += DensityMatrix[gamma][delta]*
                                                   TwoBodyInteraction(alpha,gamma,beta,delta)
                            HFmatrix[alpha][beta] = Decimal(sumFockTerm)
                            """  Adding the one-body term """
                            if beta == alpha:   HFmatrix[alpha][alpha] += singleparticleH(alpha)
		spenergies, C = np.linalg.eigh(HFmatrix)
                """ Setting up new density matrix """
                DensityMatrix = np.zeros([spOrbitals,spOrbitals])
                for gamma in range(spOrbitals):
                    for delta in range(spOrbitals):
                        sum = 0.0
                        for i in range(Nparticles):
                            sum += C[gamma][i]*C[delta][i]
                        DensityMatrix[gamma][delta] = Decimal(sum)
		newenergies = spenergies
                """ Brute force computation of difference between previous and new sp HF energies """
                sum =0.0
                for i in range(spOrbitals):
                    sum += (abs(newenergies[i]-oldenergies[i]))/spOrbitals
                difference = sum
                oldenergies = newenergies
                print "Single-particle energies, ordering may have changed "
                for i in range(spOrbitals):
                    print('{0:4d}  {1:.4f}'.format(i, Decimal(oldenergies[i])))
		hf_count += 1
\end{lstlisting}

  We end this section by rewriting the ground state energy by adding
  and subtracting $\hat{u}^{HF}$
  \[
    E_0^{HF} =\langle \Phi_0 | \hat{H} | \Phi_0\rangle = \sum_{i\le
      F}^A \langle i | \hat{h}_0 +\hat{u}^{HF}| j\rangle+
    \frac{1}{2}\sum_{i\le F}^A\sum_{j \le F}^A\left[\langle ij
      |\hat{v}|ij \rangle-\langle
      ij|\hat{v}|ji\rangle\right]-\sum_{i\le F}^A \langle i
    |\hat{u}^{HF}| i\rangle,
  \]
  which results in
  \[
    E_0^{HF} = \sum_{i\le F}^A \varepsilon_i^{HF} +
    \frac{1}{2}\sum_{i\le F}^A\sum_{j \le F}^A\langle ij\vert\hat{v}\vert ij \rangle-\sum_{i\le F}^A \langle i\vert\hat{u}^{HF}\vert i\rangle.
  \]
  Our single-particle states $ijk\dots$ are now single-particle states
  obtained from the solution of the Hartree-Fock equations.

  Using our definition of the Hartree-Fock single-particle energies we
  obtain then the following expression for the total ground-state
  energy
  \[
    E_0^{HF} = \sum_{i\le F}^A \varepsilon_i - \frac{1}{2}\sum_{i\le
      F}^A\sum_{j \le F}^A\left[\langle ij |\hat{v}|ij \rangle-\langle
      ij|\hat{v}|ji\rangle\right].
  \]
  This equation demonstrates that the total energy is not given as the
  sum of the individual single-particle energies.

  \section{Full Configuration Interaction Theory}\label{sec:chap8fci}

  Full configuration theory (FCI), which represents a discretized
  variant of the continuous eigenvalue problem, allows for, in
  principle, an exact (to numerical precision) solution of Schr\"odinger's equation
  for many interacting fermions or bosons with a given basis set. This
  basis set defines an effective Hilbert space.  For fermionic
  problems, the standard approach is to define an upper limit for the
  set of single-particle states. As an example, if we use the harmonic
  oscillator one-body Hamiltonian to generate an orthogonal
  single-particle basis, truncating the basis at some oscillator
  excitation energy provides thereby an upper limit.  Similarly,
  truncating the maximum values of $n_{x,y,z}$ for plane wave states
  with periodic boundary conditions, yields a similar upper limit.
  Table \ref{tab:table1} lists several possible truncations to the
  basis set in terms of the single-particle energies as functions of
  $n_{x,y,z}$.  This single-particle basis is then used to define all
  possible Slater determinants which can be constructed with a given
  number of fermions $A$.  The total number of Slater determinants
  determines thereafter the dimensionality of the Hamiltonian matrix
  and thereby an effective Hilbert space. 
  If we are able to set up the Hamiltonian matrix
  and solve the pertinent eigenvalue problem within this basis set,
  FCI provides numerically exact solutions to all states of interest
  for a given many-body problem. The dimensionality of the problem
  explodes however quickly. To see this it suffices to consider
  the total number of Slater determinants which can be built with say
  $N$ neutrons distributed among $n$ single-particle states. The total number is
  \[
  \left (\begin{array}{c} n \\ N\end{array} \right)
    =\frac{n!}{(n-N)!N!}.
  \]
  As an example, for a model space which comprises the first four
  major harmonic oscillator shells only, that is the $0s$, $0p$,
  $1s0d$ and $1p0f$ shells we have $40$ single particle states for
  neutrons and protons.  For the eight neutrons of oxygen-16 we would
  then have
  \[
  \left (\begin{array}{c} 40 \\ 8\end{array} \right)
    =\frac{40!}{(32)!8!}\sim 8\times 10^{7},
  \]
  possible Slater determinants. Multiplying this with the number of
  proton Slater determinants we end up with approximately $d\sim 10^{15}$ 
  possible Slater determinants and a Hamiltonian matrix of
  dimension $10^{15}\times 10^{15}$, an intractable problem if we wish to diagonalize the Hamiltonian matrix. The
  dimensionality can be reduced if we look at specific symmetries,
  however these symmetries will never reduce the problem to
  dimensionalities which can be handled by standard eigenvalue
  solvers. These are normally lumped into two main categories, direct
  solvers for matrices of dimensionalities which are smaller than
  $d\sim 10^5$, and iterative eigenvalue solvers (when only selected
  states are being sought after) for dimensionalities up to
  $10^{10}\times 10^{10}$.

  Due to its discreteness thus, the effective Hilbert space will
  always represent an approximation to the full continuous problem.
  However, with a given Hamiltonian matrix and effective Hilbert
  space, FCI provides us with true benchmarks that can convey
  important information on correlations beyond Hartree-Fock theory and
  various approximative many-body methods like many-body perturbation
  theory, coupled cluster theory, Green's function theory and the
  Similarity Renormalization Group approach. These methods are all
  discussed in this text.  Assuming that we can diagonalize the
  Hamiltonian matrix, and thereby obtain the exact solutions, this
  section serves the aim to link the exact solution obtained from FCI with various
  approximative methods, hoping thereby that eventual differences can
  shed light on which correlations  play a major role and should be
  included in the above approximative methods. The simple pairing
  model discussed in problem \ref{problem:pairingmodel} is an example
  of a system that allows us to compare exact solutions with
  those defined by many-body perturbation theory to a given order in
  the interaction, coupled cluster theory, Green's function theory and
  the Similarity Renormalization Group (SRG). 

  In order to familiarize the reader with these approximative
  many-body methods, we start with the general definition of the full
  configuration interaction problem.

  We have defined the ansatz for the ground state as
  \[
  |\Phi_0\rangle = \left(\prod_{i\le
    F}\hat{a}_{i}^{\dagger}\right)|0\rangle,
  \]
  where the variable  $i$ defines different single-particle states up to
  the Fermi level. We have assumed that we have $A$ nucleons and that the chosen single-particle basis are eigenstates of 
the one-body Hamiltonian $\hat{h}_0$ (defining thereby an orthogonal basis set).  
A given
  one-particle-one-hole ($1p1h$) state can be written as
  \[
  |\Phi_i^a\rangle = \hat{a}_{a}^{\dagger}\hat{a}_i|\Phi_0\rangle,
  \]
  while a $2p2h$ state can be written as
  \[
  |\Phi_{ij}^{ab}\rangle =
  \hat{a}_{a}^{\dagger}\hat{a}_{b}^{\dagger}\hat{a}_j\hat{a}_i|\Phi_0\rangle,
  \]
  and a general $ApAh$ state as
  \[
  |\Phi_{ijk\dots}^{abc\dots}\rangle =
  \hat{a}_{a}^{\dagger}\hat{a}_{b}^{\dagger}\hat{a}_{c}^{\dagger}\dots\hat{a}_k\hat{a}_j\hat{a}_i|\Phi_0\rangle.
  \]

  As before, we use letters $ijkl\dots$ for states below the Fermi level and
  $abcd\dots$ for states above the Fermi level. A general
  single-particle state is given by letters $pqrs\dots$.

  We can then expand our exact state function for the ground state as
  \[
  |\Psi_0\rangle=C_0|\Phi_0\rangle+\sum_{ai}C_i^a|\Phi_i^a\rangle+\sum_{abij}C_{ij}^{ab}|\Phi_{ij}^{ab}\rangle+\dots
  =(C_0+\hat{C})|\Phi_0\rangle,
  \]
  where we have introduced the so-called correlation operator
  \[
  \hat{C}=\sum_{ai}C_i^a\hat{a}_{a}^{\dagger}\hat{a}_i
  +\sum_{abij}C_{ij}^{ab}\hat{a}_{a}^{\dagger}\hat{a}_{b}^{\dagger}\hat{a}_j\hat{a}_i+\dots
  \]
  Since the normalization of $\Psi_0$ is at our disposal and since
  $C_0$ is by assumption not zero, we may arbitrarily set $C_0=1$ with
  corresponding proportional changes in all other coefficients. Using
  this so-called intermediate normalization we have
  \[
  \langle \Psi_0 | \Phi_0 \rangle = \langle \Phi_0 | \Phi_0 \rangle =
  1,
  \]
  resulting in
  \[
  |\Psi_0\rangle=(1+\hat{C})|\Phi_0\rangle.
  \]

  We rewrite
  \[
  |\Psi_0\rangle=C_0|\Phi_0\rangle+\sum_{ai}C_i^a|\Phi_i^a\rangle+\sum_{abij}C_{ij}^{ab}|\Phi_{ij}^{ab}\rangle+\dots,
  \]
  in a more compact form as
  \[
  |\Psi_0\rangle=\sum_{PH}C_H^P\Phi_H^P=\left(\sum_{PH}C_H^P\hat{A}_H^P\right)|\Phi_0\rangle,
  \]
  where $H$ stands for $0,1,\dots,n$ hole states and $P$ for
  $0,1,\dots,n$ particle states. The operator $\hat{A}_H^P$ represents
  a given set of particle-hole excitations. For a two-particle-to-hole
  excitation this operator is given by
  $\hat{A}_{2h}^{2p}=\hat{a}_{a}^{\dagger}\hat{a}_{b}^{\dagger}\hat{a}_j\hat{a}_i$.
  Our requirement of unit normalization gives
  \[
  \langle \Psi_0 | \Psi_0 \rangle = \sum_{PH}|C_H^P|^2= 1,
  \]
  and the energy can be written as
  \[
  E= \langle \Psi_0 | \hat{H} |\Psi_0 \rangle=
  \sum_{PP'HH'}C_H^{*P}\langle \Phi_H^P | \hat{H} |\Phi_{H'}^{P'}
  \rangle C_{H'}^{P'}.
  \]
  The last equation is normally solved by diagonalization, with the
  Hamiltonian matrix defined by the basis of all possible Slater
  determinants. A diagonalization is equivalent to finding the
  variational minimum of
  \[
   \langle \Psi_0 | \hat{H} |\Psi_0 \rangle-\lambda \langle \Psi_0
   |\Psi_0 \rangle,
  \]
  where $\lambda$ is a variational multiplier to be identified with
  the energy of the system.

  The minimization process results in
  \begin{align}
  0=&\delta\left[ \langle \Psi_0 | \hat{H} |\Psi_0 \rangle-\lambda
    \langle \Psi_0 |\Psi_0 \rangle\right]\\
  =& \sum_{P'H'}\left\{\delta[C_H^{*P}]\langle \Phi_H^P | \hat{H}
  |\Phi_{H'}^{P'} \rangle C_{H'}^{P'}+ C_H^{*P}\langle \Phi_H^P |
  \hat{H} |\Phi_{H'}^{P'} \rangle \delta[C_{H'}^{P'}]- \lambda(
  \delta[C_H^{*P}]C_{H'}^{P'}+C_H^{*P}\delta[C_{H'}^{P'}]\right\}.
  \end{align}
  Since the coefficients $\delta[C_H^{*P}]$ and $\delta[C_{H'}^{P'}]$
  are complex conjugates it is necessary and sufficient to require the
  quantities that multiply with $\delta[C_H^{*P}]$ to vanish.

  This leads to
  \[
  \sum_{P'H'}\langle \Phi_H^P | \hat{H} |\Phi_{H'}^{P'} \rangle
  C_{H'}^{P'}-\lambda C_H^{P}=0,
  \]
  for all sets of $P$ and $H$.

  If we then multiply by the corresponding $C_H^{*P}$ and sum over
  $PH$ we obtain
  \[ 
  \sum_{PP'HH'}C_H^{*P}\langle \Phi_H^P | \hat{H} |\Phi_{H'}^{P'}
  \rangle C_{H'}^{P'}-\lambda\sum_{PH}|C_H^P|^2=0,
  \]
  leading to the identification $\lambda = E$. This means that we have
  for all $PH$ sets
  \begin{equation}
  \sum_{P'H'}\langle \Phi_H^P | \hat{H} -E|\Phi_{H'}^{P'} \rangle =
  0. \label{eq:fullci}
  \end{equation}

  An alternative way to derive the last equation is to start from
  \[
  (\hat{H} -E)|\Psi_0\rangle = (\hat{H}
  -E)\sum_{P'H'}C_{H'}^{P'}|\Phi_{H'}^{P'} \rangle=0,
  \]
  and if this equation is successively projected against all
  $\Phi_H^P$ in the expansion of $\Psi$, we end up with
  Eq.~(\ref{eq:fullci}).

  If we are able to solve this equation by numerical diagonalization in
  a large Hilbert space (it will be truncated in terms of the number
  of single-particle states included in the definition of Slater
  determinants), it can then serve as a benchmark for other many-body
  methods which approximate the correlation operator $\hat{C}$.  Our
  pairing model discussed in problem \ref{problem:pairingmodel} is an
  example of a system which can be diagonalized exactly, providing
  thereby benchmarks for different approximative methods.

  To better understand the meaning of possible configurations and the
  derivation of a Hamiltonian matrix, we consider here a simple
  example of six fermions. We assume we can make an ansatz for the
  ground state with all six fermions below the Fermi level. We 
  label this state as a zero-particle-zero-hole state $0p-0h$. With
  six nucleons we can make at most $6p-6h$ excitations. If we have an
  infinity of single particle states above the Fermi level, we will
  obviously have an infinity of say $2p-2h$ excitations. Each specific way
  to distribute the particles represents a configuration. We will
  always have to truncate  the basis of single-particle states.
  This gives us a finite number of possible Slater determinants. Our
  Hamiltonian matrix would then look like (where each block which is
  marked with an $x$ can contain a large quantity of non-zero matrix
  elements) as shown here
  \begin{table}[h]
  \begin{center}
  \begin{tabular}{cccccccc}
  \hline \multicolumn{1}{c}{ } & \multicolumn{1}{c}{ $0p-0h$ } &
  \multicolumn{1}{c}{ $1p-1h$ } & \multicolumn{1}{c}{ $2p-2h$ } &
  \multicolumn{1}{c}{ $3p-3h$ } & \multicolumn{1}{c}{ $4p-4h$ } &
  \multicolumn{1}{c}{ $5p-5h$ } & \multicolumn{1}{c}{ $6p-6h$ }
  \\ \hline $0p-0h$ & x & x & x & 0 & 0 & 0 & 0 \\ $1p-1h$ & x & x & x
  & x & 0 & 0 & 0 \\ $2p-2h$ & x & x & x & x & x & 0 & 0 \\ $3p-3h$ &
  0 & x & x & x & x & x & 0 \\ $4p-4h$ & 0 & 0 & x & x & x & x & x
  \\ $5p-5h$ & 0 & 0 & 0 & x & x & x & x \\ $6p-6h$ & 0 & 0 & 0 & 0 &
  x & x & x \\ \hline
  \end{tabular}
  \end{center}
  \end{table}
  if the Hamiltonian contains at most a two-body interaction, as
  demonstrated in problem \ref{problem:hamiltoniansetup}.  If we use a
  so-called canonical Hartree-Fock basis \cite{shavittbartlett2009}, this corresponds to a particular unitary
  transformation where matrix elements of the type 
$\langle 0p-0h\vert \hat{H} \vert 1p-1h\rangle =\langle \Phi_0 |\hat{H}|\Phi_{i}^{a}\rangle=0$.
With a canonical Hartree-Fock basis our Hamiltonian matrix reads
\begin{table}[h]
  \begin{center}
  \begin{tabular}{cccccccc}
  \hline \multicolumn{1}{c}{ } & \multicolumn{1}{c}{ $0p-0h$ } &
  \multicolumn{1}{c}{ $1p-1h$ } & \multicolumn{1}{c}{ $2p-2h$ } &
  \multicolumn{1}{c}{ $3p-3h$ } & \multicolumn{1}{c}{ $4p-4h$ } &
  \multicolumn{1}{c}{ $5p-5h$ } & \multicolumn{1}{c}{ $6p-6h$ }
  \\ \hline $0p-0h$ & $\tilde{x}$ & 0 & $\tilde{x}$ & 0 & 0 & 0 & 0
  \\ $1p-1h$ & 0 & $\tilde{x}$ & $\tilde{x}$ & $\tilde{x}$ & 0 & 0 & 0
  \\ $2p-2h$ & $\tilde{x}$ & $\tilde{x}$ & $\tilde{x}$ & $\tilde{x}$ &
  $\tilde{x}$ & 0 & 0 \\ $3p-3h$ & 0 & $\tilde{x}$ & $\tilde{x}$ &
  $\tilde{x}$ & $\tilde{x}$ & $\tilde{x}$ & 0 \\ $4p-4h$ & 0 & 0 &
  $\tilde{x}$ & $\tilde{x}$ & $\tilde{x}$ & $\tilde{x}$ & $\tilde{x}$
  \\ $5p-5h$ & 0 & 0 & 0 & $\tilde{x}$ & $\tilde{x}$ & $\tilde{x}$ &
  $\tilde{x}$ \\ $6p-6h$ & 0 & 0 & 0 & 0 & $\tilde{x}$ & $\tilde{x}$ &
  $\tilde{x}$ \\ \hline
  \end{tabular}
  \end{center}
  \end{table}

  If we do not make any truncations in the possible sets of Slater
  determinants (many-body states) we can make by distributing $A$
  nucleons among $n$ single-particle states, we call such a
  calculation for a full configuration interaction (FCI) approach.  If
  we make truncations, we have several different possibilities to
  reduce the dimensionality of the problem.  A well-known example is
  the standard nuclear shell-model. For the nuclear shell model we define an
  effective Hilbert space with respect to a given core. The
  calculations are normally then performed for all many-body states
  that can be constructed from the effective Hilbert spaces. This
  approach requires a properly defined effective Hamiltonian.  Another
  possibility to constrain the dimensionality of the problem is to
  truncate in the number of excitations. As an example, we can limit
  the possible Slater determinants to only $1p-1h$ and $2p-2h$
  excitations. This is called a configuration interaction calculation
  at the level of singles and doubles excitations. If we truncate at
  the level of three-particle-three-hole excitations we end up with
  singles, doubles and triples excitations.  Such truncations reduce
  considerably the size of the Hamiltonian matrices to be
  diagonalized, but can lead to so-called unlinked contributions, and
  thereby wrong results, for a given expectation value
  \cite{bartlett2007}.  A third possibility is to constrain the
  number of excitations by an energy cutoff. This cutoff defines a
  maximum excitation energy. The maximum excitation energy is normally given by the 
sum of single-particle energies defined by the unperturbed one-body part of the Hamiltonian.
A commonly used basis in nuclear physics is the harmonic oscillator. The 
cutoff in energy is then defined by the maximum number of harmonic oscillator excitations.
If we do not define a core, this defines normally what is
  called the no-core shell-model approach, see for example Refs.~\cite{navratil2009,navratil2016}.

  \subsection{A non-practical way of solving the eigenvalue problem}

  For reasons to come (links with coupled cluster theory and many-body
  perturbation theory), we will rewrite Eq.~(\ref{eq:fullci}) as a set
  of coupled non-linear equations in terms of the unknown coefficients
  $C_H^P$.  To obtain the eigenstates and eigenvalues in terms of
  non-linear equations is less efficient than using standard eigenvalue solvers \cite{golubvanloan}.
  However, this digression serves the scope
  of linking full configuration interaction theory with approximative
  solutions to the many-body problem.

  To see this, we look at the contributions arising from
  \[
  \langle \Phi_H^P | = \langle \Phi_0|
  \]
  in Eq.~(\ref{eq:fullci}), that is we multiply with $\langle \Phi_0|$ from the left in
  \[
  (\hat{H} -E)\sum_{P'H'}C_{H'}^{P'}|\Phi_{H'}^{P'} \rangle=0.
  \]
  If we assume that we have a two-body operator at most, the
  Slater-Condon rule for a two-body interaction, see problem
  \ref{problem:hamiltoniansetup}, results in an expression for the
  correlation energy in terms of $C_i^a$ and $C_{ij}^{ab}$ only, namely
  \[
  \langle \Phi_0 | \hat{H} -E| \Phi_0\rangle + \sum_{ai}\langle \Phi_0
  | \hat{H} -E|\Phi_{i}^{a} \rangle C_{i}^{a}+ \sum_{abij}\langle
  \Phi_0 | \hat{H} -E|\Phi_{ij}^{ab} \rangle C_{ij}^{ab}=0,
  \]
  or
  \[
  E-E_{\mathrm{Ref}} =\Delta E=\sum_{ai}\langle \Phi_0 |
  \hat{H}|\Phi_{i}^{a} \rangle C_{i}^{a}+ \sum_{abij}\langle \Phi_0 |
  \hat{H}|\Phi_{ij}^{ab} \rangle C_{ij}^{ab},
  \]
  where the energy $E_{\mathrm{Ref}}$ is the reference energy and
  $\Delta E$ defines the so-called correlation energy.  The
  single-particle basis functions could  result from  a
  Hartree-Fock calculation or they could be the eigenstates of the one-body operator that defined the
  non-interacting part of the Hamiltonian.

  In our Hartree-Fock discussions, we have already computed
  the matrix $\langle \Phi_0 | \hat{H}|\Phi_{i}^{a}\rangle $ and
  $\langle \Phi_0 | \hat{H}|\Phi_{ij}^{ab}\rangle$.  If we are using a
  Hartree-Fock basis we have $\langle \Phi_0 | \hat{H}|\Phi_{i}^{a}\rangle=0$
  and we are left with a \emph{correlation energy} given by
  \[
  E-E_{\mathrm{Ref}} =\Delta E^{HF}=\sum_{abij}\langle \Phi_0 |
  \hat{H}|\Phi_{ij}^{ab} \rangle C_{ij}^{ab}.
  \]

  Inserting the various matrix elements we can rewrite the previous
  equation as
  \begin{equation}\label{eq:correlationenergy}
  \Delta E=\sum_{ai}\langle i| \hat{f}|a \rangle C_{i}^{a}+
  \sum_{abij}\langle ij | \hat{v}| ab \rangle C_{ij}^{ab}.
  \end{equation}
  This equation determines the correlation energy but not the
  coefficients $C$.  We need more equations. Our next step is to set
  up
  \[ 
  \langle \Phi_i^a | \hat{H} -E| \Phi_0\rangle + \sum_{bj}\langle
  \Phi_i^a | \hat{H} -E|\Phi_{j}^{b} \rangle C_{j}^{b}+
  \sum_{bcjk}\langle \Phi_i^a | \hat{H} -E|\Phi_{jk}^{bc} \rangle
  C_{jk}^{bc}+ \sum_{bcdjkl}\langle \Phi_i^a | \hat{H}
  -E|\Phi_{jkl}^{bcd} \rangle C_{jkl}^{bcd}=0,
  \]
  as this equation will allow us to find an expression for the
  coefficents $C_i^a$  through 
  \begin{equation}\label{eq:c1p1h}
  \langle i | \hat{f}| a\rangle +\langle \Phi_i^a |
  \hat{H}|\Phi_{i}^{a} \rangle C_{i}^{a}+ \sum_{bj\ne ai}\langle
  \Phi_i^a | \hat{H}|\Phi_{j}^{b} \rangle C_{j}^{b}+
  \sum_{bcjk}\langle \Phi_i^a | \hat{H}|\Phi_{jk}^{bc} \rangle
  C_{jk}^{bc}+ \sum_{bcdjkl}\langle \Phi_i^a |
  \hat{H}|\Phi_{jkl}^{bcd} \rangle C_{jkl}^{bcd}=EC_i^a.
  \end{equation}

  We see that on the right-hand side we have the energy $E$. This
  leads to a non-linear equation in the unknown coefficients since the coefficients appear also in the definition of the correlation
energy of Eq.~(\ref{eq:correlationenergy}).  These
  equations are normally solved iteratively, that is we  start
  with a guess for the coefficients $C_i^a$. A common choice is to
  use perturbation theory as a starting point for the unknown coefficients. For the one-particle-one-hole coefficients, the wave operator
(see section \ref{sec:chap8mbpt}) to first order in the interaction is given by
  \[
   C_{i}^{a}=\frac{\langle i | \hat{f}|
     a\rangle}{\epsilon_i-\epsilon_a}.
  \]

  The observant reader will however see that we need an equation for
  $C_{jk}^{bc}$ and $C_{jkl}^{bcd}$ and more complicated particle-hole excitations as well.  To find the equations for
  these coefficients we need then to continue our multiplications from
  the left with the various $\Phi_{H}^P$ terms.

  For $C_{jk}^{bc}$ we have
  \begin{align}\label{eq:c2p2h}
  0=&\langle \Phi_{ij}^{ab} | \hat{H} -E| \Phi_0\rangle +
  \sum_{kc}\langle \Phi_{ij}^{ab} | \hat{H} -E|\Phi_{k}^{c} \rangle
  C_{k}^{c}+ \\
  &\sum_{cdkl}\langle \Phi_{ij}^{ab} | \hat{H} -E|\Phi_{kl}^{cd}
  \rangle C_{kl}^{cd}+\sum_{cdeklm}\langle \Phi_{ij}^{ab} | \hat{H}
  -E|\Phi_{klm}^{cde} \rangle C_{klm}^{cde}+\sum_{cdefklmn}\langle
  \Phi_{ij}^{ab} | \hat{H} -E|\Phi_{klmn}^{cdef} \rangle
  C_{klmn}^{cdef}.
  \end{align}
  We can isolate the coefficients $C_{kl}^{cd}$ in a similar way
  as we did for the coefficients $C_{i}^{a}$.  A standard choice for
  the first iteration is to use again perturbation theory to first order in the interaction and set
  \[
  C_{ij}^{ab} =\frac{\langle ij \vert \hat{v} \vert ab
    \rangle}{\epsilon_i+\epsilon_j-\epsilon_a-\epsilon_b}.
  \]
  At the end we can rewrite our solution of the Schr\"odinger equation
  in terms of a series coupled equations for the coefficients $C_H^P$.
  This is a very cumbersome way of solving a many-body
  problem. However, by using this iterative scheme we can illustrate
  how we can compute the various terms in the wave operator or
  correlation operator $\hat{C}$. We will later identify the
  calculation of the various terms $C_H^P$ as parts of different
  many-body approximations to full configuration interaction theory.

  \subsection{Short summary}

  If we can directly diagonalize large matrices, full configuration interaction
  theory is the method of choice since  we obtain all  eigenvectors and eigenvalues. 
  The eigenvectors are obtained directly from the coefficients
    $C_H^P$ which result from the diagonalization.  We can then
compute expectation values of other operators,
    as well as transition probabilities. Moreover, correlations are easy to understand in terms of contributions
    to a given operator beyond the Hartree-Fock contribution. 
For larger dimensionalities $d$, with $d > 10^5$, iterative methods \cite{golubvanloan} like Lanczos' \cite{lanczos} or Davidson's \cite{davidson1989,davidson1993} 
algorithms are frequently used. These methods yield, with a finite number of iteration, only a subset of all eigenvalues of interest. Lanczos' algorithm converges to the extreme values, yielding the lowest-lying and highest-lying eigenstates, see for example Ref.~\cite{golubvanloan} for a proof. 

With the eigenvectors we can compute
  the correlation energy, which is defined as (with a two-body Hamiltonian)
  \[
  \Delta E=\sum_{ai}\langle i| \hat{f}|a \rangle C_{i}^{a}+
  \sum_{abij}\langle ij | \hat{v}| ab \rangle C_{ij}^{ab}.
  \]
The energy of  the ground state is then
  \[
  E=E_{\mathrm{Ref}}+\Delta E.
  \]
  However, as we have seen, even for a
  small case like the four first major shells and 
  oxygen-16 with 16 active nucleons, the dimensionality becomes quickly intractable. If we
  wish to include single-particle states that reflect weakly bound
  systems, we need a much larger single-particle basis. We need thus
  approximative methods that sum specific correlations to infinite
  order.  All these methods start normally with a Hartree-Fock basis
  as the calculational basis. In the next section we discuss one of
  these possible approximative methods, namely many-body perturbation
  theory.

  \section{Many-body perturbation theory}\label{sec:chap8mbpt}

  We assume here that we are only interested in the non-degenerate
  ground state of a given system and expand the exact wave function in
  terms of a series of Slater determinants
  \[
  \vert \Psi_0\rangle = \vert \Phi_0\rangle +
  \sum_{m=1}^{\infty}C_m\vert \Phi_m\rangle,
  \]
  where we have assumed that the true ground state is dominated by the
  solution of the unperturbed problem, that is
  \[
  \hat{H}_0\vert \Phi_0\rangle= W_0\vert \Phi_0\rangle.
  \]
  The state $\vert \Psi_0\rangle$ is not normalized and we employ again
  intermediate normalization via $\langle \Phi_0 \vert
  \Psi_0\rangle=1$.

  The Schr\"odinger equation is given by
  \[
  \hat{H}\vert \Psi_0\rangle = E\vert \Psi_0\rangle,
  \]
  and multiplying the latter from the left with $\langle \Phi_0\vert $
  gives
  \[
  \langle \Phi_0\vert \hat{H}\vert \Psi_0\rangle = E\langle
  \Phi_0\vert \Psi_0\rangle=E,
  \]
  and subtracting from this equation
  \[
  \langle \Psi_0\vert \hat{H}_0\vert \Phi_0\rangle= W_0\langle
  \Psi_0\vert \Phi_0\rangle=W_0,
  \]
  and using the fact that the  operators $\hat{H}$ and $\hat{H}_0$
  are hermitian results in
  \begin{equation}\label{eq:mbptcorrel}
  \Delta E=E-W_0=\langle \Phi_0\vert \hat{H}_I\vert \Psi_0\rangle,
  \end{equation}
  which is an exact result. This resembles our previous definition of the correlation energy except that the reference energy is now defined
by the unperturbed energy $W_0$. The reader should contrast this equation to our previous definition of the correlation energy
  \[
  \Delta E=\sum_{ai}\langle i| \hat{f}|a \rangle C_{i}^{a}+
  \sum_{abij}\langle ij | \hat{v}| ab \rangle C_{ij}^{ab},
  \]
and the total energy
  \[
  E=E_{\mathrm{Ref}}+\Delta E,
  \]
where the reference energy is given by
\[   
   E_{\mathrm{Ref}}= \langle \Phi_0 \vert \hat{H} \vert \Phi_0\rangle.
\]

  Equation (\ref{eq:mbptcorrel}) forms the starting point for all perturbative
  derivations. However, as it stands it represents nothing but a mere
  formal rewriting of Schr\"odinger's equation and is not of much
  practical use. The exact wave function $\vert \Psi_0\rangle$ is
  unknown. In order to obtain a perturbative expansion, we need to
  expand the exact wave function in terms of the interaction
  $\hat{H}_I$.

  Here we have assumed that our model space defined by the operator
  $\hat{P}$ is one-dimensional, meaning that
  \[
  \hat{P}= \vert \Phi_0\rangle \langle \Phi_0\vert ,
  \]
  and
  \[
  \hat{Q}=\sum_{m=1}^{\infty}\vert \Phi_m\rangle \langle \Phi_m\vert .
  \]

  We can thus rewrite the exact wave function as
  \[
  \vert \Psi_0\rangle= (\hat{P}+\hat{Q})\vert \Psi_0\rangle=\vert
  \Phi_0\rangle+\hat{Q}\vert \Psi_0\rangle.
  \]
  Going back to the Schr\"odinger equation, we can rewrite it as,
  adding and a subtracting a term $\omega \vert \Psi_0\rangle$ as
  \[
  \left(\omega-\hat{H}_0\right)\vert
  \Psi_0\rangle=\left(\omega-E+\hat{H}_I\right)\vert \Psi_0\rangle,
  \]
  where $\omega$ is an energy variable to be specified later.

  We assume also that the resolvent of $\left(\omega-\hat{H}_0\right)$
  exits, that is it has an inverse which defines the unperturbed
  Green's function as
  \[
  \left(\omega-\hat{H}_0\right)^{-1}=\frac{1}{\left(\omega-\hat{H}_0\right)}.
  \]

  We can rewrite Schr\"odinger's equation as
  \[
  \vert
  \Psi_0\rangle=\frac{1}{\omega-\hat{H}_0}\left(\omega-E+\hat{H}_I\right)\vert
  \Psi_0\rangle,
  \]
  and multiplying from the left with $\hat{Q}$ results in
  \[
  \hat{Q}\vert
  \Psi_0\rangle=\frac{\hat{Q}}{\omega-\hat{H}_0}\left(\omega-E+\hat{H}_I\right)\vert
  \Psi_0\rangle,
  \]
  which is possible since we have defined the operator $\hat{Q}$ in
  terms of the eigenfunctions of $\hat{H}_0$.

 Since these operators commute we have
  \[
  \hat{Q}\frac{1}{\left(\omega-\hat{H}_0\right)}\hat{Q}=\hat{Q}\frac{1}{\left(\omega-\hat{H}_0\right)}=\frac{\hat{Q}}{\left(\omega-\hat{H}_0\right)}.
  \]
  With these definitions we can in turn define the wave function as
  \[
  \vert \Psi_0\rangle=\vert
  \Phi_0\rangle+\frac{\hat{Q}}{\omega-\hat{H}_0}\left(\omega-E+\hat{H}_I\right)\vert
  \Psi_0\rangle.
  \]
  This equation is again nothing but a formal rewrite of
  Schr\"odinger's equation and does not represent a practical
  calculational scheme.  It is a non-linear equation in two unknown
  quantities, the energy $E$ and the exact wave function $\vert
  \Psi_0\rangle$. We can however start with a guess for $\vert
  \Psi_0\rangle$ on the right hand side of the last equation.

   The most common choice is to start with the function which is
   expected to exhibit the largest overlap with the wave function we
   are searching after, namely $\vert \Phi_0\rangle$. This can again
   be inserted in the solution for $\vert \Psi_0\rangle$ in an
   iterative fashion and if we continue along these lines we end up
   with
  \[
  \vert
  \Psi_0\rangle=\sum_{i=0}^{\infty}\left\{\frac{\hat{Q}}{\omega-\hat{H}_0}\left(\omega-E+\hat{H}_I\right)\right\}^i\vert
  \Phi_0\rangle,
  \]
  for the wave function and
  \[
  \Delta E=\sum_{i=0}^{\infty}\langle \Phi_0\vert
  \hat{H}_I\left\{\frac{\hat{Q}}{\omega-\hat{H}_0}\left(\omega-E+\hat{H}_I\right)\right\}^i\vert
  \Phi_0\rangle,
  \]
  which is now a perturbative expansion of the exact energy in terms
  of the interaction $\hat{H}_I$ and the unperturbed wave function
  $\vert \Psi_0\rangle$.

  In our equations for $\vert \Psi_0\rangle$ and $\Delta E$ in terms
  of the unperturbed solutions $\vert \Phi_i\rangle$ we have still an
  undetermined parameter $\omega$ and a dependecy on the exact energy
  $E$. Not much has been gained thus from a practical computational
  point of view.

  In Brilluoin-Wigner perturbation theory \cite{shavittbartlett2009} it is customary to set
  $\omega=E$. This results in the following perturbative expansion for
  the energy $\Delta E$
  \begin{align}
  \Delta E=&\sum_{i=0}^{\infty}\langle \Phi_0\vert
  \hat{H}_I\left\{\frac{\hat{Q}}{\omega-\hat{H}_0}\left(\omega-E+\hat{H}_I\right)\right\}^i\vert
  \Phi_0\rangle=\\ &\langle \Phi_0\vert
  \left(\hat{H}_I+\hat{H}_I\frac{\hat{Q}}{E-\hat{H}_0}\hat{H}_I+
  \hat{H}_I\frac{\hat{Q}}{E-\hat{H}_0}\hat{H}_I\frac{\hat{Q}}{E-\hat{H}_0}\hat{H}_I+\dots\right)\vert
  \Phi_0\rangle.
  \end{align}
  This expression depends however on the exact energy $E$ and is again
  not very convenient from a practical point of view. It can obviously
  be solved iteratively, by starting with a guess for $E$ and then
  solve till some kind of self-consistency criterion has been reached.

  Defining $e=E-\hat{H}_0$ and recalling that $\hat{H}_0$ commutes
  with $\hat{Q}$ by construction and that $\hat{Q}$ is an idempotent
  operator $\hat{Q}^2=\hat{Q}$, we can rewrite the denominator in the above
  expansion for $\Delta E$ as
  \[
  \hat{Q}\frac{1}{\hat{e}-\hat{Q}\hat{H}_I\hat{Q}}=\hat{Q}\left[\frac{1}{\hat{e}}+\frac{1}{\hat{e}}\hat{Q}\hat{H}_I\hat{Q}
    \frac{1}{\hat{e}}+\frac{1}{\hat{e}}\hat{Q}\hat{H}_I\hat{Q}
    \frac{1}{\hat{e}}\hat{Q}\hat{H}_I\hat{Q}\frac{1}{\hat{e}}+\dots\right]\hat{Q}.
  \]

  Inserted in the expression for $\Delta E$ we obtain
  \[
  \Delta E= \langle \Phi_0\vert
  \hat{H}_I+\hat{H}_I\hat{Q}\frac{1}{E-\hat{H}_0-\hat{Q}\hat{H}_I\hat{Q}}\hat{Q}\hat{H}_I\vert
  \Phi_0\rangle.
  \]
  In Rayleigh-Schr\"odinger (RS) perturbation theory \cite{shavittbartlett2009} we set $\omega = W_0$ and obtain the
  following expression for the energy difference
  \begin{align}
  \Delta E =& \sum_{i=0}^{\infty}\langle \Phi_0\vert
  \hat{H}_I\left\{\frac{\hat{Q}}{W_0-\hat{H}_0}\left(\hat{H}_I-\Delta
  E\right)\right\}^i\vert \Phi_0\rangle\\ & \langle \Phi_0\vert
  \left(\hat{H}_I+\hat{H}_I\frac{\hat{Q}}{W_0-\hat{H}_0}(\hat{H}_I-\Delta
  E)+ \hat{H}_I\frac{\hat{Q}}{W_0-\hat{H}_0}(\hat{H}_I-\Delta
  E)\frac{\hat{Q}}{W_0-\hat{H}_0}(\hat{H}_I-\Delta
  E)+\dots\right)\vert \Phi_0\rangle.
  \end{align}

  The operator $\hat{Q}$ commutes with $\hat{H_0}$ and since $\Delta
  E$ is a constant we obtain that
  \[
  \hat{Q}\Delta E\vert \Phi_0\rangle = \hat{Q}\Delta E\vert
  \hat{Q}\Phi_0\rangle = 0.
  \]
  Inserting this result in the expression for the energy gives us
  \[
  \Delta E=\langle \Phi_0\vert
  \left(\hat{H}_I+\hat{H}_I\frac{\hat{Q}}{W_0-\hat{H}_0}\hat{H}_I+
  \hat{H}_I\frac{\hat{Q}}{W_0-\hat{H}_0}(\hat{H}_I-\Delta
  E)\frac{\hat{Q}}{W_0-\hat{H}_0}\hat{H}_I+\dots\right)\vert
  \Phi_0\rangle.
  \]

\begin{figure}[hbtp]
    \includegraphics[width=0.7\linewidth,angle=90]{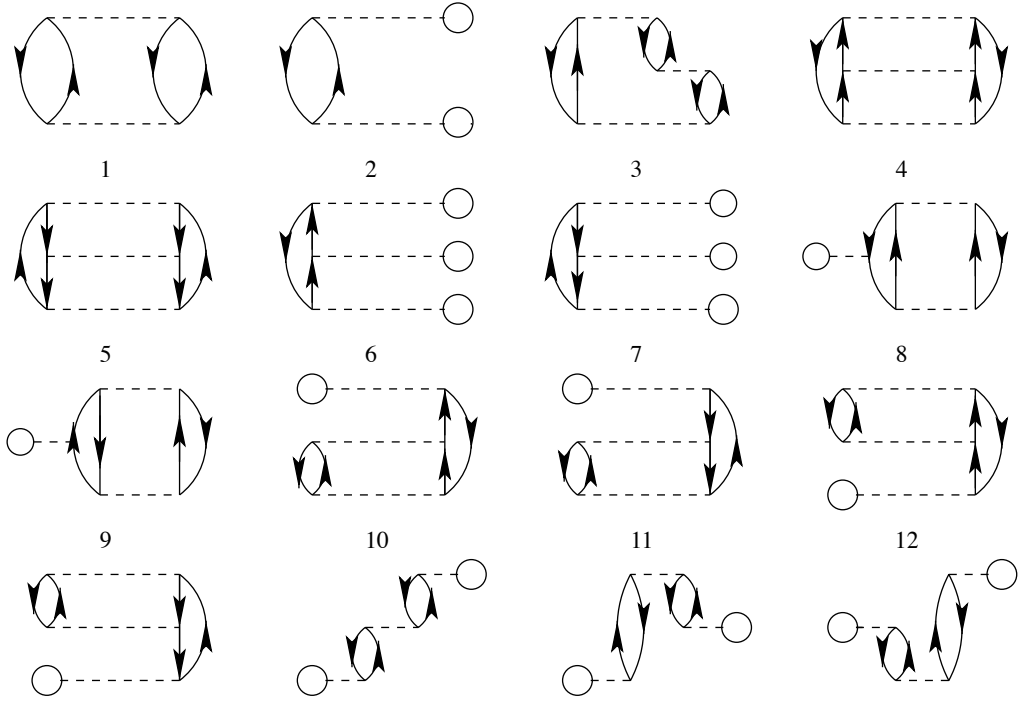}
      \caption{Diagrams which enter the definition of the ground-state
      correlation energy $\Delta E_0$ to third order in the interaction. We have not included the first-order contribution.}
      \label{fig:goldstone}
\end{figure}

  We can now perturbatively expand this expression in terms of the interaction
   $\hat{H}_I$, which is assumed to be small. We obtain then
  \[
  \Delta E=\sum_{i=1}^{\infty}\Delta E^{(i)},
  \]
  with the following expression for $\Delta E^{(i)}$
  \[
  \Delta E^{(1)}=\langle \Phi_0\vert \hat{H}_I\vert \Phi_0\rangle,
  \] 
  which is just the contribution to first order in perturbation
  theory,
  \[
  \Delta E^{(2)}=\langle\Phi_0\vert
  \hat{H}_I\frac{\hat{Q}}{W_0-\hat{H}_0}\hat{H}_I\vert \Phi_0\rangle,
  \]
  which is the contribution to second order and
  \[
  \Delta E^{(3)}=\langle \Phi_0\vert
  \hat{H}_I\frac{\hat{Q}}{W_0-\hat{H}_0}\hat{H}_I\frac{\hat{Q}}{W_0-\hat{H}_0}\hat{H}_I\Phi_0\rangle-
  \langle\Phi_0\vert \hat{H}_I\frac{\hat{Q}}{W_0-\hat{H}_0}\langle
  \Phi_0\vert \hat{H}_I\vert
  \Phi_0\rangle\frac{\hat{Q}}{W_0-\hat{H}_0}\hat{H}_I\vert
  \Phi_0\rangle,
  \]
  being the third-order contribution.
  There exists a formal theory for the calculation
  of $\Delta E_0$, see for example Ref.~\cite{shavittbartlett2009}.  According to the well-known Goldstone
  linked-diagram theory, the energy shift $\Delta E_0$ is given
  exactly by the diagrammatic expansion shown in
  Fig.~\ref{fig:goldstone}, where ground state diagrams to third order are listed. This theory is a linked-cluster
  perturbation expansion for the ground state energy of a many-body
  system, and applies equally well to both nuclear matter and
  closed-shell nuclei such as the doubly magic nucleus $^{40}$Ca.  
We assume the reader is familiar 
with the standard rules for deriving and setting up the analytical expressions for various Feymann-Goldstone diagrams \cite{shavittbartlett2009}.
In an infinite system like nuclear matter or the homogenous electron gas, all diagrams with so-called Hartree-Fock insertions like diagrams (2), (6), (7), (10-16) are zero 
due to lack of momentum conservation. They would also be zero in case a canonical \cite{shavittbartlett2009} Hartree-Fock basis is employed. 

  Using the above standard diagram rules, the various
  diagrams contained in Fig.~\ref{fig:goldstone} can be readily calculated (in
  an uncoupled scheme). Diagram  (1) results in
  \begin{equation}
     (1)=\frac{1}{2^2}\sum_{ij\leq F}\sum_{ab>F}
    \frac{\langle ij\vert\hat{v}\vert ab\rangle \langle
      ab\vert\hat{v}\vert ij\rangle}
         {\varepsilon_i+\varepsilon_j-\varepsilon_a-\varepsilon_b},
  \end{equation}
 while diagram (2) is zero due to lack of momentum conservation. We have two factors of $1/2$ since there are two equivalent pairs of fermions (two  particle states and two hole states) starting at the same vertex and ending at the same vertex. The expression for diagram (3)  is 
  \begin{equation}\label{eq:diag3}
     (3)=\sum_{ijk\leq k_F}\sum_{abc > F}
    \frac{\langle ij\vert\hat{v}\vert ab\rangle \langle
      bk\vert\hat{v}\vert ic\rangle \langle ac\vert\hat{v}\vert
      ik\rangle}
         {(\varepsilon_i+\varepsilon_j-\varepsilon_a-\varepsilon_b)
           (\varepsilon_i+\varepsilon_k-\varepsilon_a-\varepsilon_c)}.
  \end{equation}
  Diagrams (4) and (5) read
  \begin{equation}\label{eq:diag4}
     (4)=\frac{1}{2^3}\sum_{ij\le F}\sum_{abcd > F}
    \frac{\langle ij\vert\hat{v}\vert cd\rangle \langle
      cd\vert\hat{v}\vert ab\rangle \langle ab\vert\hat{v}\vert
      ij\rangle}
         {(\varepsilon_i+\varepsilon_j-\varepsilon_c-\varepsilon_d)
           (\varepsilon_i+\varepsilon_j-\varepsilon_a-\varepsilon_b)},
  \end{equation}
  \begin{equation}\label{eq:diag5}
     (5)=\frac{1}{2^3}\sum_{ijkl\le F}\sum_{ab > F}
    \frac{\langle ab\vert\hat{v}\vert kl\rangle \langle
      kl\vert\hat{v}\vert ij\rangle \langle ij\vert\hat{v}\vert
      ab\rangle}
         {(\varepsilon_i+\varepsilon_j-\varepsilon_a-\varepsilon_b)
           (\varepsilon_k+\varepsilon_l-\varepsilon_a-\varepsilon_b)},
  \end{equation}
where the factor $(1/2)^3$ arises due to three equivalent pairs of lines starting and ending at the same vertex. The last two contributions 
have an even number of hole lines and closed loops, resulting thus in a positive sign. In problem \ref{problem:diagrams}, you are asked to calculate the expressions for diagrams like (8) and (9) in the above figure. 

  In the expressions for the various diagrams the quantity $\varepsilon$ denotes the single-particle energies
  defined by $H_0$.  The steps leading to the above expressions for
  the various diagrams are rather straightforward. Though, if we wish
  to compute the matrix elements for the interaction $\hat{v}$, a serious
  problem arises. Typically, the matrix elements will contain a term
  $V(|{\mathbf r}|)$
  which represents the interaction potential $V$ between two nucleons,
  where ${\mathbf r}$ is the internucleon distance.  All modern models
  for $V$ have a strong short-range repulsive core. Hence, matrix
  elements involving $V(|{\mathbf r}|)$, will result in large (or
  infinitely large for a potential with a hard core) and repulsive
  contributions to the ground-state energy. 
  A perturbative expansion in terms of
such interaction matrix elements may thus lead to a slowly converging expansion.
A standard recipe to circumvent such problems has been to sum up 
a selected class of correlations. We discuss such possibilities in section \ref{sec:chap8cctheory}.

  \subsection{Interpreting the correlation energy and the wave operator}

  In section \ref{sec:chap8fci} we showed that we could rewrite the exact state
  function for the ground state as a linear expansion in terms of all
  possible Slater determinants.  We expanded  our
  exact state function for the ground state as
  \[
  |\Psi_0\rangle=C_0|\Phi_0\rangle+\sum_{ai}C_i^a|\Phi_i^a\rangle+\sum_{abij}C_{ij}^{ab}|\Phi_{ij}^{ab}\rangle+\dots
  =(C_0+\hat{C})|\Phi_0\rangle,
  \]
  where we introduced the so-called correlation operator
  \[
  \hat{C}=\sum_{ai}C_i^a\hat{a}_{a}^{\dagger}\hat{a}_i
  +\sum_{abij}C_{ij}^{ab}\hat{a}_{a}^{\dagger}\hat{a}_{b}^{\dagger}\hat{a}_j\hat{a}_i+\dots
  \]
  In a shell-model calculation, the unknown coefficients in $\hat{C}$
  are the eigenvectors that result from the diagonalization of the
  Hamiltonian matrix.

  How can we use perturbation theory to determine the same
  coefficients? Let us study the contributions to second order in the
  interaction, namely
  \[
  \Delta E^{(2)}=\langle\Phi_0\vert
  \hat{H}_I\frac{\hat{Q}}{W_0-\hat{H}_0}\hat{H}_I\vert \Phi_0\rangle.
  \]
  This contribution will also be discussed in connection with the
  development of a many-body program for nuclear matter, as well as
  the simple pairing model of problem \ref{problem:pairingmodel}.  The
  intermediate states given by $\hat{Q}$ can at most be of a $2p-2h$
  nature if we have a two-body Hamiltonian. This means that to second
  order in perturbation theory we can at most have $1p-1h$ and $2p-2h$
  excitations as intermediate states. When we diagonalize, these
  contributions are included to infinite order. This means that in
  order to include such correlations to higher order in the
  interaction, we need to go to higher-orders in perturbation theory.

  If we limit the attention to a Hartree-Fock basis, we have that
  $\langle\Phi_0\vert \hat{H}_I \vert 2p-2h\rangle$ is the only
  contribution since matrix elements involving $\langle\Phi_0\vert
  \hat{H}_I \vert 1p-1h\rangle$ are zero and the contribution to the
  energy from second order in Rayleigh-Schr\"odinger perturbation
  theory reduces to
  \[
  \Delta E^{(2)}=\frac{1}{4}\sum_{abij}\langle ij\vert \hat{v}\vert
  ab\rangle \frac{\langle ab\vert \hat{v}\vert
    ij\rangle}{\epsilon_i+\epsilon_j-\epsilon_a-\epsilon_b}.
  \]
Here we have used the results from problem \ref{problem:hamiltoniansetup}. 
  If we compare this to the correlation energy obtained from full
  configuration interaction theory with a Hartree-Fock basis, we found
  that
  \[
  E-E_{\mathrm{Ref}} =\Delta E=\sum_{abij}\langle ij | \hat{v}| ab
  \rangle C_{ij}^{ab},
  \]
  where the energy $E_{\mathrm{Ref}}$ is the reference energy and
  $\Delta E$ defines the so-called correlation energy.

  We see that if we set
  \[
  C_{ij}^{ab} =\frac{1}{4}\frac{\langle ab \vert \hat{v} \vert ij
    \rangle}{\epsilon_i+\epsilon_j-\epsilon_a-\epsilon_b},
  \]
  we have a perfect agreement between configuration interaction  theory  and many-body perturbation
  theory. However, configuration interaction  theory includes $2p-2h$ (and more complicated ones as
  well) correlations to infinite order. In order to make a meaningful
  comparison we would at least need to sum such correlations to
  infinite order in perturbation theory. The last equation serves
  however as a very useful comparison between configuration interaction  theory and many-body
  perturbation theory. Furthermore, for our nuclear matter studies,
  one-particle-one-hole intermediate excitations are zero due to the
  requirement of momentum conservation in infinite systems. These
  two-particle-two-hole correlations can also be summed to infinite
  order and a particular class of such excitations are given by
  two-particle excitations only. These represent in case of nuclear interactions, which
  are strongly repulsive at short interparticle distances, a
  physically intuitive way to understand the renomalization of nuclear
  forces. Such correlations are easily computed by simple matrix
  inversion techniques and have been widely employed in nuclear
  many-body theory.  Summing up two-particle excitations to infinite
  order leads to an effective two-body interaction which renomalizes
  the short-range part of the nuclear interactions. 

In summary, many-body perturbation theory introduces order-by-order
specific correlations and we can make comparisons with exact
calculations like those provided by configuration interaction theory.
The advantage of for example Rayleigh-Schr\"odinger perturbation
theory is that at every order in the interaction, we know how to
calculate all contributions. The two-body matrix elements can for example  be tabulated or
computed on the fly.  However, many-body perturbation theory suffers
from not being variational theory and there is no guarantee that
higher-order terms will improve the order-by-order convergence.  It is
also extremely tedious to compute terms beyond third order, in
particular if one is interested in effective valence space
interactions.  There are however classes of correlations which can be
summed up to infinite order in the interaction.  The hope is that such
correlations can mitigate specific convergence issues, although there
is no a priori guarantee thereof.  Examples are the so-called TDA and
RPA classes of diagrams \cite{blaizot,Ch11_Dickhoff2008,Ch11_Mattuck1992}, as well as the resummation
of two-particle-two-hole correlations discussed in chapter 11. If we limit ourselves to the
resummation of two-particle correlations only, these lead us to the
so-called $G$-matrix resummation of diagrams, see for example Ref.~\cite{day1967}.
There are however computationally inexpensive methods which
sum larger classes of correlations to infinite order in the
interaction. This leads us to section \ref{sec:chap8cctheory} and the final
many-body method of this chapter, coupled cluster theory.

\section{Coupled cluster theory}\label{sec:chap8cctheory}
  
Coester and K\"{u}mmel \cite{coester1958,coester1960,kuemmel1978} developed the ideas that led to coupled cluster
theory in the late 1950s. The correlated wave function
of a many-body system $\mid\Psi\rangle$
can be formulated as an exponential of correlation
operators $T$ acting on a reference state $\mid\Phi\rangle$,
\begin{equation*}
\mid\Psi\rangle = \exp\left(\hat{T}\right)\mid\Phi\rangle.
\end{equation*}
We will discuss how to define the operators later in this work. This simple
ansatz carries enormous power. It leads to a non-perturbative many-body
theory that includes summation of ladder diagrams \cite{brueckner1955}, ring
diagrams \cite{brandow1967}, and an infinite-order
generalization of many-body perturbation theory
\cite{bartlett1981}.
Developments and applications of coupled cluster theory took
different routes in chemistry and nuclear physics. In quantum
chemistry, coupled cluster developments and applications have proven
to be extremely useful, see for example the review by Barrett and
Musial as well as the recent textbook by Shavitt and Bartlett \cite{shavittbartlett2009}.
Many previous applications to nuclear physics struggled with the
repulsive character of the nuclear forces and limited basis sets
used in the computations \cite{kuemmel1978}. Most of these problems have been overcome
during the last decade and coupled cluster theory is one of the
computational methods of preference for doing nuclear physics, with
  applications ranging from light nuclei to medium-heavy nuclei, see
  for example the recent reviews \cite{Hagen:2015yea,hagen2016} and 
Refs.~\cite{binder2013,jansen2016}.

\subsection{A quick tour of Coupled Cluster theory}

  The ansatz for the ground state  is given by
  \begin{equation*}
     \vert \Psi_0\rangle = \vert \Psi_{CC}\rangle = e^{\hat{T}} \vert
     \Phi_0\rangle = \left( \sum_{n=1}^{A} \frac{1}{n!} \hat{T}^n
     \right) \vert \Phi_0\rangle,
  \end{equation*}
  where $A$ represents the maximum number of particle-hole excitations
  and $\hat{T}$ is the cluster operator defined as
  \begin{align*}
              \hat{T} &= \hat{T}_1 + \hat{T}_2 + \ldots + \hat{T}_A
              \\ \hat{T}_n &= \left(\frac{1}{n!}\right)^2
              \sum_{\substack{ i_1,i_2,\ldots i_n \\ a_1,a_2,\ldots
                  a_n}} t_{i_1i_2\ldots i_n}^{a_1a_2\ldots a_n}
              a_{a_1}^\dagger a_{a_2}^\dagger \ldots a_{a_n}^\dagger
              a_{i_n} \ldots a_{i_2} a_{i_1}.
          \end{align*}
      The energy is given by
      \begin{equation*}
          E_{\mathrm{CC}} = \langle\Phi_0\vert \overline{H}\vert
          \Phi_0\rangle,
      \end{equation*}
      where $\overline{H}$ is a similarity transformed Hamiltonian
      \begin{align*}
          \overline{H}&= e^{-\hat{T}} \hat{H}_N e^{\hat{T}}
          \\ \hat{H}_N &= \hat{H} - \langle\Phi_0\vert \hat{H} \vert
          \Phi_0\rangle.
      \end{align*}

      The coupled cluster energy is a function of the unknown cluster
      amplitudes $t_{i_1i_2\ldots i_n}^{a_1a_2\ldots a_n}$, given by
      the solutions to the amplitude equations
      \begin{equation}\label{eq:amplitudeeq}
          0 = \langle\Phi_{i_1 \ldots i_n}^{a_1 \ldots a_n}\vert
          \overline{H}\vert \Phi_0\rangle.
      \end{equation}
In order to set up the above equations, 
the similarity transformed Hamiltonian $\overline{H}$ is expanded
  using the Baker-Campbell-Hausdorff expression,
      \begin{equation}\label{eq:bch}
          \overline{H}= \hat{H}_N + \left[ \hat{H}_N, \hat{T} \right]
          + \frac{1}{2} \left[\left[ \hat{H}_N, \hat{T} \right],
            \hat{T}\right] + \ldots + \frac{1}{n!} \left[
            \ldots \left[ \hat{H}_N, \hat{T} \right], \ldots \hat{T}
            \right] +\dots
      \end{equation}
  and simplified using the connected cluster theorem \cite{shavittbartlett2009}
      \begin{equation*}
          \overline{H}= \hat{H}_N + \left( \hat{H}_N \hat{T}\right)_c
          + \frac{1}{2} \left( \hat{H}_N \hat{T}^2\right)_c + \dots +
          \frac{1}{n!} \left( \hat{H}_N \hat{T}^n\right)_c +\dots
      \end{equation*}
We will discuss parts of the the derivation below.
For the full derivation of these expressions, see for example Ref.~\cite{shavittbartlett2009}. 

A much used approximation is to truncate the cluster operator
$\hat{T}$ at the $n=2$ level. This defines the so-called singles and
doubles approximation to the coupled cluster state function, normally
shortened to CCSD.
The coupled cluster wavefunction is now given by
  \begin{equation*}
              \vert \Psi_{CC}\rangle = e^{\hat{T}_1 + \hat{T}_2} \vert
              \Phi_0\rangle
  \end{equation*}
  where
          \begin{align*}
              \hat{T}_1 &= \sum_{ia} t_{i}^{a} a_{a}^\dagger a_i
              \\ \hat{T}_2 &= \frac{1}{4} \sum_{ijab} t_{ij}^{ab}
              a_{a}^\dagger a_{b}^\dagger a_{j} a_{i}.
          \end{align*}

  The amplutudes $t$ play a role similar to the coefficients $C$ in
  the shell-model calculations. They are obtained by solving a set of
  non-linear equations similar to those discussed above in connection
  with the configuration interaction theory  discussion, see Eqs.~(\ref{eq:c1p1h}) and (\ref{eq:c2p2h}). 

  If we truncate our equations at the CCSD level, it corresponds to
  performing a transformation of the Hamiltonian matrix of the
  following type for the six particle problem (with a two-body
  Hamiltonian) discussed in section \ref{sec:chap8fci}
  \begin{table}
\caption{Schematic representation of blocks of matrix elements which are set to zero using the CCSD approximation.}
  \begin{center}
  \begin{tabular}{cccccccc}
  \hline \multicolumn{1}{c}{ } & \multicolumn{1}{c}{ $0p-0h$ } &
  \multicolumn{1}{c}{ $1p-1h$ } & \multicolumn{1}{c}{ $2p-2h$ } &
  \multicolumn{1}{c}{ $3p-3h$ } & \multicolumn{1}{c}{ $4p-4h$ } &
  \multicolumn{1}{c}{ $5p-5h$ } & \multicolumn{1}{c}{ $6p-6h$ }
  \\ \hline $0p-0h$ & $\tilde{x}$ & $\tilde{x}$ & $\tilde{x}$ & 0 & 0
  & 0 & 0 \\ $1p-1h$ & 0 & $\tilde{x}$ & $\tilde{x}$ & $\tilde{x}$ & 0
  & 0 & 0 \\ $2p-2h$ & 0 & $\tilde{x}$ & $\tilde{x}$ & $\tilde{x}$ &
  $\tilde{x}$ & 0 & 0 \\ $3p-3h$ & 0 & $\tilde{x}$ & $\tilde{x}$ &
  $\tilde{x}$ & $\tilde{x}$ & $\tilde{x}$ & 0 \\ $4p-4h$ & 0 & 0 &
  $\tilde{x}$ & $\tilde{x}$ & $\tilde{x}$ & $\tilde{x}$ & $\tilde{x}$
  \\ $5p-5h$ & 0 & 0 & 0 & $\tilde{x}$ & $\tilde{x}$ & $\tilde{x}$ &
  $\tilde{x}$ \\ $6p-6h$ & 0 & 0 & 0 & 0 & $\tilde{x}$ & $\tilde{x}$ &
  $\tilde{x}$ \\ \hline
  \end{tabular}
  \end{center}
  \end{table}

  In our configuration interaction theory discussion the correlation energy is defined as (with a two-body Hamiltonian) 
  \[
  \Delta E=\sum_{ai}\langle i| \hat{f}|a \rangle C_{i}^{a}+
  \sum_{abij}\langle ij | \hat{v}| ab \rangle C_{ij}^{ab}.
  \]
We can obtain a similar expression for the correlation energy using coupled cluster theory. 
Using Eq.~(\ref{eq:bch}) we can write the expression for the coupled cluster  ground state energy as an infinite sum over nested commutators
        \begin{align*}
            E_{CC} &= \bra{\Phi_0}\Bigl( \hat{H}_N + \left[ \hat{H}_N, \hat{T} \right] +
                \frac{1}{2} \left[\left[ \hat{H}_N, \hat{T} \right], \hat{T}\right] \\
                & \qquad + \frac{1}{3!} \left[ \left[\left[ \hat{H}_N, \hat{T} \right], \hat{T}\right], \hat{T} \right] \\
                & \qquad + \frac{1}{4!} \left[ \left[ \left[\left[ \hat{H}_N, \hat{T} \right], \hat{T}\right], \hat{T} \right], \hat{T} \right] +... \Bigr) \ket{\Phi_0}. \\
        \end{align*}
One can show that this infinite series truncates naturally at a given order of nested commutators \cite{shavittbartlett2009}. 
Let us demonstrate briefly how we can construct the expressions for the correlation 
energy by approximating $\hat{T}$ at the CCSD level, that is $\hat{T}\approx \hat{T}_1+\hat{T}_2$.
        The first term is zero by construction
        \begin{equation*}
            \bra{\Phi_0} \hat{H}_N \ket{\Phi_0} = 0.
        \end{equation*}       
     The second term can be split into the following contributions
        \begin{align*}
        \bra{\Phi_0}\left[ \hat{H}_N, \hat{T} \right]\ket{\Phi_0} & = 
            \bra{\Phi_0} \Bigl(\left[ \hat{F}_N, \hat{T}_1 \right] + \left[ \hat{F}_N, \hat{T}_2 \right]
            + \left[ \hat{V}_N, \hat{T}_1 \right] + \left[ \hat{V}_N, \hat{T}_2 \right] \Bigr) \ket{\Phi_0}.
    \end{align*}
Let us start with $\left[ \hat{F}_N, \hat{T}_1 \right]$, where the one-body operator $\hat{F}_N$ is defined in Eq.~(\ref{eq:hfn}). 
In the equations below we employ the shorthand $f^p_q = \langle p\vert \hat{f} \vert q\rangle$.  
We write out the commutator  as
    \begin{align*}
        \left[ \hat{F}_N, \hat{T}_1 \right] &= \sum_{pqia} \left(f_q^p \normord{a_p^\dagger a_q} 
            t_i^a \normord{a_a^\dagger a_i} - t_i^a \normord{a_a^\dagger a_i} f_q^p \normord{a_p^\dagger a_q} \right) \\
        &= \sum_{pqia} f_q^p t_i^a \left( \normord{a_p^\dagger a_q} \normord{a_a^\dagger a_i} -
                \normord{a_a^\dagger a_i} \normord{a_p^\dagger a_q} \right).
    \end{align*}
We have kept here the curly brackets that indicate that the chains of operators are normal ordered with respect to the new reference state. 
If we consider the second set of operators and rewrite them with curly brackets (bringing back the normal ordering) we have
        \begin{align*}
            \left\{a_a^\dagger a_i\right\} \left\{a_p^\dagger a_q \right\} &= \normord{a_a^\dagger a_i a_p^\dagger a_q}
                = \normord{a_p^\dagger a_q a_a^\dagger a_i} \\ 
        \normord{a_p^\dagger a_q}\normord{a_a^\dagger a_i} &= \normord{a_p^\dagger a_q a_a^\dagger a_i}  \\
            & \quad + \normord{
               \contraction{}{a}{{}^\dagger_pa_q a^\dagger_a}{a}
                a^\dagger_p a_q a^\dagger_a a_i} +
            \normord{
                \contraction{a^\dagger_p}{a}{{}_q}{a}
                a^\dagger_p a_q a^\dagger_a a_i} \\ 
            & \quad + \normord{
                \contraction[1.5ex]{}{a}{{}^\dagger_pa_q a^\dagger_a}{a}
                \contraction{a^\dagger_p}{a}{{}_q}{a}
                a^\dagger_p a_q a^\dagger_a a_i} \\ 
            &=  \normord{a_p^\dagger a_q a_a^\dagger a_i} + \delta_{qa} \normord{a_p^\dagger a_i} + \delta_{pi} \normord{a_q a_a^\dagger}
            + \delta_{qa}\delta_{pi}.
        \end{align*}
We can then rewrite the two sets of operators as
        \begin{align*}
            \left\{a_p^\dagger a_q \right\}\left\{a_a^\dagger a_i\right\} - \left\{a_a^\dagger a_i\right\} \left\{a_p^\dagger a_q \right\} &= \delta_{qa} \left\{ a_p^\dagger a_i\right\} + \delta_{pi} \left\{ a_q a_a^\dagger \right\} + \delta_{qa}\delta_{pi}.
    \end{align*}

        Inserted into the original expression, we arrive at the explicit value of the commutator
        \[
        \left[ \hat{F}_N, \hat{T}_1 \right] = \sum_{pai} f_a^p t_i^a \normord{a_p^\dagger a_i} + 
                \sum_{qai} f_q^i t_i^a \normord{a_q a_a^\dagger} + \sum_{ai} f_a^i t_i^a.
        \]
We are now ready to compute the expectation value with respect to our reference state. Since the two first terms require the ground state linking to 
a one-particle-one-hole state, the first two terms are zero and we are left with 
\begin{equation}\label{eq:firsttermE}
\langle \Phi_0 \vert \left[ \hat{F}_N, \hat{T}_1 \right] \vert \Phi_0\rangle = \sum_{ai} f_a^i t_i^a.
\end{equation}
The two first terms will however contribute to the calculation of the Hamiltonian  matrix element which connects the ground state and a one-particle-one-hole excitation. 
  
Let us next look at the term $\left[ \hat{F}_N, \hat{T}_2 \right]$. We have
    \begin{align*}
        \left[ \hat{F}_N, \hat{T}_2 \right] 
        &= \left[\sum_{pq} f_q^p \normord{a^\dagger_p a_q},
            \frac{1}{4}\sum_{ijab} t_{ij}^{ab} \normord{a^\dagger_a a^\dagger_b a_j a_i} \right] \\
        &= \frac{1}{4}\sum_{\substack{pq\\ijab}}
            f_q^p t_{ij}^{ab}
        \left( \normord{a^\dagger_p a_q} \normord{a^\dagger_a a^\dagger_b a_j a_i}
            - \normord{a^\dagger_a a^\dagger_b a_j a_i} \normord{a^\dagger_p a_q} \right).
    \end{align*}
The last set of operators can be rewritten as 
    \begin{align*}
        \normord{a^\dagger_a a^\dagger_b a_j a_i} \normord{a^\dagger_p a_q} &= 
            \normord{a^\dagger_a a^\dagger_b a_j a_i a^\dagger_p a_q} \\ 
        &= \normord{a^\dagger_p a_q a^\dagger_a a^\dagger_b a_j a_i} 
        \\
        \normord{a^\dagger_p a_q} \normord{a^\dagger_a a^\dagger_b a_j a_i} &= 
            \normord{a^\dagger_p a_q a^\dagger_a a^\dagger_b a_j a_i}
        + \left\{ 
        \contraction{}{a}{{}^\dagger_pa_q a^\dagger_a a^\dagger_b}{a}
        a^\dagger_p a_q a^\dagger_a a^\dagger_b a_j a_i\right\}
        + \left\{ 
        \contraction{}{a}{{}^\dagger_pa_q a^\dagger_a a^\dagger_b a_j}{a}
        a^\dagger_p a_q a^\dagger_a a^\dagger_b a_j a_i\right\} \\
        & \quad 
        + \left\{ 
        \contraction{a^\dagger_p}{a}{{}_q}{a}
        a^\dagger_p a_q a^\dagger_a a^\dagger_b a_j a_i\right\}
        + \left\{ 
        \contraction{a^\dagger_p}{a}{{}_q a^\dagger_a}{a}
        a^\dagger_p a_q a^\dagger_a a^\dagger_b a_j a_i\right\}
        + \left\{ 
        \contraction{a^\dagger_p}{a}{{}_q}{a}
        \contraction[1.5ex]{}{a}{{}^\dagger_pa_q a^\dagger_a a^\dagger_b}{a}
        a^\dagger_p a_q a^\dagger_a a^\dagger_b a_j a_i\right\} \\
        & \quad 
        + \left\{ 
        \contraction{a^\dagger_p}{a}{{}_q}{a}
        \contraction[1.5ex]{}{a}{{}^\dagger_pa_q a^\dagger_a a^\dagger_b a_j}{a}
        a^\dagger_p a_q a^\dagger_a a^\dagger_b a_j a_i\right\}
        + \left\{ 
        \contraction{a^\dagger_p}{a}{{}_q a^\dagger_a}{a}
        \contraction[1.5ex]{}{a}{{}^\dagger_pa_q a^\dagger_a a^\dagger_b}{a}
        a^\dagger_p a_q a^\dagger_a a^\dagger_b a_j a_i\right\}
        + \left\{ 
        \contraction{a^\dagger_p}{a}{{}_q a^\dagger_a}{a}
        \contraction[1.5ex]{}{a}{{}^\dagger_pa_q a^\dagger_a a^\dagger_b a_j}{a}
        a^\dagger_p a_q a^\dagger_a a^\dagger_b a_j a_i\right\} \\ 
        &= \normord{a^\dagger_p a_q a^\dagger_a a^\dagger_b a_j a_i}
        - \delta_{pj} \normord{a_q a^\dagger_a a^\dagger_b a_i}
        + \delta_{pi} \normord{a_q a^\dagger_a a^\dagger_b a_j} \\
        & \quad + \delta_{qa} \normord{a^\dagger_p a^\dagger_b a_j a_i}
        - \delta_{qb}\normord{a^\dagger_p a^\dagger_a a_j a_i} 
        - \delta_{pj} \delta_{qa} \normord{a^\dagger_b a_i} \\
        & \quad + \delta_{pi} \delta_{qa} \normord{a^\dagger_b a_j}
        + \delta_{pj} \delta_{qb} \normord{a^\dagger_a a_i}
        - \delta_{pi} \delta_{qb} \normord{a^\dagger_a a_j}.
    \end{align*}
 We can then rewrite the two sets of operators as
    \begin{align*}
        & \Bigl( \normord{a^\dagger_p a_q} \normord{a^\dagger_a a^\dagger_b a_j a_i}
            - \normord{a^\dagger_a a^\dagger_b a_j a_i} \normord{a^\dagger_p a_q} \Bigr) = \\
        & \qquad - \delta_{pj} \normord{a_q a^\dagger_a a^\dagger_b a_i}
        + \delta_{pi} \normord{a_q a^\dagger_a a^\dagger_b a_j}
        + \delta_{qa} \normord{a^\dagger_p a^\dagger_b a_j a_i} \\
        & \qquad - \delta_{qb}\normord{a^\dagger_p a^\dagger_a a_j a_i} 
        - \delta_{pj} \delta_{qa} \normord{a^\dagger_b a_i}
        + \delta_{pi} \delta_{qa} \normord{a^\dagger_b a_j}
        + \delta_{pj} \delta_{qb} \normord{a^\dagger_a a_i} \\
        & \qquad - \delta_{pi} \delta_{qb} \normord{a^\dagger_a a_j},
    \end{align*}
which, when    inserted into the original expression gives us
    \begin{align*}
        \left[ \hat{F}_N, \hat{T}_2 \right]
        &= \frac{1}{4} \sum_{\substack{pq\\abij}} f_q^p t_{ij}^{ab} \Bigl(
        - \delta_{pj} \normord{a_q a^\dagger_a a^\dagger_b a_i}
        + \delta_{pi} \normord{a_q a^\dagger_a a^\dagger_b a_j} \\
        & \quad + \delta_{qa} \normord{a^\dagger_p a^\dagger_b a_j a_i}
        - \delta_{qb}\normord{a^\dagger_p a^\dagger_a a_j a_i} 
        - \delta_{pj} \delta_{qa} \normord{a^\dagger_b a_i} \\
        & \quad + \delta_{pi} \delta_{qa} \normord{a^\dagger_b a_j}
        + \delta_{pj} \delta_{qb} \normord{a^\dagger_a a_i}
        - \delta_{pi} \delta_{qb} \normord{a^\dagger_a a_j} \Bigr).
    \end{align*}
    After renaming indices and changing the order of operators, we arrive at the explicit expression
    \[
        \left[ \hat{F}_N, \hat{T}_2 \right]
        = \frac{1}{2} \sum_{qijab} f_q^i t_{ij}^{ab} \normord{a_q a^\dagger_a a^\dagger_b a_j}
        + \frac{1}{2} \sum_{pijab} f_a^p t_{ij}^{ab} \normord{a^\dagger_p a^\dagger_b a_j a_i}+ \sum_{ijab} f_a^i t_{ij}^{ab} \normord{a^\dagger_b a_j}. 
    \]
In this case we have two sets of two-particle-two-hole operators and one-particle-one-hole operators and all these terms result in zero expectation values. 
However, these terms are important for the amplitude equations.  
In a similar way we can compute the terms involving the interaction $\hat{V}_N$.  
We obtain then
    \begin{align*}
        \bra{\Phi_0} \left[ \hat{V}_N, \hat{T}_1 \right] \ket{\Phi_0} &= 
            \bra{\Phi_0}
                \left[ \frac{1}{4} \sum_{pqrs} \bra{pq}\hat{v}\ket{rs} \normord{a^\dagger_p a^\dagger_q a_s  a_r},
                    \sum_{ia} t_i^a \normord{a^\dagger_a a_i} \right] \ket{\Phi_0} \\ 
            &= \frac{1}{4}\sum_{\substack{
                pqr \\
                sia}} \bra{pq}\ket{rs} t_i^a \bra{\Phi_0} 
                \left[ \normord{a^\dagger_p a^\dagger_q a_s  a_r}, \normord{a^\dagger_a a_i} \right]
                \ket{\Phi_0} \\ 
        &= 0,
    \end{align*}
and
    \begin{align*}
        & \bra{\Phi_0} \left[ \hat{V}_N, \hat{T}_2 \right] \ket{\Phi_0}= \\
            & \quad \bra{\Phi_0}
                \left[ \frac{1}{4} \sum_{pqrs} \bra{pq}\hat{v}\ket{rs} \normord{a^\dagger_p a^\dagger_q a_s  a_r},
                    \frac{1}{4}\sum_{ijab} t_{ij}^{ab} \normord{a^\dagger_a a^\dagger_b a_j a_i} \right] \ket{\Phi_0} \\ 
            &= \frac{1}{16}\sum_{\substack{
                    pqr \\
                    sijab}} \bra{pq}\hat{v}\ket{rs}t_{ij}^{ab} \bra{\Phi_0} 
                \left[ \normord{a^\dagger_p a^\dagger_q a_s  a_r}, \normord{a^\dagger_a a^\dagger_b a_j a_i} \right]
                \ket{\Phi_0} \\ 
            &= \frac{1}{16}\sum_{\substack{
                    pqr \\
                    sijab}} \bra{pq}\hat{v}\ket{rs}t_{ij}^{ab} \bra{\Phi_0}
            \Bigl(
            \left\{
            \contraction{a^\dagger_p a^\dagger_q a_s}{a}{{}_r}{a}
            \contraction[1.25ex]{a^\dagger_p a^\dagger_q}{a}{{}_s a_r a^\dagger_a}{a}
            \contraction[1.50ex]{a^\dagger_p}{a}{{}^\dagger_q a_s a_r a^\dagger_a a^\dagger_b}{a}
            \contraction[1.75ex]{}{a}{{}^\dagger_p a^\dagger_q a_s a_r a^\dagger_a a^\dagger_b a_j}{a}
            a^\dagger_p a^\dagger_q a_s  a_r a^\dagger_a a^\dagger_b a_j a_i \right\}
            + \left\{
            \contraction{a^\dagger_p a^\dagger_q}{a}{{}_s a_r}{a}
            \contraction[1.25ex]{a^\dagger_p a^\dagger_q a_s}{a}{{}_r a^\dagger_p}{a}
            \contraction[1.50ex]{a^\dagger_p}{a}{{}^\dagger_q a_s a_r a^\dagger_a a^\dagger_b}{a}
            \contraction[1.75ex]{}{a}{{}^\dagger_p a^\dagger_q a_s a_r a^\dagger_a a^\dagger_b a_j}{a}
            a^\dagger_p a^\dagger_q a_s  a_r a^\dagger_a a^\dagger_b a_j a_i \right\} \\
            & \quad \left\{
            \contraction{a^\dagger_p a^\dagger_q a_s}{a}{{}_r}{a}
            \contraction[1.25ex]{a^\dagger_p a^\dagger_q}{a}{{}_s a_r a^\dagger_a}{a}
            \contraction[1.5ex]{}{a}{{}^\dagger_p a^\dagger_q a_s a_r a^\dagger_a a^\dagger_b}{a}
            \contraction[1.75ex]{a^\dagger_p}{a}{{}^\dagger_q a_s a_r a^\dagger_a a^\dagger_b a_j}{a}
            a^\dagger_p a^\dagger_q a_s  a_r a^\dagger_a a^\dagger_b a_j a_i \right\}
            + \left\{
            \contraction{a^\dagger_p a^\dagger_q}{a}{{}_s a_r}{a}
            \contraction[1.25ex]{a^\dagger_p a^\dagger_q a_s}{a}{{}_r a^\dagger_p}{a}
            \contraction[1.5ex]{}{a}{{}^\dagger_p a^\dagger_q a_s a_r a^\dagger_a a^\dagger_b}{a}
            \contraction[1.75ex]{a^\dagger_p}{a}{{}^\dagger_q a_s a_r a^\dagger_a a^\dagger_b a_j}{a}
            a^\dagger_p a^\dagger_q a_s  a_r a^\dagger_a a^\dagger_b a_j a_i \right\}
            \Bigr) \ket{\Phi_0} \\ 
            &= \frac{1}{4} \sum_{ijab} \bra{ij}\hat{v}\ket{ab} t_{ij}^{ab}.
    \end{align*}
The final contribution to the correlation energy comes from the non-linear terms with the amplitudes squared. 
The contribution from the $\hat{T}^2$ is given by
    \begin{align*}
        & \bra{\Phi_0} \frac{1}{2} \left( \hat{V}_N \hat{T}_1^2 \right) \ket{\Phi_0} = \\
            & \quad \frac{1}{8} \sum_{pqrs} \sum_{ijab} \bra{pq}\hat{v}\ket{rs} t_i^a t_j^b 
            \bra{\Phi_0} \left(\normord{a^\dagger_p a^\dagger_q a_s  a_r} 
            \normord{a^\dagger_a a_i} \normord{a^\dagger_b a_j} \right)_c\ket{\Phi_0} \\ 
        &= \frac{1}{8} \sum_{pqrs} \sum_{ijab} \bra{pq}\hat{v}\ket{rs} t_i^a t_j^b \bra{\Phi_0} \\
        & \quad \Bigl( 
        \left\{
        \contraction{a^\dagger_p a^\dagger_q a_s}{a}{{}_r}{a}
        \contraction[1.25ex]{a^\dagger_p}{a}{{}^\dagger_q a_s a_r a^\dagger_a}{a}
        \contraction[1.5ex]{a^\dagger_p a^\dagger_q }{a}{{}_s a_r a^\dagger_a a_i}{a}
        \contraction[1.75ex]{}{a}{{}^\dagger_p a^\dagger_q a_s a_r a^\dagger_a a_i a^\dagger_b}{a}
        a^\dagger_p a^\dagger_q a_s  a_r a^\dagger_a a_i a^\dagger_b a_j \right\}
        +\left\{
        \contraction{a^\dagger_p a^\dagger_q}{a}{{}_s a_r}{a}
        \contraction[1.25ex]{a^\dagger_p}{a}{{}^\dagger_q a_s a_r a^\dagger_a}{a}
        \contraction[1.5ex]{a^\dagger_p a^\dagger_q a_s}{a}{{}_r a^\dagger_a a_i}{a}
        \contraction[1.75ex]{}{a}{{}^\dagger_p a^\dagger_q a_s a_r a^\dagger_a a_i a^\dagger_b}{a}
        a^\dagger_p a^\dagger_q a_s  a_r a^\dagger_a a_i a^\dagger_b a_j \right\}
        + \left\{
        \contraction{a^\dagger_p a^\dagger_q a_s}{a}{{}_r}{a}
        \contraction[1.25ex]{}{a}{{}^\dagger_p q^\dagger_q a_s a_r a^\dagger_a}{a}
        \contraction[1.5ex]{a^\dagger_p a^\dagger_q }{a}{{}_s a_r a^\dagger_a a_i}{a}
        \contraction[1.75ex]{a^\dagger_p}{a}{{}^\dagger_q a_s a_r a^\dagger_a a_i a^\dagger_b}{a}
        a^\dagger_p a^\dagger_q a_s  a_r a^\dagger_a a_i a^\dagger_b a_j \right\} \\
        & \quad +\left\{
        \contraction{a^\dagger_p a^\dagger_q}{a}{{}_s a_r}{a}
        \contraction[1.25ex]{}{a}{{}^\dagger_p q^\dagger_q a_s a_r a^\dagger_a}{a}
        \contraction[1.5ex]{a^\dagger_p a^\dagger_q a_s}{a}{{}_r a^\dagger_a a_i}{a}
        \contraction[1.75ex]{a^\dagger_p}{a}{{}^\dagger_q a_s a_r a^\dagger_a a_i a^\dagger_b}{a}
        a^\dagger_p a^\dagger_q a_s  a_r a^\dagger_a a_i a^\dagger_b a_j \right\}
        \Bigr) \ket{\Phi_0} \\ 
        &= \frac{1}{2} \sum_{ijab} \bra{ij}\hat{v}\ket{ab} t_i^a t_j^b.
    \end{align*}
 Collecting all terms we have   the final expression for the correlation energy with a two-body interaction given by
  \begin{equation}\label{eq:energyccsd}
  \Delta E=\sum_{ai}\langle i| \hat{f}|a \rangle t_{i}^{a}+\frac{1}{2} \sum_{ijab} \bra{ij}\hat{v}\ket{ab} t_i^a t_j^b+
  \frac{1}{4}\sum_{ijab}\langle ij | \hat{v}| ab \rangle t_{ij}^{ab}.
  \end{equation}
We leave it as a challenge to the reader to derive the corresponding equations for the Hamiltonian matrix elements of Eq.~(\ref{eq:amplitudeeq}).

  There are several interesting features with the coupled cluster
  equations. With a truncation like CCSD or even with the inclusion of
  triples (CCSDT), we can include to infinite order correlations based
  on one-particle-one-hole and two-particle-two-hole contributions.
  We can include a large basis of single-particle states, normally not
  possible in standard FCI calculations. Typical FCI calculations for
  light nuclei $A\le 16$ can be performed in at most some few harmonic
  oscillator shells. For heavier nuclei, at most two major shells can
  be included due to too large dimensionalities.  However, coupled
  cluster theory is non-variational and if we want to find properties
  of excited states, additional calculations via for example equation
  of motion methods are needed \cite{shavittbartlett2009,Hagen:2015yea}.
  If correlations are strong, a single-reference ansatz may not be the
  best starting point and a multi-reference approximation is needed
  \cite{jansen2015}. Furthermore, we cannot quantify properly the
  error we make when truncations are made in the cluster operator.

  \subsection{The CCD approximation}

  We will now approximate the cluster operator $\hat{T}$ to include
  only $2p-2h$ correlations. This leads to the so-called CCD
  approximation, that is
  \[
  \hat{T}\approx
  \hat{T}_2=\frac{1}{4}\sum_{abij}t_{ij}^{ab}a^{\dagger}_aa^{\dagger}_ba_ja_i,
  \]
  meaning that we have
  \[
  \vert \Psi_0 \rangle \approx \vert \Psi_{CCD} \rangle =
  \exp{\left(\hat{T}_2\right)}\vert \Phi_0\rangle.
  \]

  Inserting these equations in the expression for the computation of
  the energy we have, with a Hamiltonian defined with respect to a
  general reference vacuum
  \[
  \hat{H}=\hat{H}_N+E_{\mathrm{ref}},
  \]
  with
  \[
  \hat{H}_N=\sum_{pq}\langle p \vert \hat{f} \vert q \rangle
  a^{\dagger}_pa_q + \frac{1}{4}\sum_{pqrs}\langle pq \vert \hat{v}
  \vert rs \rangle a^{\dagger}_pa^{\dagger}_qa_sa_r,
  \]
  we obtain that the energy can be written as
  \[
  \langle \Phi_0 \vert
  \exp{-\left(\hat{T}_2\right)}\hat{H}_N\exp{\left(\hat{T}_2\right)}\vert
  \Phi_0\rangle = \langle \Phi_0 \vert \hat{H}_N(1+\hat{T}_2)\vert
  \Phi_0\rangle = E_{CCD}.
  \]
  This quantity becomes
  \[
  E_{CCD}=E_{\mathrm{ref}}+\frac{1}{4}\sum_{abij}\langle ij \vert
  \hat{v} \vert ab \rangle t_{ij}^{ab},
  \]
  where the latter is the correlation energy from this level of
  approximation of coupled cluster  theory.  Similarly, the expression for the
  amplitudes reads (see problem \ref{problem:amplitudes})
  \[
  \langle \Phi_{ij}^{ab} \vert
  \exp{\left(-\hat{T}_2\right)}\hat{H}_N\exp{\left(\hat{T}_2\right)}\vert
  \Phi_0\rangle = 0.
  \]
  These equations can be reduced to (after several applications of
  Wick's theorem), for all $i > j$ and all $a > b$,
  \begin{align}
  0 = \langle ab \vert \hat{v} \vert ij \rangle +
  \left(\epsilon_a+\epsilon_b-\epsilon_i-\epsilon_j\right)t_{ij}^{ab}+\frac{1}{2}\sum_{cd} \langle ab \vert \hat{v} \vert
  cd \rangle t_{ij}^{cd}+\frac{1}{2}\sum_{kl} \langle kl \vert \hat{v}
  \vert ij \rangle t_{kl}^{ab}+\hat{P}(ij\vert ab)\sum_{kc} \langle kb
  \vert \hat{v} \vert cj \rangle t_{ik}^{ac} & \nonumber
  \\ +\frac{1}{4}\sum_{klcd} \langle kl \vert \hat{v} \vert cd \rangle
  t_{ij}^{cd}t_{kl}^{ab}+\hat{P}(ij)\sum_{klcd} \langle kl \vert
  \hat{v} \vert cd \rangle t_{ik}^{ac}t_{jl}^{bd}-\frac{1}{2}\hat{P}(ij)\sum_{klcd} \langle kl \vert \hat{v} \vert
  cd \rangle t_{ik}^{dc}t_{lj}^{ab}-\frac{1}{2}\hat{P}(ab)\sum_{klcd}
  \langle kl \vert \hat{v} \vert cd \rangle t_{lk}^{ac}t_{ij}^{db},&
  \label{eq:ccd}
  \end{align}
  where we have defined
  \[
  \hat{P}\left(ab\right)= 1-\hat{P}_{ab},
  \]
  where $\hat{P}_{ab}$ interchanges two particles occupying the
  quantum numbers $a$ and $b$.  The operator $\hat{P}(ij\vert ab)$ is
  defined as
  \[
  \hat{P}(ij\vert ab) = (1-\hat{P}_{ij})(1-\hat{P}_{ab}).
  \]
  Recall also that the unknown amplitudes $t_{ij}^{ab}$ represent
  anti-symmetrized matrix elements, meaning that they obey the same
  symmetry relations as the two-body interaction, that is
  \[
  t_{ij}^{ab}=-t_{ji}^{ab}=-t_{ij}^{ba}=t_{ji}^{ba}.
  \]
  The two-body matrix elements are also anti-symmetrized, meaning that
  \[
  \langle ab \vert \hat{v} \vert ij \rangle = -\langle ab \vert
  \hat{v} \vert ji \rangle= -\langle ba \vert \hat{v} \vert ij
  \rangle=\langle ba \vert \hat{v} \vert ji \rangle.
  \]
  The non-linear equations for the unknown amplitudes $t_{ij}^{ab}$
  are solved iteratively. We discuss the implementation of these
  equations below.

  \subsection{Approximations to the full CCD equations.}
  It is useful to make approximations to the equations for the
  amplitudes. These serve as important benchmarks when we are to develop a many-body code.
The standard method for solving these equations is to
  set up an iterative scheme where method's like Newton's method or
  similar root searching methods are used to find the amplitudes, see for example Ref.~\cite{baran2008}.

  Iterative solvers need a guess for the amplitudes. A good starting
  point is to use the correlated wave operator from perturbation
  theory to first order in the interaction.  This means that we define
  the zeroth approximation to the amplitudes as
  \[
  t^{(0)}=\frac{\langle ab \vert \hat{v} \vert ij
    \rangle}{\left(\epsilon_i+\epsilon_j-\epsilon_a-\epsilon_b\right)},
  \]
  leading to our first approximation for the correlation energy at the
  CCD level to be equal to second-order perturbation theory without
  $1p-1h$ excitations, namely
  \[
  \Delta E_{\mathrm{CCD}}^{(0)}=\frac{1}{4}\sum_{abij} \frac{\langle
    ij \vert \hat{v} \vert ab \rangle \langle ab \vert \hat{v} \vert
    ij
    \rangle}{\left(\epsilon_i+\epsilon_j-\epsilon_a-\epsilon_b\right)}.
  \]

  With this starting point, we are now ready to solve
  Eq. (\ref{eq:ccd}) iteratively. Before we attack the full equations,
  it is however instructive to study a truncated version of the
  equations. We will first study the following approximation where we
  take away all terms except the linear terms that involve the
  single-particle energies and the two-particle intermediate
  excitations, that is
  \begin{equation}
  0 = \langle ab \vert \hat{v} \vert ij \rangle +
  \left(\epsilon_a+\epsilon_b-\epsilon_i-\epsilon_j\right)t_{ij}^{ab}+\frac{1}{2}\sum_{cd}
  \langle ab \vert \hat{v} \vert cd \rangle t_{ij}^{cd}.
  \label{eq:ccd1}
  \end{equation}

  Setting the single-particle energies for the hole states equal to an
  energy variable $\omega = \epsilon_i+\epsilon_j$,
  Eq.~(\ref{eq:ccd1}) reduces to the well-known equations for the
  so-called $G$-matrix, widely used in infinite matter and finite
  nuclei studies, see for example Refs.~\cite{day1967,hh2000}.  The equation can then be reordered
  and solved by matrix inversion.  To see this let us define the
  following quantity
  \[
  \tau_{ij}^{ab}=
  \left(\omega-\epsilon_a-\epsilon_b\right)t_{ij}^{ab},
  \]
  and inserting
  \[
  1=\frac{\left(\omega-\epsilon_c-\epsilon_d\right)}{\left(\omega-\epsilon_c-\epsilon_d\right)},
  \]
  in the intermediate sums over $cd$ in Eq.~(\ref{eq:ccd1}), we can
  rewrite the latter equation as
  \[
  \tau_{ij}^{ab}(\omega)= \langle ab \vert \hat{v} \vert ij \rangle +
  \frac{1}{2}\sum_{cd} \langle ab \vert \hat{v} \vert cd \rangle
  \frac{1}{\omega-\epsilon_c-\epsilon_d}\tau_{ij}^{cd}(\omega),
  \]
  where we have inserted an explicit energy dependence via the parameter $\omega$. This
  equation, transforming a two-particle configuration into a single
  index, can be rewritten as  a matrix inversion problem.  Solving
  the equations for a fixed energy $\omega$ allows us to compare
  directly with results from Green's function theory when only
  two-particle intermediate states are included.

  To solve Eq.~(\ref{eq:ccd1}), we start with a guess for
  the unknown amplitudes, normally using the wave operator defined by
  first order in perturbation theory, leading to a zeroth-order
  approximation for the correlation energy given by second-order perturbation
  theory.  A simple approach to the
  solution of Eq.~(\ref{eq:ccd1}), is to thus to
  \begin{enumerate}
  \item Start with a guess for the amplitudes and compute the zeroth
    approximation to the correlation energy

  \item Use the ansatz for the amplitudes to solve Eq. (\ref{eq:ccd1})
    via for example your root-finding method of choice (Newton's
    method or modifications thereof can be used) and continue these
    iterations till the correlation energy does not change more than a
    prefixed quantity $\lambda$; $\Delta E_{\mathrm{CCD}}^{(i)}-\Delta E_{\mathrm{CCD}}^{(i-1)} \le \lambda$.

  \item It is common during the iterations to scale the amplitudes
    with a parameter $\alpha$, with $\alpha \in (0,1]$ as
      $t^{(i)}=\alpha t^{(i)}+(1-\alpha)t^{(i-1)}$.
  \end{enumerate}

  \noindent
  The next approximation is to include the two-hole term in
  Eq.~(\ref{eq:ccd}), a term which allow us to make a link with
  Green's function theory with two-particle and two-hole
  correlations discussed in chapter 11. This means that we solve
  \begin{equation}
  0 = \langle ab \vert \hat{v} \vert ij \rangle +
  \left(\epsilon_a+\epsilon_b-\epsilon_i-\epsilon_j\right)t_{ij}^{ab}+\frac{1}{2}\sum_{cd}
  \langle ab \vert \hat{v} \vert cd \rangle
  t_{ij}^{cd}+\frac{1}{2}\sum_{kl} \langle kl \vert \hat{v} \vert ij
  \rangle t_{kl}^{ab}.
  \label{eq:ccd2}
  \end{equation}
  This equation is solved the same way as we would do for
  Eq. (\ref{eq:ccd1}). The final step is then to include all terms in
  Eq. (\ref{eq:ccd}).

\section{Developing a numerical project}\label{sec:chap8numproject}
A successful numerical project relies on us having expertise in several scientific and
engineering disciplines. We need a thourough understanding of the relevant scientific
domain to ask the right questions and interpret the results, but the tools we
use require a proficiency in mathematics to develop models and work out
analytical results, in numerics to choose the correct algorithms, in computer
science to understand what can go wrong with our algorithms when the problem is
discretized and solved on a digital computer, and in software engineering to
develop and maintain a computer program that solves our problem.

Indenpendent of your scientific background, you are probably also educated in
mathematics and numerics. Unfortunately, the computer science and
software engeneering aspects of computing are often neglected and thought of as
skills you pick up along the way. This is a problem for many reasons. First,
running a numerical project is very similar to running a
physical experiment. Your codes are the blueprints the compiler uses to build
the experiment from the components of the computer. It is unthinkable to publish
results from a physical experiment without a thorough understanding of the
experimental
equipment. Second, the blueprints are not only used to tell the compiler what to
build, but also by humans to understand what is beeing built, how to fix
it if something goes wrong, and how to improve it. If the blueprints are not
properly written and readable, human understanding is lost. Last, components of an experiment
are always tested individually to establish tolerances and that they work
according to specification. In software engineering, this corresponds to writing
testable code where you can be confident of the quality of each piece. These are
skills many writers of scientific software never learn and as a consequence
many numerical experiments are not properly understood and are never independently
verified.

In this section we will focus on some key tools and strategies that we feel are
important for developing and running a numerical experiment. Our main concerns
are that our results can be validated, independently verified, and run
efficiently. In addition, we will discuss tools that make the whole process somewhat easier. We will
cover testing, tracking changes with version control software, public code
repositories, and touch upon simple profiling tools to guide the optimization
process. Finally we will present a numerical project where we have developed a
code to calculate properties of nuclear matter using coupled-cluster
theory. Here, we will make extensive use of the simple pairing model of problem
\ref{problem:pairingmodel}. This model allows for benchmarks against exact
results. In addition, it provides analytical answers to several approximations,
from perturbation theory to specific terms in the solution of the coupled
cluster equations, the in-medium similarity renormalization group approach of
chapter 10 and the Green's function approach of chapter 11.

\subsection{Validation and verification}
The single most important thing in a numerical experiment is to get the correct
answer. A close second is to be confident that the answer is correct and why. A
lucky coincidence must be distinguishable from a consistently correct result. The
only way to do this is to validate the code by writing and running tests -- lots
of tests. Ideally, every aspect of a code should be tested and it should be
possible to run the tests automatically. As most of you have probably
experienced, it's very easy to introduce errors into a project. And very often,
symptoms of the errors are not visible where the errors have been introduced. By
having a large set of automated tests and running them often, symptoms of errors
can be discovered quickly and the errors tracked down while recent changes are
fresh in memory.

As scientists we are trained to validate our methods and findings. By applying
the same rigourous process to our software, we can achieve the same level of
confidence in our code as we have for the rest of our work. We advocate testing
at three distinct levels. Let's start with discussing validation tests, as this
is the type of test you are probably most familiar with. In a validation test,
your application is run as in production mode and the test fails if it cannot
reproduce a known result. The known result could be a published benchmark, a
simplified model where analytical results are available, an approximate result
from a different method, or even an earlier result from the same code. We will
discuss this type of testing further in section \ref{sec:ccdcode}.

Analogous to testing individual components in a physical experiment, is a type
of test called a unit test. This is a very fine-grained test that will typically
only test a class or a procedure, or even a small part of your code  at a time. This is where you test that a data
structure has been correctly filled, that an algorithm works appropriately, that
a file has been read correctly, and basically every other component test you can
think of. It does take a little more work to setup as testing needs to be done
outside of your normal program flow. Typically this involves writing different
executables that create the necessary dependencies before testing a component.
The advantages of writing unit tests are many. First, because you know that
the individual pieces of your code work independently, you will achieve a higher
degree of confidence in your results. Second, you will develop a programming
style that favors highly decoupled units because such units are easier to test.
This allows talking about the code at a higher level of abstraction, which helps
understanding. Last, your tests become the documentation of how your code is
supposed to be used. This might not seem important while your are actively
working on a project, but it will be invaluable down the line when you want to add
new features. Also, when you share your code as part of the
scientific process, these tests will be the way your peers will start to
understand your work.  This means that your final production code will also include various tests.

While validation tests test your code at the coursest level, and unit tests test
your code at the finest level, integration tests test how your components work
together. If, for example, your program solves differential equations as parts
of a larger problem, the components that make up your differential-equation
solver can be tested alone. If your solver can solve a set of representative
problems that either have anlytical solutions or can be worked out using some
other tool, you can be more confident that it will work on your specific
problem.  Moreover, writing integration tests pushes you to develop more general
components. Instead of writing a routine that only solves the differential
equations you need, you write a solver that can solve many different types of
differential equations. This allows your components to be reused in other
projects and by other people.

To many, this rigourous approach to testing software might seem like a waste of
time. Our view is that testing software is crucial to the scientific process and
we should strive to apply the same level of rigour to our software as we do to
every other aspect of our work. On a more pragmatic level, you can either spend
your time writing tests and make sure your components work, or you can spend
your time debugging when something goes wrong and worry that your results are
not valid. We definitely prefer, from own and other people's experience, the first approach.

\subsection{Tracking changes}
If you're not using a tool to track the changes you make to your code, now is
the time to start. There are several tools available, but the
authors are using git~(\url{https://git-scm.com/}), an open-source version-control system
that can run on Linux, OSX, and Windows. By tracking changes, it is easier to
correct a mistake when it inevitably creeps into the code. It is possible to go
back to a previously validated version and by using branches, you can work on
different versions of the code simultaneously. For example, you can create a
production branch where everything is validated and ready to run, and you can
create a development branch to implement new features.  There are also code
repositories where you can store a copy of your code for free, without worrying
about things getting lost. The source codes discussed in this book are hosted on
for example GitHub~(\url{https://github.com/}), which uses git to track all changes to the code. By
using a service like this, it is easier to synchronize code between multiple
machines. Multiple developers can work on the same code at the same time and
share changes without worrying about losing contributions. It can also become
the official public repository of your software to enable your peers to verify
your work. The software discussed in this chapter is available from our GitHub repository 
\url{https://github.com/ManyBodyPhysics/LectureNotesPhysics/tree/master/Programs/Chapter8-programs/}.
\subsection{Profile-guided optimization}\label{subsec:profiling}
The aim of this subsection is to discuss in more detail how we can make the
computations discussed in connection with equations
Eqs.~(\ref{eq:bruteforceMBPT}) and (\ref{eq:smartMBPT}) more efficient using
physical constraints, algorithm improvements, and parallel processing. For
pedagocical reasons, we will use the MBPT parts of the program due to their
simplicity while still containing the important elements of a larger, more
complicated CCD calculation.  The codes can be found at the github link
\url{https://github.com/ManyBodyPhysics/LectureNotesPhysics/tree/master/Programs/Chapter8-programs/cpp/MBPT/src}.
We will demonstrate the use of a simple profiler to help
guide our development efforts. Our starting points are  naive implementations of
many-body perturbation theory to second (MBPT2) and third order (MBPT3) in the interaction. For reference, we calculate
properties of nuclear matter and construct our Hamiltonian in a free-wave basis
using the Minnesota~\cite{minnesota} potential discussed in section \ref{sec:interaction}. As the model
is not as important as the performance in this section, we postpone a discussion of
the model to section \ref{sec:ccdcode}.

\begin{lstlisting}[language=c++,numbers=left,firstline=14,lastline=34,
caption=Trivial implementation of a MBPT2 diagram. \label{code:mbpt2V06}]
  double energy = 0.0;
  for(int i = 0; i < modelspace.indhol; ++i){
    for(int j = 0; j < modelspace.indhol; ++j){
      if(i == j){ continue; }
      for(int a = modelspace.indhol; a < modelspace.indtot; ++a){
        for(int b = modelspace.indhol; b < modelspace.indtot; ++b){
          if(a == b){ continue; }
          energy0 = potential->get_element(modelspace.qnums, i, j, a, b);
          energy0 *= energy0;
          energy0 /= (modelspace.qnums[i].energy + modelspace.qnums[j].energy -
                          modelspace.qnums[a].energy - modelspace.qnums[b].energy);
          energy += energy0;
        }
      }
    }
  }
  energy *= 0.25;
  return energy;
}
\end{lstlisting}

Listing \ref{code:mbpt2V06} shows a possible early implementation to solve Eq.~\ref{eq:bruteforceMBPT} from MBPT2. This function has a loop over all single-particle indices and calls the
V\_Minnesota function to calculate the two-body interaction for each set of
indices. The energy denominators are calculated from the single-particle
energies stored in the modelspace structure and partial results are accumulated into
the energy variable.
This function represents a straightforward implementation of MBPT2. We normally recommend, when developing a code, to write
the first implementation in a way which is as close as possible to the mathematical expressions, in this particular case Eq.~(\ref{eq:bruteforceMBPT}).
\begin{table}
\caption{Total runtime for the MBPT2 implementation in Listing 
    \ref{code:mbpt2V06} for different model spaces and particle numbers.}\label{tab:mbpt2V06_runtime}
  \begin{center}
      \begin{tabular}{cccc}
      \hline
      Number of & Number of & Number of & Runtime (s) \\
      states & protons & neutrons & Listing \ref{code:mbpt2V06}\\
      \hline
      \hline
        342 & 0 & 2 & $< 0.01$ \\
         & 0 & 14 & $0.15$ \\
         & 0 & 38 & $1.00$ \\
         & 0 & 54 & $1.81$ \\
         684 & 2 & 2 & $0.88$ \\
         & 14 & 14 & $2.43$ \\
         & 38 & 38 & $15.7$ \\
         & 54 & 54 & $28.2$ \\
         1598 & 0 & 2 & $0.04$ \\
         & 0 & 14 & $3.32$ \\
         & 0 & 38 & $25.0$ \\
         & 0 & 54 & $58.9$ \\
         3196 & 2 & 2 & $0.88$ \\
         & 14 & 14 & $54.3$ \\
         & 38 & 38 & $399$ \\
         & 54 & 54 & $797$ \\
      \hline
      \end{tabular}
  \end{center}
\end{table}

Table \ref{tab:mbpt2V06_runtime} shows the total execution time for this
application for different model spaces (defined by the number of single-particle states) and number of particles on a local
workstation. Your runtimes will be different. Our goals are converged
calculations of pure neutron matter as well as nuclear matter, where the number
of states and the number of protons and neutrons goes to infinity. It suffices
to say that we cannot reach our goals with this code.

We want to decrease the run time of this application, but it can be
difficult to decide where we should spend our time improving this
code. Our first approach is  to observe what goes on inside
the program. For that we will use one the simplest possible profiling
tools called
gprof~(\url{https://sourceware.org/binutils/docs/gprof/}). Alternatively,
software like Valgrind is also highly recommended
\url{http://valgrind.org}.  If you are using integrated development
environments (IDEs) like Qt \url{https://www.qt.io/}, performance and
debugging tools are integrated with the IDE.

To use gprof the code must first be compiled and linked with the -pg flag. This flag enables
the collection of runtime information so that a call graph and a profile can be
constructed when your program is run.

\begin{table}
\caption{Flat profile for the MBPT2 implementation in Listing \ref{code:mbpt2V06}
using 1598 states calculating pure neutron matter with 54 neutrons}\label{profile:mbpt2V06}
\begin{verbatim}
Flat profile:

Each sample counts as 0.01 seconds.
  %   cumulative   self              self     total           
 time   seconds   seconds    calls  ms/call  ms/call  name    
 58.01     26.26    26.26 2523524846     0.00     0.00  V_Minnesota(...)
 39.64     44.21    17.95                             mbpt2V00::getEnergy()
  2.35     45.27     1.07                             spinExchangeTerm(...)
\end{verbatim}
\end{table}
Table \ref{profile:mbpt2V06} shows the top few lines of the flat profile
generated for MBPT2 version in Listing \ref{code:mbpt2V06}. The leftmost column
shows the percentage of run time spent in the different functions and it shows
that about $58\%$ of the time is spent calculating the potential while about $40\%$ 
is spent in the loops in the actual MBPT2 function. The remaining part 
is spent in the spinExchangeTerm function which is called from the
potential function.  Even though the application spends most of its time
generating the potential, we don't want to spend too much time on improving this
code. We use the Minnesota potential for testing and benchmark purposes only. 
For more realistic calculations, one should employ the chiral interaction models discussed earlier.
It is, however, possible to reduce the
number of times this function is called.  The 4th column in table
\ref{profile:mbpt2V06} shows that for this particular instance, the potential
function was called 2.5 billion times. However, due to known symmetries of the
nuclear interaction we know that most of these calls result in matrix elements that
are zero. If we can exploit this structure to reduce the number of calls to the
potential function we will greatly reduce the total run time of this program.
The details of how this is done is presented in section \ref{sec:ccdcode}.

\begin{lstlisting}[language=c++,numbers=left,firstline=15,lastline=41,
caption=Block-sparse implementation of a MBPT2 diagram. \label{code:mbpt2V02}]
double mbpt2V02::getEnergy() {
  double energy = 0.0;
  double energy0;
  int nhh, npp, i, j, a, b;
  for(int chan = 0; chan < channels.size; ++chan){
    nhh = channels.nhh[chan];
    npp = channels.npp[chan];
    if(nhh*npp == 0){ continue; }

    for(int hh = 0; hh < nhh; ++hh){
      i = channels.hhvec[chan][2*hh];
      j = channels.hhvec[chan][2*hh + 1];
      for(int pp = 0; pp < npp; ++pp){
        a = channels.ppvec[chan][2*pp];
        b = channels.ppvec[chan][2*pp + 1];
        energy0 = V_Minnesota(modelspace, i, j, a, b, L);
        energy0 *= energy0;
        energy0 /= (modelspace.qnums[i].energy + modelspace.qnums[j].energy -
                      modelspace.qnums[a].energy - modelspace.qnums[b].energy);
        energy += energy0;
      }
    }
  }
  energy *= 0.25;
  return energy;

}
\end{lstlisting}

Listing \ref{code:mbpt2V02} shows a version of this code where the potential function is
not called when we know that the matrix element is zero. This code loops over
channels, which are the dense blocks of the full interaction. We have
pre-computed the two-body configurations allowed in each channel and store them
in the channels structure. The potential is computed in the same way as before,
but for fewer combinations of indices.

\begin{table}
\caption{Flat profile for the MBPT2 implementation in Listing \ref{code:mbpt2V02}
using 1598 states calculating pure neutron matter with 54 neutrons}\label{profile:mbpt2V02}
\begin{verbatim}
Flat profile:

Each sample counts as 0.01 seconds.
  %   cumulative   self              self     total           
 time   seconds   seconds    calls  ms/call  ms/call  name    
 66.69      0.08     0.08  2520526     0.00     0.00  V_Minnesota(...)
 16.67      0.10     0.02  4770508     0.00     0.00  Chan_2bInd(...)
  8.34      0.11     0.01        1    10.00    10.00  Build_Model_Space(...)
  8.34      0.12     0.01        1    10.00    30.01  Setup_Channels_MBPT(...)
\end{verbatim}
\end{table}

\begin{table}
\caption{Total runtime for different MBPT2 implementations for different model spaces.}\label{tab:mbpt2V02_runtime}
  \begin{center}
      \begin{tabular}{ccccc}
      \hline
      Number of & Number of & Number of & \multicolumn{2}{c}{Runtime (s)} \\
      states & protons & neutrons & Listing \ref{code:mbpt2V06} & Listing \ref{code:mbpt2V02}\\
      \hline
      \hline
        342 & 0 & 2 & $< 0.01$ & $< 0.01$ \\
         & 0 & 14 & $0.15$ & $< 0.01$\\
         & 0 & 38 & $1.00$ & $0.03$\\
         & 0 & 54 & $1.81$ & $0.05$\\
         684 & 2 & 2 & $0.88$ & $< 0.01$\\
         & 14 & 14 & $2.43$ & $0.04$ \\
         & 38 & 38 & $15.7$ & $0.19$ \\
         & 54 & 54 & $28.2$ & $0.31$ \\
         1598 & 0 & 2 & $0.04$ & $< 0.01$\\
         & 0 & 14 & $3.32$ & $0.03$ \\
         & 0 & 38 & $25.0$ & $0.23$ \\
         & 0 & 54 & $58.9$ & $0.44$\\
         3196 & 2 & 2 & $0.88$ & $< 0.01$ \\
         & 14 & 14 & $54.3$ & $0.21$ \\
         & 38 & 38 & $399$ & $1.40$ \\
         & 54 & 54 & $797$ & $2.67$ \\
      \hline
      \end{tabular}
  \end{center}
\end{table}

The profile in Table \ref{profile:mbpt2V02} shows that the
potential function is now only called 2.5 million times, a reduction of three
orders of magnitude. Table \ref{tab:mbpt2V02_runtime} summarizes the execution
times of these two versions of MBPT2.

\begin{lstlisting}[language=c++,numbers=left,firstline=15,lastline=47,
caption=Block-sparse implementation of a MBPT3 diagram. \label{code:mbpt3V02}]
double mbpt3V02::getEnergy() {
  double energy = 0.0;
  double energy0, energy1;
  int nhh, npp, i, j, a, b, c, d;
  for(int chan = 0; chan < channels.size; ++chan){
    nhh = channels.nhh[chan];
    npp = channels.npp[chan];
    if(nhh*npp == 0){ continue; }

    for(int hh = 0; hh < nhh; ++hh){
      i = channels.hhvec[chan][2*hh];
      j = channels.hhvec[chan][2*hh + 1];
      for(int ab = 0; ab < npp; ++ab){
        a = channels.ppvec[chan][2*ab];
        b = channels.ppvec[chan][2*ab + 1];
        energy0 = V_Minnesota(modelspace, i, j, a, b, L);
        energy0 /= (modelspace.qnums[i].energy + modelspace.qnums[j].energy -
                      modelspace.qnums[a].energy - modelspace.qnums[b].energy);
        for(int cd = 0; cd < npp; ++cd){
          c = channels.ppvec[chan][2*cd];
          d = channels.ppvec[chan][2*cd + 1];
          energy1 = V_Minnesota(modelspace, a, b, c, d, L);
          energy1 *= V_Minnesota(modelspace, c, d, i, j, L);
          energy1 /= (modelspace.qnums[i].energy + modelspace.qnums[j].energy -
                        modelspace.qnums[c].energy - modelspace.qnums[d].energy);
          energy += energy0*energy1;
        }
      }
    }
  }
  energy *= 0.125;
  return energy;
}
\end{lstlisting}

\begin{table}[hbt]
\caption{Flat profile for the MBPT3 implmentation in Listing \ref{code:mbpt3V02}
using 3196 states calculating nuclear matter with 14 protons and 14 neutrons}\label{profile:mbpt3V02}
\begin{verbatim}
Flat profile:

Each sample counts as 0.01 seconds.
  %   cumulative   self              self     total           
 time   seconds   seconds    calls  ms/call  ms/call  name    
 90.76    177.21   177.21 6068445596     0.00     0.00  V_Minnesota(...)
  8.76    194.31    17.11                             mbpt3V02::getEnergy()
\end{verbatim}
\end{table}
Listing \ref{code:mbpt3V02} shows an implementation of MBPT3 that uses a
block-sparse representation of the interaction. Compared to MBPT2 it loops over
two additional particle indices which increases the computational complexity
by several orders of magnitude. However, we are now calculating the interaction
many more times than what is necessary. The profile in Table
\ref{profile:mbpt3V02} shows that we calculated over six 
billion matrix elements. By moving the construction of the interaction out of
the main loops and storing the elements, we can eliminate these redundant calls
to the potential function at the expense of using memory to store the elements. 

\begin{lstlisting}[language=c++,numbers=left,firstline=16,lastline=65,
caption=Block-sparse implementation of a MBPT3 diagram with interaction stored
in memory. \label{code:mbpt3V05}]
double mbpt3V05::getEnergy() {
  double energy = 0.0;
  double *V1, *S1, *V2, *S2;
  char N = 'N';
  char T = 'T';
  double fac0 = 0.0;
  double fac1 = 1.0;
  int nhh, npp, i, j, a, b, c, d, idx;
  double energy0;
  for(int chan = 0; chan < channels.size; ++chan){
    nhh = channels.nhh[chan];
    npp = channels.npp[chan];
    if(nhh*npp == 0){ continue; }

    V1 = new double[nhh * npp];
    S1 = new double[nhh * npp];
    V2 = new double[npp * npp];
    S2 = new double[nhh * nhh];
    for(int ab = 0; ab < npp; ++ab){
      a = channels.ppvec[chan][2*ab];
      b = channels.ppvec[chan][2*ab + 1];
      for(int hh = 0; hh < nhh; ++hh){
        i = channels.hhvec[chan][2*hh];
        j = channels.hhvec[chan][2*hh + 1];
        idx = hh * npp + ab;
        energy0 = V_Minnesota(modelspace, i, j, a, b, L);
        energy0 /= (modelspace.qnums[i].energy + modelspace.qnums[j].energy -
                    modelspace.qnums[a].energy - modelspace.qnums[b].energy);
        V1[idx] = energy0;
      }
      for(int cd = 0; cd < npp; ++cd){
        c = channels.ppvec[chan][2*cd];
        d = channels.ppvec[chan][2*cd + 1];
        idx = ab * npp + cd;
        V2[idx] = V_Minnesota(modelspace, a, b, c, d, L);
      }
    }

    RM_dgemm(V1, V2, S1, &nhh, &npp, &npp, &fac1, &fac0, &N, &N);
    RMT_dgemm(S1, V1, S2, &nhh, &nhh, &npp, &fac1, &fac0, &N, &T);
    delete V1; delete V2; delete S1;
    for(int hh = 0; hh < nhh; ++hh){
      energy += S2[nhh*hh + hh];
    }
    delete S2;
  }
  energy *= 0.125;

  return energy;
}
\end{lstlisting}

Listing \ref{code:mbpt3V05} shows the new version of the code that stores matrix
elements of the interaction. The explicit summation to calculate the energy can now be done by
using matrix products by calling the BLAS (Basic Linear Algebra Subprograms) \cite{blas}
dgemm wrappers \verb;RM_dgemm; and
\verb;RMT_dgemm;. Note that the code has grown more complicated for every new
optimization we have introduced. This increases the possibility of introducing
errors significantly. It is a good thing we have tests to make sure that the
results haven't changed between versions. Table \ref{profile:mbpt3V05} shows the
profile for this version and we have reduced considerably the number of calls to the function which sets up the interaction.
    
\begin{table}[h]
\caption{Flat profile for the MBPT3 implmentation in Listing \ref{code:mbpt3V05}
using 3196 states calculating nuclear matter with 14 protons and 14 neutrons}\label{profile:mbpt3V05}
\begin{verbatim}
Flat profile:

Each sample counts as 0.01 seconds.
  %   cumulative   self              self     total           
 time   seconds   seconds    calls  ms/call  ms/call  name    
 91.16     14.59    14.59 484191644     0.00     0.00  V_Minnesota(...)
  7.68     15.82     1.23                             mbpt3V05::getEnergy()
\end{verbatim}
\end{table}
    
\begin{table}
\caption{Total runtime for the MBPT3 implementation in Listing 
    \ref{code:mbpt3V02} for different model spaces.}\label{tab:mbpt3V02_runtime}
  \begin{center}
      \begin{tabular}{ccccc}
      \hline
      Number of & Number of & Number of & \multicolumn{2}{c}{Runtime (s)} \\
      states & protons & neutrons & Listing \ref{code:mbpt3V02} & Listing \ref{code:mbpt3V05} \\
      \hline
      \hline
        342 & 0 & 2 & $0.08$ & $0.02$\\
         & 0 & 14 & $2.38$ & $0.31$\\
         & 0 & 38 & $9.20$ & $0.49$\\
         684 & 2 & 2 & $1.05$ & $0.18$\\
         & 14 & 14 & $27.8$ & $2.00$\\
         & 38 & 38 & $107$ & $3.18$\\
         1598 & 0 & 2 & $1.81$ & $0.46$\\
         & 0 & 14 & $80.0$ & $10.5$\\
         & 0 & 38 & $456$ & $29.8$\\
         3196 & 2 & 2 & $23.1$ & $4.00$\\
         & 14 & 14 & $884$ & $69.8$\\
         & 38 & 38 & $> 10^3$ & $190$\\
      \hline
      \end{tabular}
  \end{center}
\end{table}

Table \ref{tab:mbpt3V02_runtime} summarizes the execution times so far. 

Table \ref{profile:mbpt3V05} shows that the potential function is still the most
expensive function in our program, but we would like to get a more detailed
profile of this function. The code to calculate the potential is filled with
calls to the exponential function which is part of the standard library. Since
we have linked to the standard library dynamically, gprof is not able to show
time spent in these functions. We can get a little bit more detail by linking
statically. This is done by introducing the -static flag to the compiler. Table
\ref{profile:mbpt3V05_v2} shows the new profile.

\begin{table}[h]
\caption{Flat profile for the MBPT3 implmentation in Listing
\ref{code:mbpt3V05} compiled with the -static flag enabled using 3196 states
calculating nuclear matter with 14 protons and 14
neutrons}\label{profile:mbpt3V05_v2}
\begin{verbatim}
Flat profile:

Each sample counts as 0.01 seconds.
  %   cumulative   self              self     total           
 time   seconds   seconds    calls  ms/call  ms/call  name    
 68.06     44.68    44.68                             __ieee754_exp_avx
 21.26     58.64    13.96 484191644     0.00     0.00  V_Minnesota(...)
  5.37     62.17     3.53                             exp
  2.18     63.60     1.43                             mbpt3V05::getEnergy()
  0.94     64.22     0.62                             dgemm_otcopy
  0.55     64.58     0.36                             dgemm_kernel
  0.53     64.93     0.35                             __mpexp_fma4
  0.25     65.09     0.17                             __floor_sse41
\end{verbatim}
\end{table}

The profile is now a lot more busy and it shows a longer runtime
than the previous profile. This is because gprof doesn't sample time spent
in dynamically linked libraries. The total runtime in this profile
corresponds better with Table \ref{tab:mbpt3V02_runtime}, but it is also more
difficult to read. What is clear is that the call to the function labelled
\verb;__ieee754_exp_avx; takes up almost $70\%$ of the total run time. This function represents
the calls to the exponential function in the potential code. If we can reduce the
number of evaluations of the exponential function, we can further reduce the run
time of this application. We leave that as an exercise to the reader.

\begin{figure}
\includegraphics[width=\linewidth]{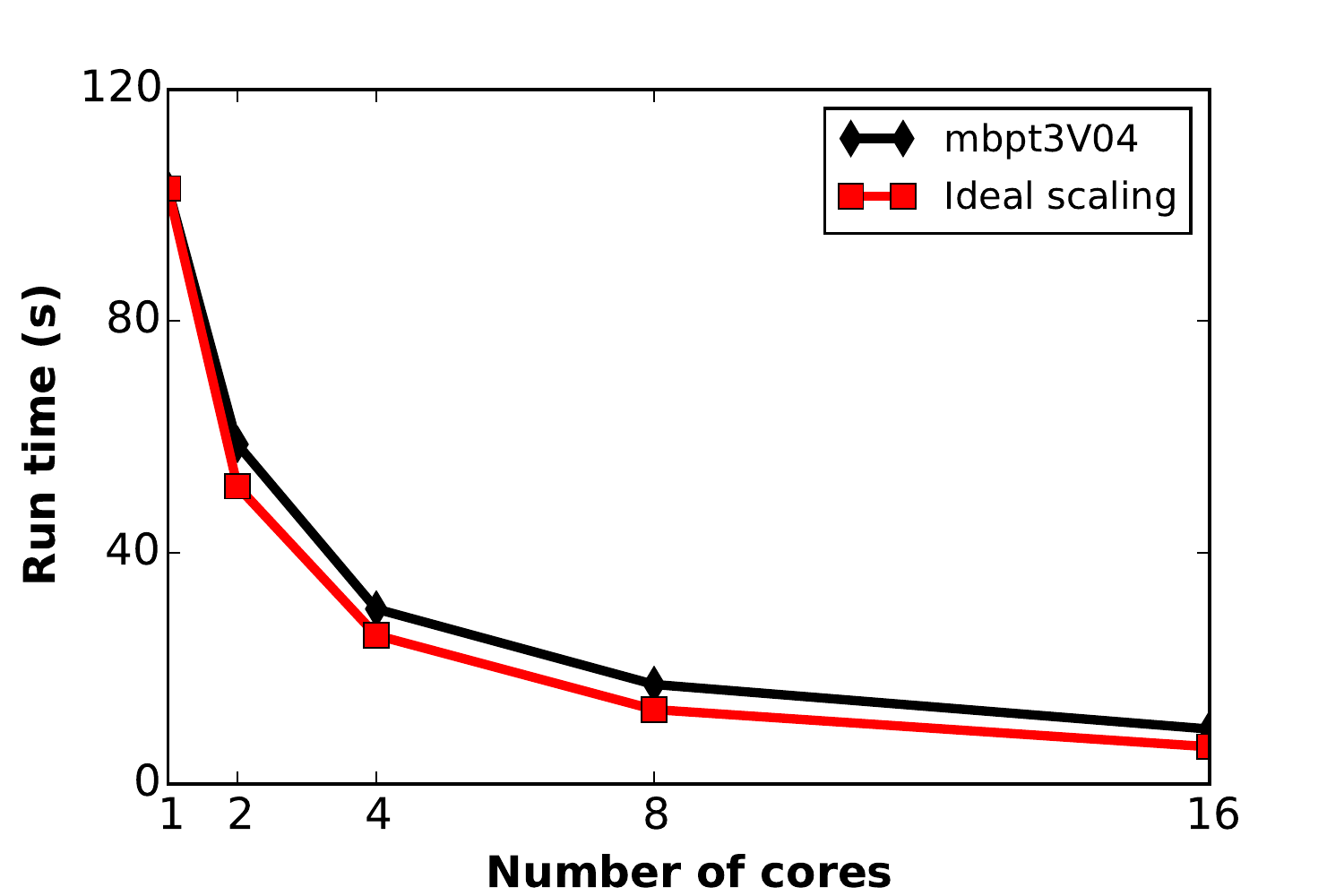}
\caption{Total runtime for the mbpt3 code in in Listing \ref{code:mbpt3V04}
uwing different number of OpenMP threads on a 16 code node of a Cray XK7.}
\label{fig:mbpt3_omp}
\end{figure}

The next level of optimization that we will discuss here is the introduction of
parallelism. Most modern computers have more than one cpu core available for
computation, but the codes we have presented so far will only run on one of
these cores. The simplest way to make this code run in parallel is to introduce
OpenMP~(\url{http://www.openmp.org/}) directives. This will split the work between multiple
execution streams that all share the same view of memory. Listing
\ref{code:mbpt3V04} shows a new version of the MBPT3 function where we have
introduced OpenMP directives in lines 19 and 21. The first line marks the start
of a parallel region and defines which variables the cores can share and
which must be duplicated. The second line defines a parallel loop, where each
core is responsible for only a section of the loop. As long as we have enough
work in the outermost loop, this strategy will work quite well as shown in
Figure \ref{fig:mbpt3_omp}. Here we show the total run time of this code using
different number of cores compared to the best run times we could have gotten
with this approach if our parallel regions scaled perfectly with the number of
cores. In reality this never happens. In this particular case, we could have
made the potential function more cache friendly. With this version the different
cores are fighting each other for access to memory and cache. This reduces
performance somewhat. 
        
\begin{lstlisting}[language=c++,numbers=left,firstline=16,lastline=69,
caption=Block-sparse implementation of a MBPT3 diagram with interaction stored
in memory and openmp directives. \label{code:mbpt3V04}]
double mbpt3V04::getEnergy() {
  double energy = 0.0;
  double *V1, *S1, *V2, *S2;
  char N = 'N';
  char T = 'T';
  double fac0 = 0.0;
  double fac1 = 1.0;
  int nhh, npp, i, j, a, b, c, d, idx;
  double energy0;
  for(int chan = 0; chan < channels.size; ++chan){
    nhh = channels.nhh[chan];
    npp = channels.npp[chan];
    if(nhh*npp == 0){ continue; }

    V1 = new double[nhh * npp];
    S1 = new double[nhh * npp];
    V2 = new double[npp * npp];
    S2 = new double[nhh * nhh];
    #pragma omp parallel shared(V1, V2) private(i, j, a, b, c, d, idx, energy0)
    {
      #pragma omp for schedule(static)
      for(int ab = 0; ab < npp; ++ab){
        a = channels.ppvec[chan][2*ab];
        b = channels.ppvec[chan][2*ab + 1];
        for(int hh = 0; hh < nhh; ++hh){
          i = channels.hhvec[chan][2*hh];
          j = channels.hhvec[chan][2*hh + 1];
          idx = hh * npp + ab;
          energy0 = V_Minnesota(modelspace, i, j, a, b, L);
          energy0 /= (modelspace.qnums[i].energy + modelspace.qnums[j].energy -
                        modelspace.qnums[a].energy - modelspace.qnums[b].energy);
          V1[idx] = energy0;
        }
        for(int cd = 0; cd < npp; ++cd){
          c = channels.ppvec[chan][2*cd];
          d = channels.ppvec[chan][2*cd + 1];
          idx = ab * npp + cd;
          V2[idx] = V_Minnesota(modelspace, a, b, c, d, L);
        }
      }
    }

    RM_dgemm(V1, V2, S1, &nhh, &npp, &npp, &fac1, &fac0, &N, &N);
    RMT_dgemm(S1, V1, S2, &nhh, &nhh, &npp, &fac1, &fac0, &N, &T);
    delete V1; delete V2; delete S1;
    for(int hh = 0; hh < nhh; ++hh){
      energy += S2[nhh*hh + hh];
    }
    delete S2;
  }
  energy *= 0.125;

  return energy;
}
\end{lstlisting}
In this function we have used the matrix-matrix multiplication function $dgemm$ of BLAS \cite{blas}. 
Finally, the above codes can easily be extended upon by including MPI \cite{mpi,openmpi} and/or a mix of OpenMP and MPI commands for distributed memory architectures. We leave this as a challenge to the reader. The coding practices and examples developed in this section, are reused in our development of the  coupled cluster code discussed in the next section. There  we discuss however in more detail how to develop an efficient coupled cluster code for infinite matter, with a focus on validation and verification and simplifications of the equations.

\subsection{Developing a CCD code for infinite matter\label{sec:ccdcode}}

  This section focuses on writing a working CCD code from
  scratch. Based on the previous discussion, what follows serves also the scope of outlining how to start a larger numerical project.  
We will in particular pay attention to possible benchmarks that can be used to validate our codes. 

We will assume that you have 
opted for a specific mathematical method for solving Schr\"odinger's equation. Here the mathematics is  given by the CCD equations.  
Our basic steps can then be split up as follows 
  \begin{itemize}
\item Write a first version of the CCD  code which is as close as possible to the mathematics of your equations. In this stage we will not 
focus on high-performance computing aspects and code efficiency.
\item Try to find possible benchmarks you can test your code against. In our case, the pairing model serves as an excellent testcase. 
\item With a functioning code that reproduces possible analytical and/or numerical results, we can start to analyze our code. In particular, if there are mathematical operations which can be simplified and/or can be represented in simpler ways etc etc. The modified code can hopefully reduce memory needs and time spent on computations.  The usage of specific symmetries of the interaction will turn out particularly useful. 
\end{itemize}
In this specific section, we will try to follow the above three steps,
with less attention on speed and numerical efficiency.  Our aim is to
have a code which passes central tests and can be properly validated
and verified.  If you are familiar with high-performance computing
topics, you are obviously not limited to follow the basic steps
outlined here.  However, when developing a numerical project we have
often found it easier and less error-prone to start with the basic
mathematical expressions. 
With a first functioning code, we will 
delve into high-performance computing topics. A good read on developing numerical projects and clear code is Martin's recent text \cite{martin2015}.
We recommend it highly and have borrowed many ideas and coding philosophies therefrom. 

We start with implementing the CCD equations as they stand here 
  \begin{align}
  \left(\epsilon_i+\epsilon_j-\epsilon_a-\epsilon_b\right)t_{ij}^{ab}
  = \langle ab \vert \hat{v} \vert ij \rangle+\frac{1}{2}\sum_{cd} \langle ab \vert \hat{v} \vert cd \rangle
  t_{ij}^{cd}+\frac{1}{2}\sum_{kl} \langle kl \vert \hat{v} \vert ij
  \rangle t_{kl}^{ab}+\hat{P}(ij\vert ab)\sum_{kc} \langle kb \vert
  \hat{v} \vert cj \rangle t_{ik}^{ac} & \nonumber
  \\ +\frac{1}{4}\sum_{klcd} \langle kl \vert \hat{v} \vert cd \rangle
  t_{ij}^{cd}t_{kl}^{ab}+\hat{P}(ij)\sum_{klcd} \langle kl \vert
  \hat{v} \vert cd \rangle t_{ik}^{ac}t_{jl}^{bd}-\frac{1}{2}\hat{P}(ij)\sum_{klcd} \langle kl \vert \hat{v} \vert
  cd \rangle t_{ik}^{dc}t_{lj}^{ab}-\frac{1}{2}\hat{P}(ab)\sum_{klcd}
  \langle kl \vert \hat{v} \vert cd \rangle t_{lk}^{ac}t_{ij}^{db},&
  \label{eq:ccd2b}
  \end{align}
  for all $i < j$ and all $a < b$, using the standard notation that
  $a,b,...$ are particle states and $i,j,...$ are hole states. The CCD correlation energy is given by
  \begin{equation}
  \Delta E_{CCD} = \frac{1}{4} \sum_{ijab}
  \braket{ij|\hat{v}|ab}t^{ab}_{ij}.
  \label{eq:ccdcorr}
  \end{equation}
  One way to solve these equations, is to write equation
  (\ref{eq:ccd2b}) as a series of iterative nonlinear algebraic
  equations
  \begin{align}
  t_{ij}^{ab}{}^{(n+1)}& = \frac{1}{\epsilon^{ab}_{ij}} \bigg(\langle
  ab \vert \hat{v} \vert ij \rangle +\frac{1}{2}\sum_{cd} \langle ab \vert \hat{v} \vert cd \rangle
  t_{ij}^{cd}{}^{(n)}+\frac{1}{2}\sum_{kl} \langle kl \vert \hat{v}
  \vert ij \rangle t_{kl}^{ab}{}^{(n)}+\hat{P}(ij\vert ab)\sum_{kc}
  \langle kb \vert \hat{v} \vert cj \rangle t_{ik}^{ac}{}^{(n)} & 
  \nonumber \\ & +\frac{1}{4}\sum_{klcd} \langle kl \vert \hat{v} \vert
  cd \rangle
  t_{ij}^{cd}{}^{(n)}t_{kl}^{ab}{}^{(n)}+\hat{P}(ij)\sum_{klcd}
  \langle kl \vert \hat{v} \vert cd \rangle
  t_{ik}^{ac}{}^{(n)}t_{jl}^{bd}{}^{(n)} \nonumber \\ &-\frac{1}{2}\hat{P}(ij)\sum_{klcd} \langle kl \vert \hat{v} \vert
  cd \rangle
  t_{ik}^{dc}{}^{(n)}t_{lj}^{ab}{}^{(n)}-\frac{1}{2}\hat{P}(ab)\sum_{klcd}
  \langle kl \vert \hat{v} \vert cd \rangle
  t_{lk}^{ac}{}^{(n)}t_{ij}^{db}{}^{(n)} \bigg),
  \label{eq:ccd3}
  \end{align}
  for all $i < j$ and all $a < b$, where $\epsilon^{ab}_{ij} =
  \left(\epsilon_i+\epsilon_j-\epsilon_a-\epsilon_b\right)$, and
  $t_{ij}^{ab}{}^{(n)}$ is the $t$ amplitude for the nth iteration of
  the series. This way, given some starting guess
  $t_{ij}^{ab}{}^{(0)}$, we can generate subsequent $t$ amplitudes
  that converge to some value. Discussions of the mathematical details
  regarding convergence will be presented below; for now we will mainly focus
  on implementing these equations into a computer program and
  assume convergence. In pseudocode, the function that updates the $t$
  amplitudes looks like
\begin{svgraybox}
  \begin{algorithmic} 
  \State CCD\_Update() \For{$i \in \{0,N_{\mathrm{Fermi}}-1\}$ } \For{$j \in
    \{0,N_{\mathrm{Fermi}}-1\}$ } \For{$a \in \{N_{\mathrm{Fermi}},N_{sp}-1\}$ } \For{$b
    \in \{N_{\mathrm{Fermi}},N_{sp}-1\}$ } \State $\text{sum} \gets
  \text{TBME}[\text{index}(a,b,i,j)$] \For{$c \in
    \{N_{\mathrm{Fermi}},N_{sp}-1\}$ } \For{$d \in \{N_{\mathrm{Fermi}},N_{sp}-1\}$ }
  \State $\text{sum} \gets \text{sum} +
  0.5\times\text{TBME}[\text{index}(a,b,c,d)] \times
  t\_\text{amplitudes}\_\text{old}[\text{index}(c,d,i,j)]$ \EndFor
  \EndFor \State ...  \State sum $\gets$ sum + (all other terms)
  \State ...  \State energy\_denom =
  SP\_energy[$i$]+SP\_energy[$j$]-SP\_energy[$a$]-SP\_energy[$b$]
  \State t\_amplitudes[index($a,b,i,j$)] = sum/energy\_denom \EndFor
  \EndFor \EndFor \EndFor
  \end{algorithmic}
\end{svgraybox} 
 Here we have defined $N_{\mathrm{Fermi}}$ to be the fermi level while $N_{sp}$ is the total
  number of single particle (s.p.) states, indexed from 0 to
  $N_{sp}-1$. At the most basic level, the CCD equations are just the
  addition of many products containing $t_{ij}^{ab}$ amplitudes and
  two-body matrix elements (TBMEs) $\braket{ij|\hat{v}|ab}$.
  Care should thus be placed into how we store these objects. These are
  objects with four indices and a  sensible first implementation
  of the CCD equations would be to create two four-dimensional arrays to store the
  objects. However, it is often more convenient to work with simple
  one-dimensional arrays instead. The function $index()$ maps the four
  indices onto one index so that a one-dimensional array can be used. An example of a brute force implementation 
  of such a function is
\begin{svgraybox} 
 \begin{algorithmic}
  \Function{index}{$p,q,r,s$} \State \textbf{return} $p\times N_{sp}^3 +
  q\times N_{sp}^2 + r\times N_{sp} + s$ \EndFunction
  \end{algorithmic}
\end{svgraybox}
  Because elements with repeated indices vanish,
  $t_{ii}^{ab}=t_{ij}^{aa}=0$ and
  $\braket{pp|\hat{v}|rs}=\braket{pq|\hat{v}|rr}=0$, data structures
  using this index function will contain many elements that are
  automatically zero. This means that we need to discuss more efficient storage
  strategies later. Notice also that we are looping over all
  indices $i,j,a,b$, rather than the restricted indices. This means that we
  are doing redundant work. The reason for presenting the equations this way is merely pedagogical. When developing a program, we would recommend to write a code which is as close as possible to the mathematical expressions. The first version of our code will then often be slow, as discussed in subsection \ref{subsec:profiling}.  
Below we will however unrestrict these indices in order to achieve a better speed up of our code. 

\begin{figure}
    \includegraphics[width=\linewidth]{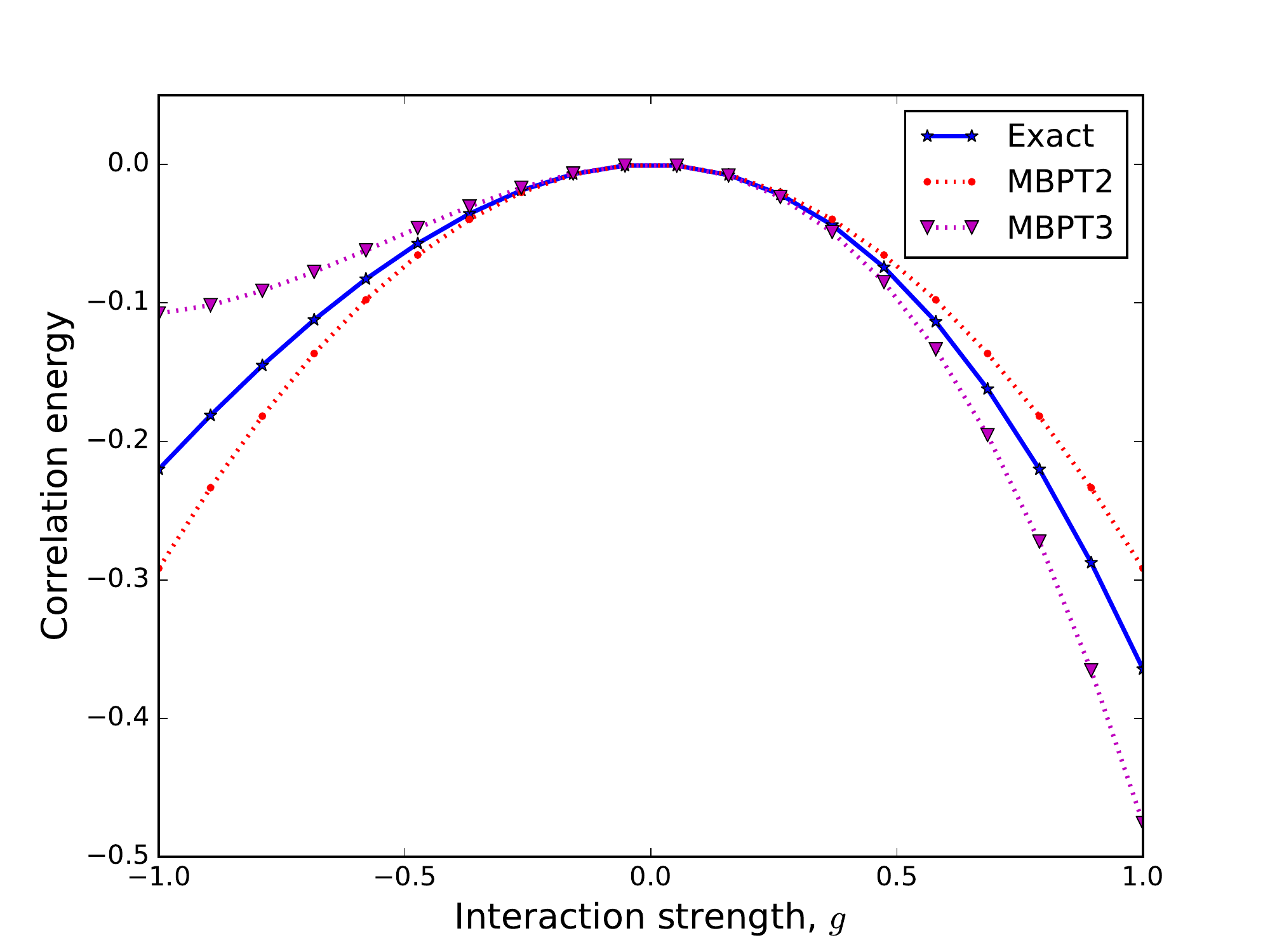}
    \caption{Correlation energy for the pairing model with exact diagonalization, MBPT2 and perturbation theory to third order MBPT3 for a range of interaction values.}
    \label{fig:diagpairing}
  \end{figure}

\begin{figure}
    \includegraphics[width=\linewidth]{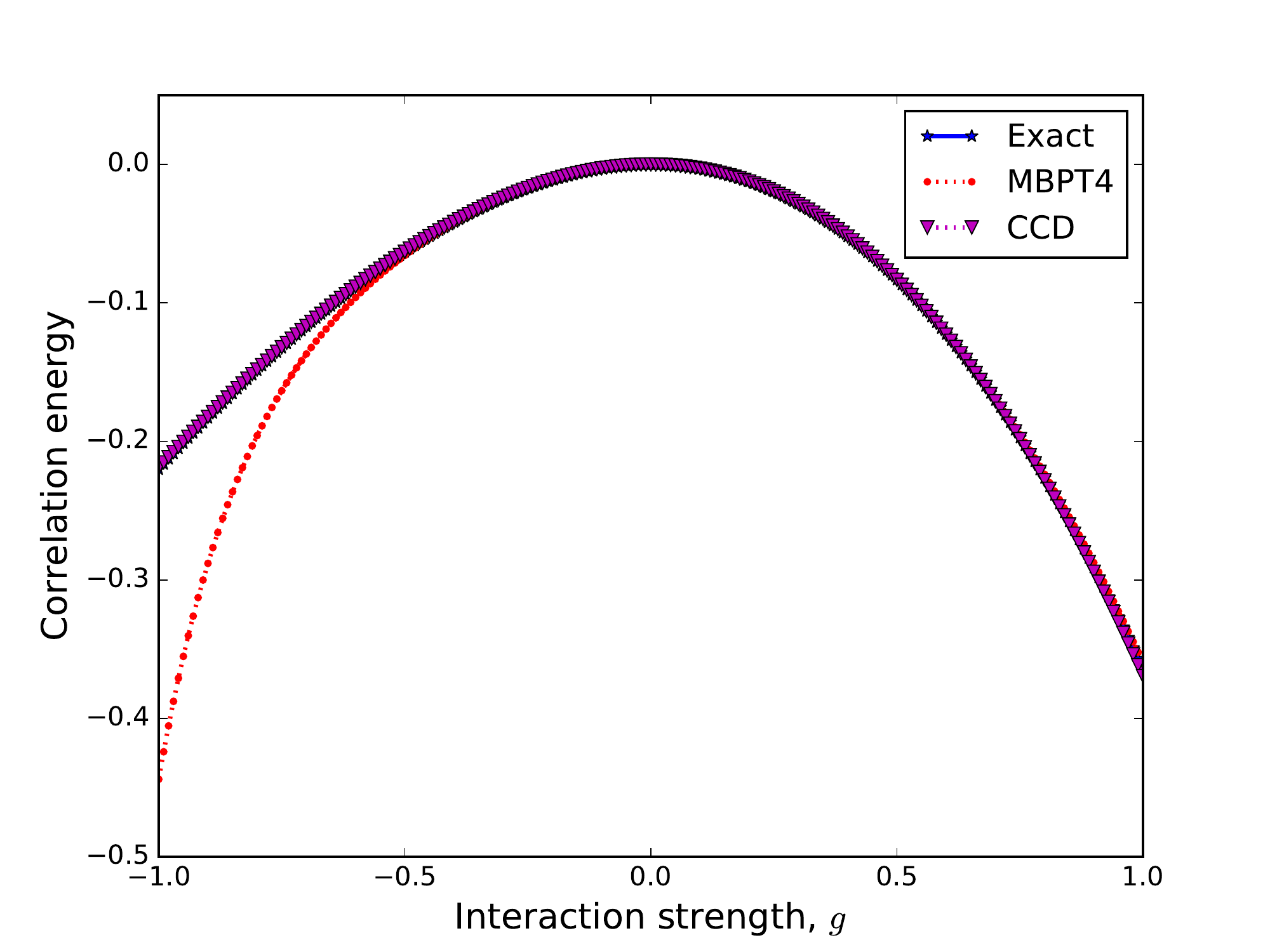}
    \caption{Correlation energy for the pairing model with exact diagonalization, CCD and perturbation theory to fourth order (MBPT4) for a range of interaction values.}
    \label{fig:pairingccmbpt4}
  \end{figure}

   The goal of our code is to calculate the correlation energy,
   $\Delta E_{CCD}$, meaning that after each iteration of our equations, we use
   our newest $t$ amplitudes to update the correlation energy
  \begin{equation}
  \Delta E_{CCD}^{(n)} = \frac{1}{4} \sum_{ijab}
  \braket{ij|\hat{v}|ab}t^{ab}_{ij}{}^{(n)}.
  \end{equation}
  We check that our result is converged by testing whether the
  most recent iteration has changed the correlation energy by less
  than some tolerance threshold $\eta$,
  \begin{equation}
  \eta > | \Delta E_{CCD}^{(n+1)} - \Delta E_{CCD}^{(n)} |.
  \end{equation}
  The basic structure of the iterative process looks like
\begin{svgraybox}
  \begin{algorithmic}
    \While {(abs(energy\_Diff) $>$ tolerance)} \State CCD\_Update()
    \State correlation\_Energy $\gets$ CCD\_Corr\_Energy() \State
    energy\_Diff $\gets$ correlation\_Energy -
    correlation\_Energy\_old \State correlation\_Energy\_old $\gets$
    correlation\_Energy \State t\_amplitudes\_old $\gets$
    t\_amplitudes \EndWhile
  \end{algorithmic}
\end{svgraybox}
  Prior to this algorithm, the $t$ amplitudes should be initalized,
  $t_{ij}^{ab}{}^{(0)}$. A particularly convenient choice, as discussed above, is to 
use many-body perturbation theory for the wave operator with
  \begin{equation}
  t_{ij}^{ab}{}^{(0)} = \frac{\langle ab \vert \hat{v} \vert ij\rangle}{\epsilon^{ab}_{ij}},
  \label{eq:ccdGuess}
  \end{equation}
which results in the correlation energy
  \begin{equation}
  \Delta E_{CCD}^{(1)} = \frac{1}{4} \sum_{ijab}\frac{\langle ij \vert \hat{v} \vert ab \rangle\langle ab \vert \hat{v} \vert ij \rangle}{\epsilon^{ab}_{ij}}.
  \end{equation}
  This is the familiar result from many-body perturbation theory to second order (MBPT2).  It is a very useful result, as one iteration
  of the CCD equations can be ran, and checked against MBPT2 to give
  some confidence that everything is working correctly. Additionally, running a program using a minimal test case is another useful way to make sure that a program is working correctly. For this purpose, we turn our attention to the simple pairing model Hamiltonian of problem \ref{problem:pairingmodel},
  \begin{equation}
  \hat{H}_0 = \delta \sum_{p \sigma} (p-1) a^{\dagger}_{p \sigma} a_{p
    \sigma}\label{eq:sppairing}
  \end{equation}
  \begin{equation}
  \hat{V} = -\frac{1}{2}g \sum_{pq} a^{\dagger}_{p+}a^{\dagger}_{p-}
  a_{q-}a_{q+}\label{eq:intpairing}
  \end{equation}
  which represents a basic pairing model with $p$ levels, each having a
  spin degeneracy of 2. The form of the coupled cluster equations 
  uses single-particle states that are eigenstates of the
  Hartree-Fock operator, $\left(\hat{u}+\hat{u}_{\text{HF}}\right)\vert
  p\rangle=\epsilon_{p}\vert p\rangle$. In the pairing model, this
  condition is already fulfilled. All we have to do is define the
  lowest $N_{\mathrm{\mathrm{Fermi}}}$ states as holes and  redefine the single-particle
  energies,
  \begin{equation}\label{eq:pairingsp}
  \epsilon_q = h_{qq} + \sum_{i} \braket{qi|\hat{v}|qi}.
  \end{equation}
  To be more specific, let us look at the pairing model with four
  particles and eight single-particle states. These states (with $\delta =1.0$) could be labeled as shown in 
Table \ref{tab:pairingmodelsp}.
\begin{table}
\caption{Single-particle states and their quantum numbers and their energies from Eq.~(\ref{eq:pairingsp}). The degeneracy for every quantum number $p$ is equal to 2 due to the two possible spin values.} \label{tab:pairingmodelsp}
  \begin{center}
      \begin{tabular}{| l | l | l | l | l |}
      \hline State Label & p & 2s$_z$ & E & type\\ \hline 0 & 1 & 1 &
      -g/2 & hole \\ \hline 1 & 1 & -1 & -g/2 & hole \\ \hline 2 & 2 &
      1 & 1-g/2 & hole \\ \hline 3 & 2 & -1 & 1-g/2 & hole \\ \hline 4
      & 3 & 1 & 2 & particle \\ \hline 5 & 3 & -1 & 2 & particle
      \\ \hline 6 & 4 & 1 & 3 & particle \\ \hline 7 & 4 & -1 & 3 &
      particle \\ \hline
      \end{tabular}
  \end{center}
\end{table}
The Hamiltonian matrix for this   four-particle problem with no broken pairs is defined by six possible Slater determinants,
one representing the ground state and zero-particle-zero-hole excitations $0p-0h$, four representing various $2p-2h$ excitations and finally one representing a $4p-4h$ excitation. Problem \ref{problem:pairingmodel} gives us for this specific problem
  \[
  H = \begin{bmatrix}
  2\delta -g & -g/2 & -g/2 & -g/2 & -g/2 & 0 \\ -g/2 & 4\delta -g &
  -g/2 & -g/2 & -0 & -g/2 \\ -g/2 & -g/2 & 6\delta -g & 0 & -g/2 &
  -g/2 \\ -g/2 & -g/2 & 0 & 6\delta-g & -g/2 & -g/2 \\ -g/2 & 0 & -g/2
  & -g/2 & 8\delta-g & -g/2 \\ 0 & -g/2 & -g/2 & -g/2 & -g/2 &
  10\delta -g
  \end{bmatrix}
  \]
  The python program (included for pedagogical purposes only) at \url{https://github.com/ManyBodyPhysics/LectureNotesPhysics/tree/master/Programs/Chapter8-programs/python/mbpt.py} diagonalizes the above Hamiltonian
  matrix for a given span of interaction strength values, performing
  a full configuration interaction calculation. It plots the correlation energy, that is the difference between the ground state energy and the reference energy. Furthermore, for the pairing model we have added results from perturbation theory to second order (MBPT2)
and third order in the interaction MBPT3. Second order perturbation theory includes diagram (2) of Fig.~\ref{fig:goldstone}
while MBPT3 includes diagrams (3), (4), (5), (8) and (9) as well. Note that diagram (3) is zero for the pairing model and that diagrams (8) and (9) contribute as well with a canonical Hartree-Fock basis. 
  In the case of the simple pairing model it is easy to calculate
  $\Delta E_{MBPT2}$ anyltically. This is a very useful  check of our codes since this analytical expression  can  also be used to check our first CCD iteration.
We restate this expression here but restrict the sums over single-particle states
  \[
  \Delta E_{MBPT2} = \frac{1}{4} \sum_{abij} \frac{\braket{ij|\hat{v}|ab}
    \braket{ab|\hat{v}|ij}}{ \epsilon_{ij}^{ab}} = \sum_{a<b,i<j}
  \frac{\braket{ij|\hat{v}|ab} \braket{ab|\hat{v}|ij}}{ \epsilon_{ij}^{ab}}
  \]
  For our pairing example we obtain the following result
  \[
  \Delta E_{MBPT2} = \frac{\braket{01|\hat{v}|45}^2}{\epsilon_{01}^{45}} +
  \frac{\braket{01|\hat{v}|67}^2}{\epsilon_{01}^{67}} +
  \frac{\braket{23|\hat{v}|45}^2}{\epsilon_{23}^{45}} +
  \frac{\braket{23|\hat{v}|67}^2}{\epsilon_{23}^{67}},
  \]
which translates into
  \[
  \Delta E_{MBPT2} = -\frac{g^2}{4} \bigg( \frac{1}{ 4 + g} +
  \frac{1}{ 6 + g} + \frac{1}{ 2 + g} + \frac{1}{ 4 + g} \bigg).
  \]
 This expression can be used to check the results
  for any value of $g$ and provides thereby an important test of  our codes.
Figure \ref{fig:diagpairing} shows the resulting correlation energies for the exact case, MBPT2 and MBPT3.
We note from the above program that we have coded the expressions for
the various diagrams following strictly the mathematical expressions
of for example Eqs.~(\ref{eq:diag3})-(\ref{eq:diag5}).  This means
that for every diagram we loop explicitely over every single-particle
state. As we will see later, this is extremely inefficient from a
computational point of view. In our discussions below, we will rewrite
the computations of most diagrams in terms of efficient matrix-matrix
multiplications or matrix-vector multiplications.  Figure
\ref{fig:diagpairing} shows us that the approximation to both second
and third order are very good when the interaction strength is small
and contained in the interval $g\in[-0.5,0.5]$, but as the interaction
gets stronger in absolute value the agreement with the exact reference
energy for MBPT2 and MBPT3 worsens. We also note that the third-order
result is actually worse than the second order result for larger
values of the interaction strength, indicating that there is no
guarantee that higher orders in many-body perturbation theory may
reduce the relative error in a systematic way.  Adding fourth order
contributions as shown in Fig.~\ref{fig:pairingccmbpt4} for negative
interaction strengths gives a better result than second and third
order, while for $g>0$ the relative error is worse.  The fourth order
contributions (not shown in the above figure) include also
four-particle-four-hole correaltions.  However, the disagreement for
stronger interaction values hints at the possibility that many-body
perturbation theory may not converge order by order.  Perturbative
studies of nuclear systems may thus be questionable, unless selected
contributions that soften the interactions are properly softened.  We
note also the non-variational character of many-body perturbation
theory, with results at different levels of many-body perturbation
theory either overshooting or undershooting the true ground state
correlation energy.  The coupled cluster results are included in
Fig.~\ref{fig:pairingccmbpt4} where we display the difference between
the exact correlation energy and the correlation energy obtained with
many-body perturbation theory to fourth order.  The MBPT4 results show
a better agreement with the exact solution, but here as well for
larger positive values of the interaction, we see clear signs of
deviations. Coupled cluster theory with doubles only shows a very good
agreement with the exact results. For larger values of $g$ one will
however observe larger discrepancies.
In Table \ref{tab:selectedbenchmarks} we list for the sake of completeness also our coupled cluster results at the CCD level for the same system.
\begin{table}
\caption{Coupled cluster and MBPT2 results for the simple pairing model with eight single-particle levels and four spin $1/2$ fermions
for different values of the interaction strength $g$.} \label{tab:selectedbenchmarks}
  \begin{center}
      \begin{tabular}{| l | l | l | l |}
      \hline g & E$_{ref}$ & $\Delta E_{MBPT2}$ & $\Delta E_{CCD}$
      \\ \hline -1.0 & 3 & -0.46667 & -0.21895* \\ \hline -0.5 & 2.5 &
      -0.08874 & -0.06306 \\ \hline 0.0 & 2 & 0 & 0 \\ \hline 0.5 &
      1.5 & -0.06239 & -0.08336 \\ \hline 1.0 & 1 & -0.21905 &
      -0.36956 \\ \hline
      \end{tabular}
  \end{center}
\end{table}
  The $g=-1.0$ case diverges without implementing iterative
  mixing. Sometimes iterative solvers run into oscillating solutions,
  and mixing can help the iterations break this cycle.
  \begin{equation}
  t^{(i)} = \alpha t^{(i)}_{no\_mixing} + (1 - \alpha) t^{(i-1)}
  \end{equation}

  Once a working pairing model has been implemented, improvements can
  start to be made, all the while using the pairing model to make sure
  that the code is still working and giving correct answers. Realistic
  systems will be much larger than this small pairing example.

  One limitation that will be ran into while trying to do realistic
  CCD calculations is that of memory. The four-indexed two-body matrix elements (TBMEs) and
  $t$-amplitudes have to store a lot of elements, and the size of these
  arrays can quickly exceed the available memory on
  a machine. If a calculation wants to use 500 single-particle basis states, then
  a structure like $\braket{pq|v|rs}$ will need a  length of 500 for each of
  its four indices, which means it will have $500^4 = 625\times 10^8$
  elements. To get a handle on how much memory is used, consider the
  elements as double-precision floating point type. One double takes
  up 8 bytes of memory. This specific array would take up $8\times 625\times 10^8$ bytes
  = $5000 \times 10^8$ bytes = $500$ Gbytes of memory. Most personal
  computers in 2016 have 4-8 Gbytes of RAM, meaning that this calculation would
  be way out of reach. There are supercomputers that can handle
  applications using 500 Gbytes of memory, but we can quickly reduce
  the total memory required by applying some physical arguments. In
  addition to vanishing elements with repeated indices, mentioned
  above, elements that do not obey certain symmetries are also
  zero. Almost all realistic two-body forces preserve some quantities
  due to symmetries in the interaction. For example, an interaction
  with rotational symmetry will conserve angular momentum. This means
  that a two-body ket state $\ket{rs}$, which has some set of quantum
  numbers, will retain quantum numbers corresponding to the
  interaction symmetries after being acted on by $\hat{v}$. This state
  is then projected onto $\ket{pq}$ with its own set of quantum
  numbers. Thus $\braket{pq|v|rs}$ is only non-zero if $\ket{pq}$ and
  $\ket{rs}$ share the same quantum numbers that are preserved by
  $\hat{v}$. In addition, because the cluster operators represent
  excitations due to the interaction, $t_{ij}^{ab}$ is only non-zero
  if $\ket{ij}$ has the same relevant quantum numbers as $\ket{ab}$.

  To take advantage of this, these two-body ket states can be
  organized into symmetry ``channels'' of shared quantum numbers. In the case
  of the pairing model, the interaction preserves the total spin
  projection of a two-body state, $S_{z}=s_{z1}+s_{z2}$. The single
  particle states can have spin of +1/2 or -1/2, so there can be three
  two-body channels with $S_{z}=-1,0,+1$. These channels can then be
  indexed with a unique label in a similar way to the single particle
  index scheme. In more complicated systems, there will be many more
  channels involving multiple symmetries, so it is useful to create a
  data structure that stores the relevant two-body quantum numbers to
  keep track of the labeling scheme.

  It is more efficient to use two-dimensional array data
  structures, where the first index refers to the channel number and
  the second refers to the element within that channel. So to access
  matrix elements or $t$ amplitudes, you can loop over the channels
  first, then the indices within that channel. To get an idea of the
  savings using this block diagonal structure, let's look at a case
  with a plane wave basis, with three momentum and one spin quantum
  numbers, with an interaction that conserves linear momentum in all
  three dimensions, as well as the total spin projection. Using 502
  basis states, the TBME's require about 0.23 Gb of memory in block
  diagonal form, which is an enormous saving from the 500 Gb mentioned
  earlier in the na\"ive storage scheme.

  Since the calculation of all  zeros can now be avoided,
  improvements in speed and memory will now follow. To get a handle on
  how these CCD calculations are implemented we need only to look at the
  most expensive sum in equation \ref{eq:ccd2b}. This corresponds to
  the sum over $klcd$. Since this sum is repeated for all $i < j$ and
  $a < b$, it means that these equations will scale as
  $\mathcal{O}(n_{p}^{4} n_{h}^{4})$. However,
  they can be rewritten using intermediates as

  \begin{align}
  0 &= \braket{ab|\hat{v}|ij} + \hat{P}(ab) \sum_{c} \braket{b| \chi
    |c} \braket{ac| t |ij} - \hat{P}(ij) \sum_{k} \braket{k| \chi |j}
  \braket{ab| t |ik} \nonumber \\ &+ \frac{1}{2}\sum_{cd} \braket{ab|
    \chi |cd} \braket{cd| t |ij} + \frac{1}{2} \sum_{kl} \braket{ab| t
    |kl} \braket{kl| \chi |ij} \\ &+ \hat{P}(ij)\hat{P}(ab) \sum_{kc}
  \braket{ac| t |ik}\braket{kb| \chi |cj}, \nonumber
  \end{align}
  for all $i,j,a,b$, the reason why these indices are now unrestricted
  will be explained later. The intermediates $\chi$ are defined as
  \begin{equation}
  \braket{b| \chi |c} = \braket{b|f|c} - \frac{1}{2} \sum_{kld}
  \braket{bd|t|kl} \braket{kl|v|cd},
  \end{equation}
  \begin{equation}
  \braket{k| \chi |j} = \braket{k|f|j} + \frac{1}{2} \sum_{cdl}
  \braket{kl|v|cd} \braket{cd|t|jl},
  \end{equation}
  \begin{equation}
  \braket{kl| \chi |ij} = \braket{kl|v|ij} + \frac{1}{2} \sum_{cd}
  \braket{kl|v|cd} \braket{cd|t|ij},
  \label{eq:mtxEx}
  \end{equation}
  \begin{equation}
  \braket{kb| \chi |cj} = \braket{kb|v|cj} + \frac{1}{2} \sum_{dl}
  \braket{kl|v|cd} \braket{db|t|lj},
  \end{equation}
  \begin{equation}
  \braket{ab| \chi |cd} = \braket{ab|v|cd}.
  \end{equation}

\begin{figure}
    \includegraphics[width=\linewidth]{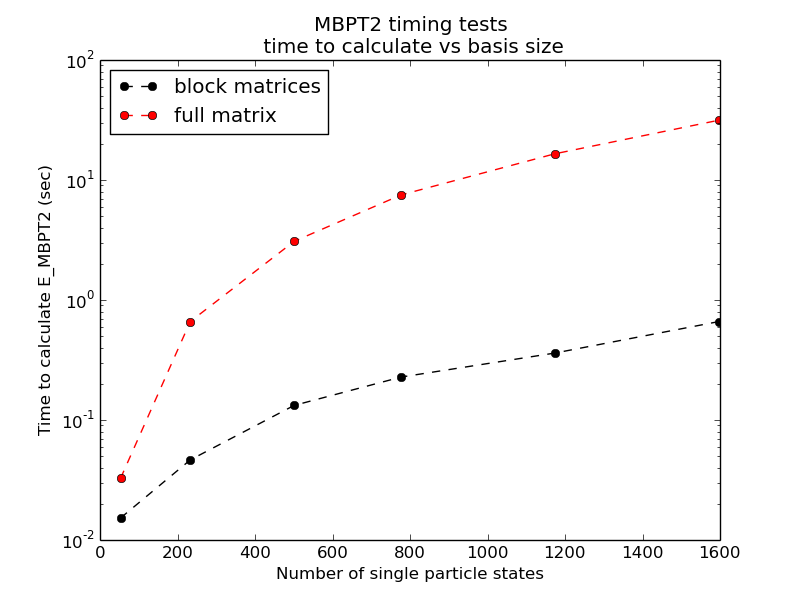}
    \caption{MBPT2 contribution to the correlation for pure neutron matter with $N=14$ neutrons and periodic boundary conditions. Up to approximately 1600 single-particle states have been included in the sums over intermediate states in Eqs.~(\ref{eq:bruteforceMBPT}) and (\ref{eq:smartMBPT})}. 
    \label{fig:speedup1}
  \end{figure}

  With the introduction of the above intermediates, the CCD equations scale now as $\mathcal{O}(n_{h}^{2}
  n_{p}^{4})$, which is an important improvement. This is of
  course at the cost of computing the intermediates at the beginning
  of each iteration, where the most expensive one, $\braket{kb| \chi |cj}$ scales as $\mathcal{O}(n_{h}^{3} n_{p}^{3})$. To
  further speed up these computations, we see that these sums can be
  written in terms of  matrix-matrix multiplications. It is not obvious how to
  write all of these sums in such a way, but it is useful to first
  recall that the expression for the multiplication of two matrices $\hat{C} =
  \hat{A}\times \hat{B}$ can be written as
  \begin{equation}
  C_{ij} = \sum_{k} A_{ik} \times B_{kj}.
  \end{equation}
  We see then  that equation (\ref{eq:mtxEx}) can be written as
  \[
  \braket{K| \chi |I} = \braket{K|v|I} + \frac{1}{2} \sum_{C}
  \braket{K|v|C} \braket{C|t|I}
  \]
  by mapping the two index pairs $kl \to K, ij \to I, cd \to C$. The sum looks now 
like a matrix-matrix multiplication. This is
  useful because there are packages like BLAS (Basic Linear Algebra
  Subprograms) \cite{blas} which have extremely fast implementations of
  matrix-matrix multiplication, as discussed in connection with the listing \label{code:mbpt3V04}.
  The simplest example to consider is the expression for the correlation energy from MBPT2. We rewrite 
  \begin{equation}\label{eq:bruteforceMBPT}
  \Delta E_{MBPT2} = \frac{1}{4}\sum_{abij} \frac{\braket{ij|\hat{v}|ab}\braket{ab|\hat{v}|ij}}{ \epsilon_{ij}^{ab}},
\end{equation}
by defining the matrices $\hat{A}$ and $\hat{B}$ with new indices $I=(ij)$ and $A=(ab)$. The individual matrix elements of these matrices are 
\[
A_{IA} = \langle I \vert \hat{v} \vert A \rangle,
\]
and 
\[
B_{AI} = \frac{\langle A \vert \hat{v} \vert I \rangle}{\epsilon^A_I}.
\]
We can then rewrite the correlation energy from MBPT2 as
\begin{equation}\label{eq:smartMBPT}
  \Delta E_{MBPT2} = \frac{1}{4}\hat{A}\times \hat{B}.
\end{equation}
Figure \ref{fig:speedup1} shows the difference between the brute force summation over single-particle states
of Eq.~(\ref{eq:bruteforceMBPT}) and the smarter set up in terms of indices including two-body configurations only, that is Eq.~(\ref{eq:smartMBPT}).
In these calculations we have only considered pure neutron matter with $N=14$ neutrons and a density $n=0.08$ fm$^{-3}$ and plane wave single-particle states with periodic boundary conditions, allowing for up to 
$1600$ single-particle basis states. The Minnesota interaction model \cite{minnesota} has been used in these calculations.  
With $40$ single-particle shells, see Table \ref{tab:table1} for
example, we have in total $2713$ single-particle states.  Using the
smarter algorithm the final calculation time is $2.4$ s (this is the
average time from ten numerical experiments).  The total time using
the brute force summation over single-particle indices is $100.6$ s
(again the average of ten numerical experiments), resulting in a speed
up of 42. It is useful to dissect the final time in terms of different
operations.  For the smarter algorithm most of the time is spent
setting up the matrix elements for the two-body channels and to load
the matrix elements. The final matrix-matrix multiplication takes only
$1\%$ of the total time. For the brute force algorithm, the
multiplication and summation over the various single-particle states
takes almost half of the total time. Here we have deliberately only
focused on the difference between the two ways of computing
Eqs.~(\ref{eq:bruteforceMBPT}) and (\ref{eq:smartMBPT}). We have, on
purpose, not performed a proper timing analysis since this was done in the previous subsection. 
In this section we
have chosen to focus on the development of a program which produces
the correct results.  As mentioned above, the true elephant in the room, in terms of
computational time, is our computation of the matrix elements of the
nucleon-nucleon potential. We have deliberately omitted the time spent on setting up the interactions here.
For the rest of this section we wil focus on various physics
applications of our newly developed CCD code.

With the definition of the intermediates and appropriate matrix-matrix
multiplications and a working CCD program, we can move on to more
realistic cases. One such case is infinite nuclear matter using a
plane-wave basis. These states are solutions to the free-particle
Hamiltonian,
  \begin{equation}
  \frac{-\hbar^2}{2m}\nabla^2\mathop{\phi(\vec{x})}=\epsilon\mathop{\phi(\vec{x})}.
  \end{equation}
  For a finite basis, we approximate the problem by constructing a box
  with sides of length $L$, which quantizes the momentum, and impose
  periodic boundary conditions in each direction.
  \begin{align}
  \mathop{\phi(x_{i})}=\mathop{\phi(x_{i}+L)}
  \\ \mathop{\phi_{\vec{k}}(\vec{x})}=\frac{1}{\sqrt{L^{3}}}e^{i\vec{k}\cdot\vec{x}},\hspace{0.5cm}\vec{k}=\frac{2\pi\vec{n}}{L}
  \end{align}

\begin{figure}
    \includegraphics[width=\linewidth]{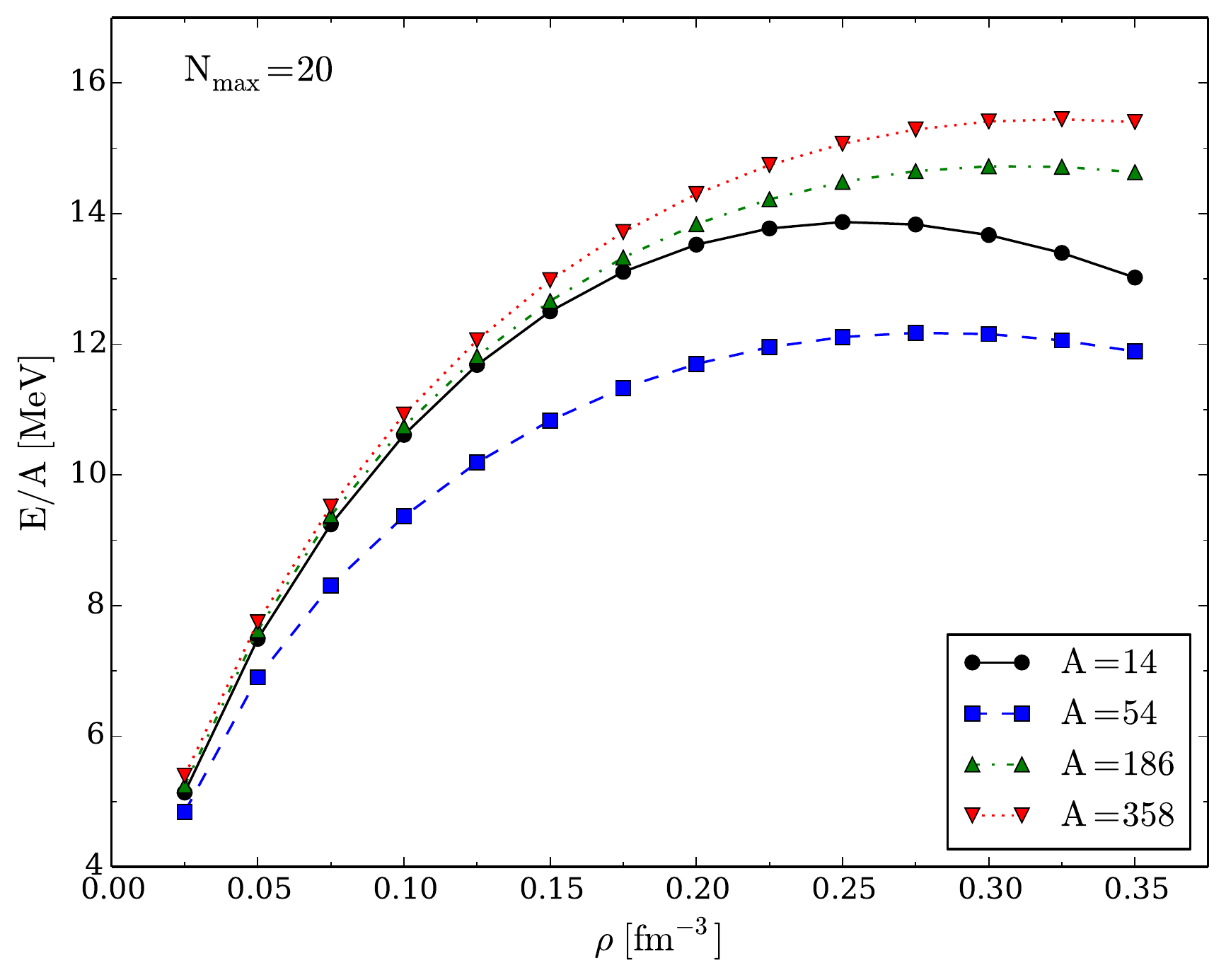}
    \caption{Energy per particle of pure neutron matter computed in
      the CCD approximation with the Minnesota interaction model  \cite{minnesota} for different
      numbers of particles with $\mathrm{N_{max}=20}$.}
    \label{fig:fig1}
  \end{figure}

\begin{figure}
    \includegraphics[width=\linewidth]{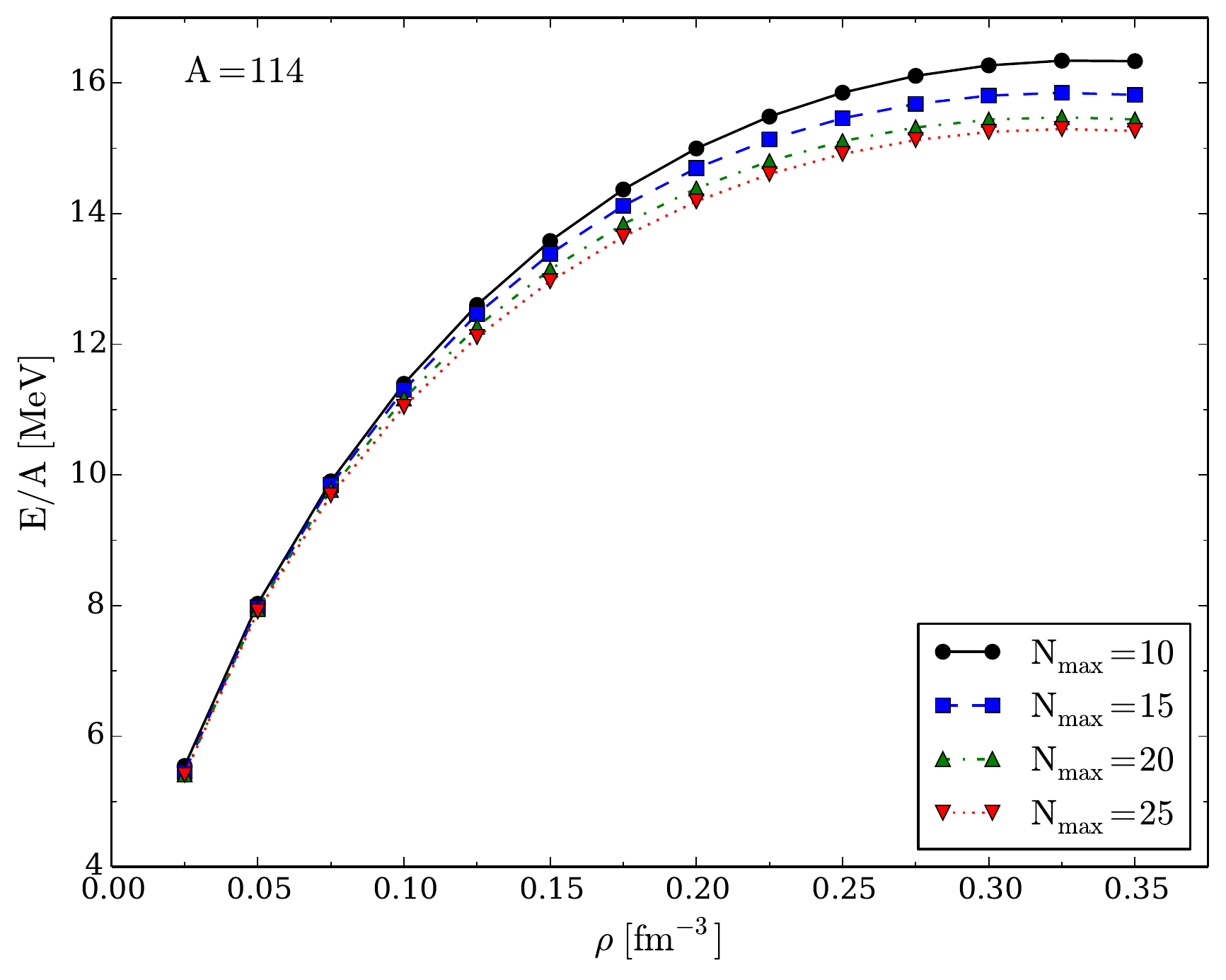}
    \caption{Energy per particle of pure neutron matter computed in
      the CCD approximation with the Minnesota interaction model  \cite{minnesota} for different
      model space sizes with $\mathrm{A=114}$.}
    \label{fig:fig2}
  \end{figure}

  The first step in calculating infinite matter is to construct a
  model space by finding every single-particle state relevant to a
  given problem. In our case, this amounts to looping over the quantum
  numbers for spin, isospin, and the three momentum directions. To
  control the model space size, the momentum can be truncated to give
  a cubic space, where $n_{i}\leq n_{\text{max}}$, or a spherical
  space, where $n_{x}^{2}+n_{y}^{2}+n_{z}^{2}\leq N_{\text{max}}$. The
  number of single-particle states in a cubic space increases rapidly
  with $n_{\text{max}}$ compared to the spherical case with
  $N_{\text{max}}$. For example, in pure neutron matter a cubic space
  with $n_{\text{max}}=3$ has $668$ states while the spherical space
  with $N_{\text{max}}=17$ has $610$ states. Therefore, the spherical
  case will be used for the rest of the calculations here. The loop
  increases in energy by counting the number of shells, so states can
  be 'filled' by labeling the first $P$ proton and $N$ neutron states
  as holes. The following loop is for pure neutron matter and requires
  the number of neutrons, $N$ and density, $\rho=N/L^{3}$, as
  input. Symmetric nuclear matter requires an extra loop over isospin.
\begin{svgraybox}
  \begin{algorithmic}
    \State $n=0$ \For{$\text{shell}\in\{ 0,...,N_{\text{max}}\}$}
    \For{$-\sqrt{N_{\text{max}}}\leq n_{x}\leq\sqrt{N_{\text{max}}}$}
    \For{$-\sqrt{N_{\text{max}}}\leq n_{y}\leq\sqrt{N_{\text{max}}}$}
    \For{$-\sqrt{N_{\text{max}}}\leq n_{z}\leq\sqrt{N_{\text{max}}}$}
    \For{$s_{z}\in\{-\frac{1}{2},\frac{1}{2}\}$}
    \If{$n_{x}^{2}+n_{y}^{2}+n_{z}^{2}=\text{shell}$} \State
    $\text{Energy}=\frac{4\pi^{2}\hbar^{2}}{2m}\times\text{shell}$
    \If{$n<N$} \State $\text{type}=\text{``hole''}$ \Else \State
    $\text{type}=\text{``particle''}$ \EndIf \State STATES $\gets$
    ($n$, $n_{x}$, $n_{y}$, $n_{z}$, $s_{z}$, Energy, type) \State
    $n\gets n+1$ \EndIf \EndFor \EndFor \EndFor \EndFor \EndFor
  \end{algorithmic}
\end{svgraybox}
  The next step is to build every two-body state in the model space
  and separate them by their particle/hole character and combined
  quantum numbers. While each single-particle state is unique,
  two-body states can share quantum numbers with other members of a
  particular two-body channel. These channels allow us to remove
  matrix elements and cluster amplitudes that violate the symmetries
  of the interaction. This reduces greatly the size and speed of the
  calculation. Our structures will depend on direct two-body channels,
  $T$, where the quantum numbers are added, and cross two-body
  channels, $X$, where the quantum numbers are subtracted. Before
  filling the channels, it is helpful to order them with an index
  function which returns a unique index for a given set of two-body
  quantum numbers. Without an index function, one has to loop over all
  the channels for each two-body state, adding a substantial amount
  of time to this algorithm. An example of an index function for the
  direct channels in symmetric nuclear matter is, for
  $N_{x}=n_{x,1}+n_{x,2}$, $N_{y}=n_{y,1}+n_{y,2}$,
  $N_{z}=n_{z,1}+n_{z,2}$, $S_{z}=s_{z,1}+s_{z,2}$,
  $T_{z}=t_{z,1}+t_{z,2}$, $m=2\lfloor\sqrt{N_{\text{max}}}\rfloor$,
  and $M=2m+1$,
  \begin{equation}
  \text{Ind}\left( N_{x},N_{y},N_{z},S_{z},T_{z}\right)=2\left(
  N_{x}+m\right)M^{3}+2\left( N_{y}+m\right)M^{2}+2\left(
  N_{z}+m\right)M+2\left( S_{z}+1\right)+\left(T_{z}+1\right).
  \end{equation}
  This function, which can also be used for the cross-channel index
  function, is well suited for a cubic model space but can be applied
  in either case. An additional restriction for two-body states is
  that they must be composed of two different states to satisfy the
  Pauli-exclusion principle.
\begin{svgraybox}
  \begin{algorithmic}
    \For{$\text{sp1}\in STATES$} \For{$\text{sp2}\in STATES$}
    \If{$sp1\neq sp2$} \State $N_{i}\gets n_{i,1}+n_{i,2}$ \State
    $S_{z}\gets s_{z,1}+s_{z,2}$ \State $T_{z}\gets t_{z,1}+t_{z,2}$
    \State
    $\text{i\_dir}\gets\text{Ind}\left(N_{x},N_{y},N_{z},S_{z},T_{z}\right)$
    \State $T\gets$ (sp1, sp2, i\_dir) \State $N'_{i}\gets
    n_{i,1}-n_{i,2}$ \State $S'_{z}\gets s_{z,1}-s_{z,2}$ \State
    $T'_{z}\gets t_{z,1}-t_{z,2}$ \State
    $\text{i\_cross}\gets\text{Ind}\left(N'_{x},N'_{y},N'_{z},S'_{z},T'_{z}\right)$
    \State $X\gets$ (sp1, sp2, i\_cross) \EndIf \EndFor \EndFor
  \end{algorithmic}
\end{svgraybox}
  From the cross channels, one can construct the cross channel
  complements, $X'$, where $X\left( pq\right)\equiv X'\left(
  qp\right)$. Also from the direct channels, one can construct
  one-body, and corresponding three-body, channels for each
  single-particle state, $K$ by finding every combination of two
  two-body states within a direct channel that contains that single
  particle state, $T\left( pq\right)=T\left( rs\right)\Rightarrow
  K_{p}\gets\left( qrs\right)$.
\begin{svgraybox}
  \begin{algorithmic}
    \For{$\text{Chan}\in T$} \For{$\text{tb1}\in\text{Chan}$}
    \For{$\text{tb2}\in\text{Chan}$} \State $K\gets\text{tb1}_{1}$
    \State
    $K_{\text{tb1}_{1}}\gets\mathop{\text{tb1}_{2},\text{tb2}_{1},\text{tb2}_{2}}$
    \EndFor \EndFor \EndFor
  \end{algorithmic}
\end{svgraybox}
  These different structures can be further categorized by a two-body
  state's particle-hole character, $\braket{pp| t |hh}, \braket{hh| v
    |hh}, \braket{pp| v |pp}, \braket{hh| v |pp}$, and $\braket{hp| v
    |hp}$, which greatly simplifies the matrix-matrix multiplications
  of the CCD iterations by indexing the summed variables in a
  systematic way. Summations are constructed by placing two
  structures next to each other in such a way that the inner summed
  indices are of the same channel. The resulting structure is indexed
  by the outer channels as shown here for several of the intermediates defined above
  \[
  \braket{b| \chi |c} = \braket{b|f|c} - \frac{1}{2} \sum_{kld}\braket{bd|t|kl} \braket{kl|v|cd} \rightarrow f_{c}^{b}\left(
  K\left( b\right),K\left( c\right)\right) -\frac{1}{2}t_{kl}^{bd}\left( K\left( b\right),K_{b}\left(kld\right)\right) v_{cd}^{kl}
\left( K_{c}\left(kld\right),K\left( c\right)\right),
\]
\[
\braket{kl| \chi |ij} = \braket{kl|v|ij} + \frac{1}{2} \sum_{cd}
  \braket{kl|v|cd} \braket{cd|t|ij} \rightarrow v_{ij}^{kl}\left(
  T\left( kl\right),T\left( ij\right)\right)+ 
  \frac{1}{2}v_{cd}^{kl}\left( T\left( kl\right),T\left(
  cd\right)\right) t_{ij}^{cd}\left( T\left( cd\right),T\left(
  ij\right)\right),
\]
\[
\braket{kb| \chi |cj} = \braket{kb|v|cj} +\frac{1}{2} \sum_{dl} \braket{kl|v|cd} \braket{db|t|lj} \rightarrow
  v_{cj}^{kb}\left( X\left( kc\right),X\left( jb\right)\right)+\frac{1}{2}v_{cd}^{kl}\left( X\left( kc\right),X\left(
  dl\right)\right) t_{lj}^{db}\left( X\left( dl\right),X\left(jb\right)\right),
\]
\[
\braket{ab| \chi |cd} = \braket{ab|v|cd}
  \rightarrow v_{cd}^{ab}\left( T\left( ab\right),T\left(
  cd\right)\right),
\]
\[
\sum_{c} \braket{b| \chi |c} \braket{ac| t |ij}
  \rightarrow \chi_{c}^{b}\left( K\left( b\right),K\left(
  c\right)\right)\cdot t_{ij}^{ac}\left( K\left( c\right), K_{c}\left(
  ija\right)\right),
\]
\[
\sum_{k} \braket{k| \chi |j} \braket{ab| t |ik}
  \rightarrow \chi_{j}^{k}\left( K\left( j\right),K\left(
  k\right)\right)\cdot t_{ik}^{ab}\left( K\left( c\right), K_{c}\left(
  ija\right)\right),
\] 
\[
\sum_{cd} \braket{ab| \chi |cd} \braket{cd| t
    |ij} \rightarrow \chi_{cd}^{ab}\left( T\left( ab\right),T\left(
  cd\right)\right)\cdot t_{ij}^{cd}\left( T\left( cd\right),T\left(
  ij\right)\right),
\] 
\[
\sum_{kl} \braket{ab| t |kl} \braket{kl| \chi |ij} \rightarrow t_{kl}^{ab}\left( T\left( ab\right),T\left(
  kl\right)\right)\cdot \chi_{ij}^{kl}\left( T\left( kl\right),T\left(ij\right)\right),
\]
and finally
\[
\sum_{kc} \braket{ac| t |ik}\braket{kb| \chi
    |cj} = \sum_{kc} \braket{ai^{-1}| t |kc^{-1}}\braket{kc^{-1}| \chi
    |jb^{-1}} \rightarrow t_{ik}^{ac}\left( X\left( ia\right),X\left(
  kc\right)\right)\cdot \chi_{cj}^{kb}\left( X\left( kc\right),X\left(
  jb\right)\right).
\]
Figure \ref{fig:fig1} displays the convergence of the energy per particle for pure neutron matter as function of number particles 
for  the CCD approximation with the Minnesota interaction model  \cite{minnesota} for different with $\mathrm{N_{max}=20}$.
Similarly, Fig.~\ref{fig:fig2} shows the convergence in terms of different model space sizes $\mathrm{N_{max}=20}$ with 
a fixed number of neutrons $N=114$. 
We see from the last figure that at the CCD level and neutron matter only there is a good convergence with  $\mathrm{N_{max}=25}$.
This results depends however on the type of interaction and many-body approximation. 

\begin{figure}
    \includegraphics[width=\linewidth]{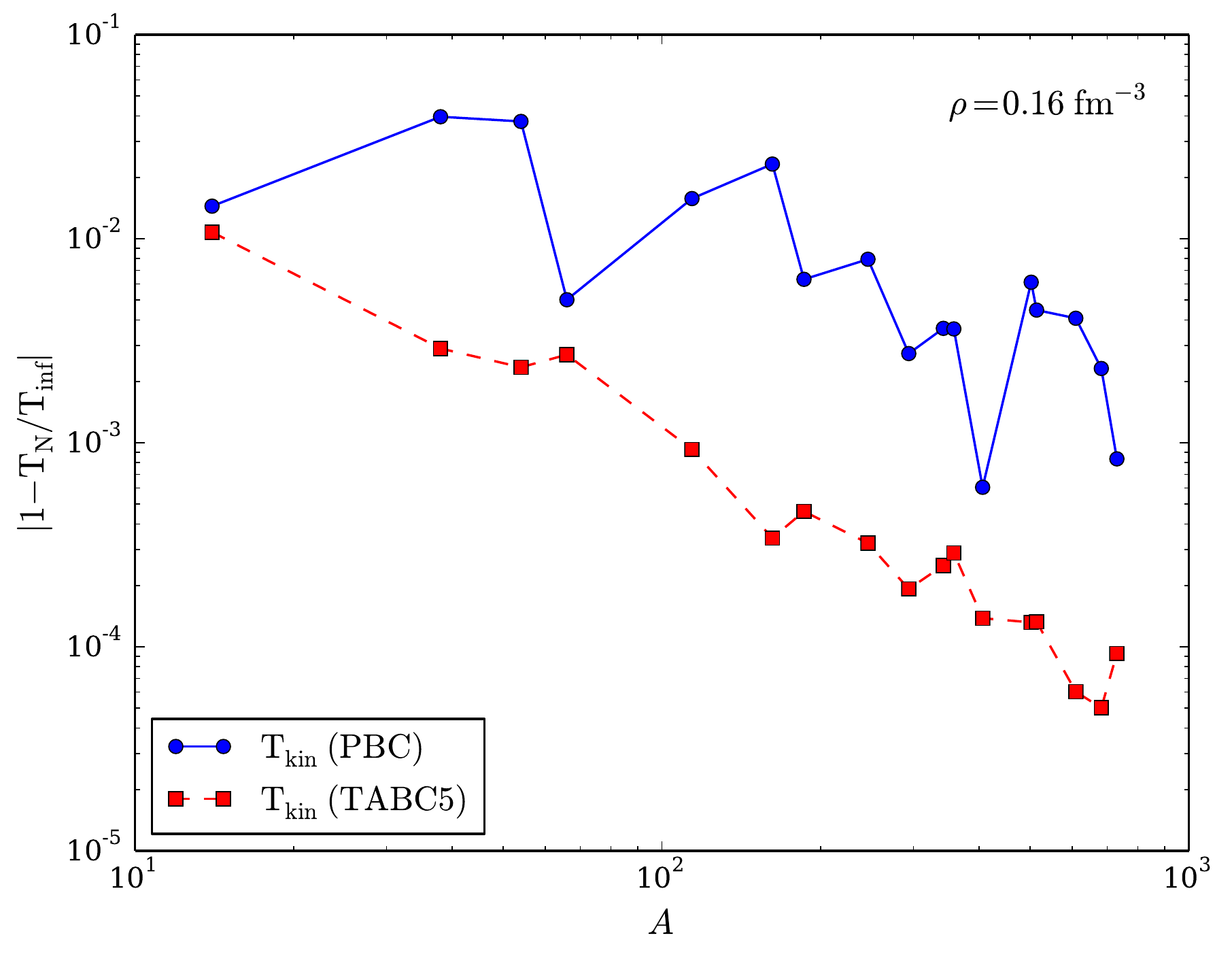}
    \caption{Finite-size effects in the kinetic energy of pure neutron
      matter computed with the Minnesota interaction model  \cite{minnesota} as a function of
      the number of particles for both periodic boundary conditions (PBC)
      and twist-averaged boundary conditions (TABC5).}
    \label{fig:fig3}
  \end{figure}

\begin{figure}
    \includegraphics[width=\linewidth]{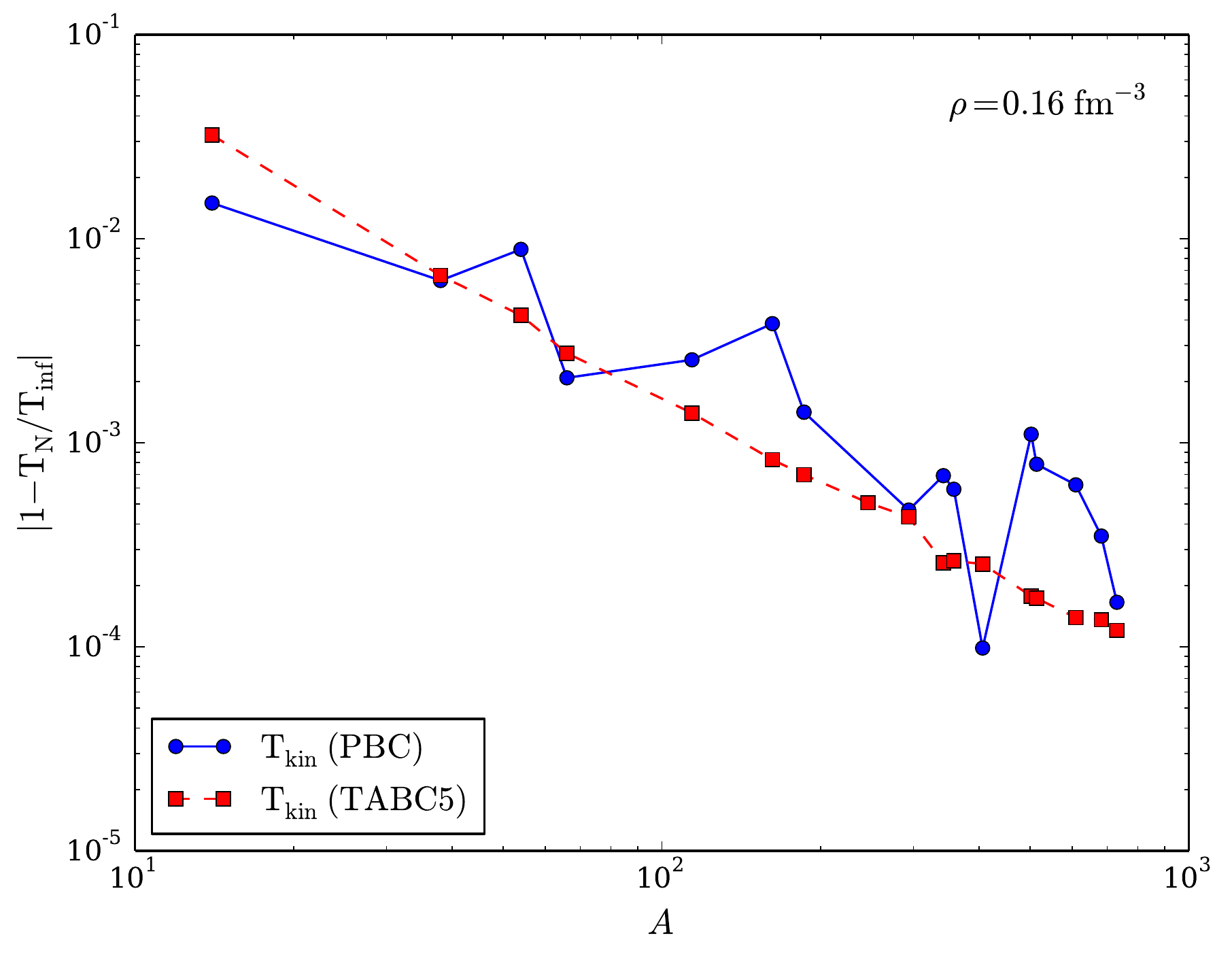}
    \caption{Finite-size effects in the Hartree-Fock energy of pure
      neutron matter computed with the Minnesota interaction model  \cite{minnesota} as a
      function of the number of particles for both periodic boundary (PBC)
      conditions and twist-averaged boundary conditions (TABC5).}
    \label{fig:fig4}
  \end{figure}

In these calculations we
approximated our problem with periodic boundary conditions,
  $\mathop{\phi(x_{i})}=\mathop{\phi(x_{i}+L)}$, but we could have
  chosen anti-periodic boundary conditions,
  $\mathop{\phi(x_{i})}=-\mathop{\phi(x_{i}+L)}$. The difference
  between these two shows how the correlation energy contains
  finite-size effects \cite{gros1992,gros1996,shepherd2012,shepherd2016}. One solution to this problem is by integrating
  over solutions between periodic and anti-periodic conditions, known
  as twist-averaging \cite{twistaverage}. First, we multiply the single-particle states by
  a phase for each direction, characterized by a twist-angle,
  $\theta_{i}$.
  \begin{equation}
    \mathop{\phi_{\vec{k}}(\vec{x}+\vec{L})}\rightarrow\mathop{e^{i\vec{\theta}}\phi_{\vec{k}}(\vec{x})}
  \end{equation}
  $\theta_{i}=0$ for PBC and $\theta_{i}=\pi$ for APBC
  \begin{align}
  \vec{k}\rightarrow\vec{k}+\frac{\vec{\theta}}{L}
  \\ \epsilon_{\vec{k}}\rightarrow\epsilon_{\vec{k}}+\frac{\pi}{L}\vec{k}\cdot\vec{\theta}+\frac{\pi^{2}}{L^{2}}
  \end{align}
  Adding these phases changes the single-particle energies, the
  correction of which disappear as $L\rightarrow\infty$, depending on
  $\vec{\theta}$ and thus changes the shell structure so that hole
  states can jump up to particle states and {\em vice versa}. It is thence
  necessary to fill hole states separately for each
  $\vec{\theta}$. Integration over a quantitiy is approximated by a
  weighted sum, such as Gauss-Legendre quadrature, over the quantity
  for each set of twist angles. The algorithm becomes then
\begin{svgraybox}
  \begin{algorithmic}
    \State Build mesh points and weights for each direction $i$:
    $\{\theta_{i},w_{i}\}$ \State $E_{\text{twist}}=0$
    \For{$\mathop{(\theta_{x},w_{x})}\in\mathop{\{\theta_{x},w_{x}\}}$}
    \For{$\mathop{(\theta_{y},w_{y})}\in\mathop{\{\theta_{y},w_{y}\}}$}
    \For{$\mathop{(\theta_{z},w_{z})}\in\mathop{\{\theta_{z},w_{z}\}}$}
    \State Build Basis States with $k_{i}\rightarrow
    k_{i}+\frac{\theta_{i}}{L}$ \State Order States by Energy and Fill
    Holes \State Get Result $E$ (T,HF,CCD) \State
    $E_{\text{twist}}=E_{\text{twist}}+\frac{1}{\pi^{3}}w_{x}w_{y}w_{z}E$
    \EndFor \EndFor \EndFor
  \end{algorithmic}
\end{svgraybox}
  This technique gives results which depend much less on the particle
  number, but requires a full calculation for each set of twist
  angles, which can grow very quickly. For example, using 10 twist
  angles in each direction requires 1000 calculations. To see the
  effects of twist averaging, it is easy to calculate the kinetic
  energy per particle and the Hartree-Fock energy per particle, which
  avoids the full CCD calculation. These calculations can be compared
  to the exact values for infinite matter, which are calculated by
  integrating the the relevent values up to the fermi surface. The kinetic energy is given by
  \[
\text{T}_{\text{inf}}=\frac{3\hbar^{2}k_{f}^{2}}{10m},
\]
while the potential energy to first order (corresponding to the Hartree-Fock contribution) reads
\[
\text{HF}_{\text{inf}}=\frac{1}{\mathop{(2\pi)^{6}}}\frac{L^{3}}{2\rho}\int_{0}^{k_{f}}d\vec{k}_{1}\int_{0}^{k_{f}}d\vec{k}_{2}\braket{\vec{k}_{1}\vec{k}_{2}|\hat{v}|\vec{k}_{1}\vec{k}_{2}}.
\]
Figure \ref{fig:fig3} shows the CCD kinetic energy of pure neutron
      matter computed with the Minnesota interaction model  \cite{minnesota} as a function of
      the number of particles for both periodic boundary conditions (PBC)
      and twist-averaged boundary conditions (TABC5). We see clearly that the 
twist-averaged boundary conditions soften the dependence on finite size effects. 
Similarly, Fig.~\ref{fig:fig4} displays the corresponding Hartree-Fock energy (the reference energy) 
obtained with Minnesota interaction using both periodic and twist-average boundary conditions.  
The results show again a weaker dependence on finite size effects.

\begin{figure}
    \includegraphics[width=\linewidth]{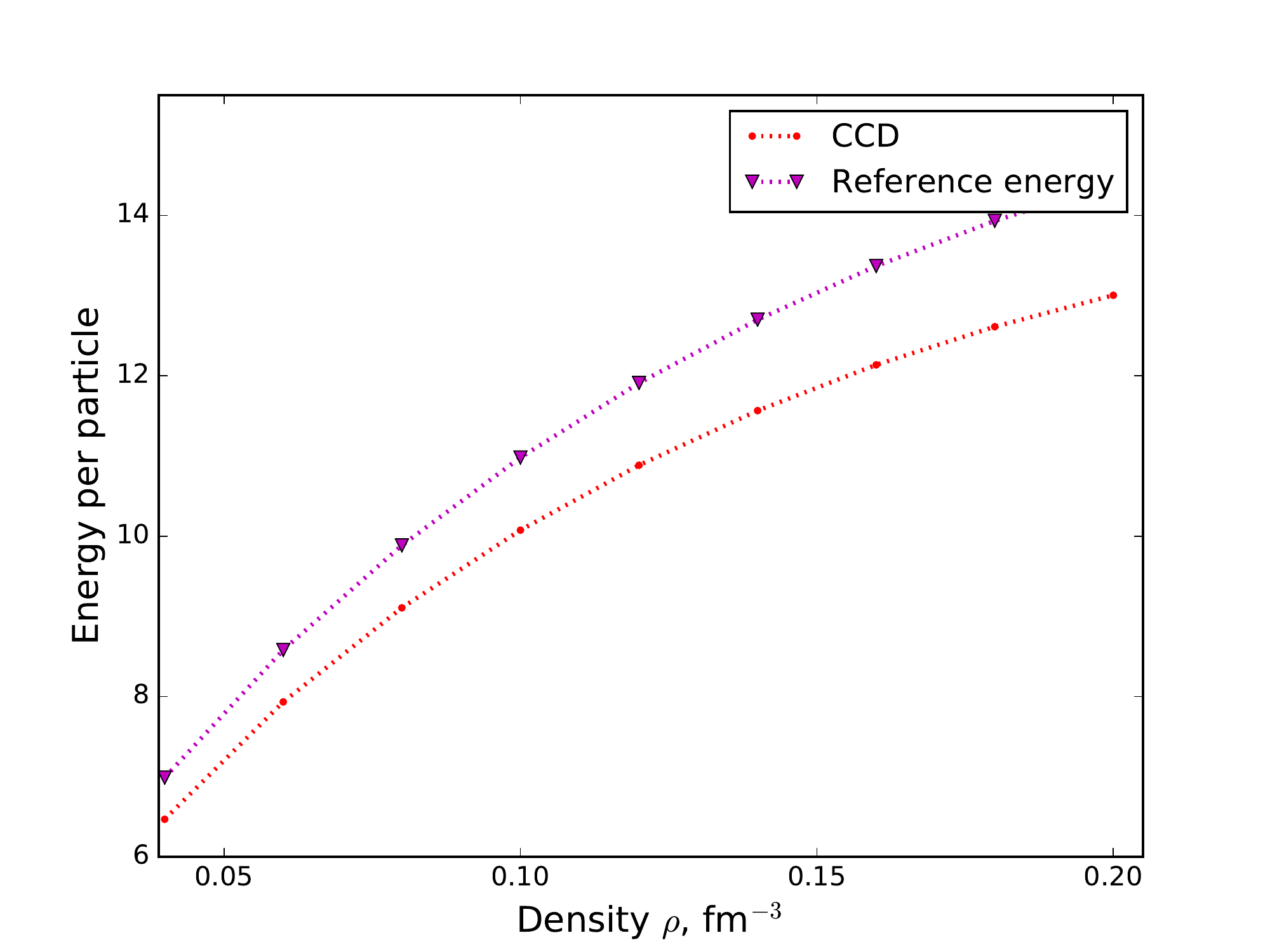}
    \caption{Energy per particle for pure neutron matter as function of density from coupled cluster calculations with doubles correlations  only. The reference energy is included for comparison. The results have been obtained with Minnesota interaction model using periodic  boundary conditions and $N=66$ neutrons and single-particle states up to $N_{max}=36$, resulting in a total of $2377$ single-particle states. }  
    \label{fig:finalccdenergy}
  \end{figure}

With all these ingredients, we can now compute the final CCD energy and thereby the equation of state for infinite neutron matter.
Figure \ref{fig:finalccdenergy} displays the total CCD energy (including the reference energy) as well as the reference energy
obtained with the Minnesota interaction model. The computations have been performed with $N=66$ neutrons and a maximum number of single-particle states constrained by $N_{max}=36$. This corresponds to $2377$ single-particle states. 
We see from this figure that the correlations brought by coupled cluster theory are at the order of $10\%$ roughly of the reference energy. It means that for this system (neutrons only) with the Minnesota potential, higher-order correlations can most likely be treated in a perturbative way. Many-body perturbation theory to second order gives results which are very close to our CCD results, as seen in 
Table \ref{tab:ecomparison}.
\begin{table}
\caption{CCD and MBPT2 results for infinite neutron matter with $N=66$ neutrons and a maximum number of single-particle states constrained by $N_{max}=36$. }\label{tab:ecomparison}
  \begin{center}
      \begin{tabular}{| l | l | l |}
      \hline Density $\rho$ fm$^{-3}$ & $E_{MBPT2}$ & $E_{CCD}$
      \\ \hline 0.04 & 6.472   & 6.468    \\ \hline
      0.06 & 7.919  & 7.932   \\ \hline
      0.08 & 9.075  & 9.136    \\ \hline
      1.0 & 9.577 &  10.074  \\  \hline
      1.2 & 10.430 & 10.885   \\ \hline
      1.4 & 11.212 & 11.565   \\  \hline
      1.6 & 11.853  & 12.136    \\  \hline
      1.8 & 12.377 & 12.612   \\ \hline
      2.0 & 12.799 & 13.004   \\ \hline
      \end{tabular}
  \end{center}
\end{table}
For low densities we observe a good agreement while higher densities
bring in particle-particle correlations that become more important as
the density increases. Coupled cluster theory sums to infinite order for
example particle-particle correlations and with increasing densities
this is reflected in differences between the two many-body
approximations.  The above results agree well with the recent coupled
cluster calculations of Refs.~\cite{baardsen2013,hagenmatter}, obtained with
interaction models from effective field theory.  With the inclusion of
proton correlations as well as other potential models we may expect 
larger differences between
different methods and interactions.  In chapters 9, 10 and 11 we compare the above results with 
those obtained with Monte Carlo methods, the in-medium renormalization group approach and Green's function theory.

The discussions hitherto have focused on the
development of an efficient and flexible many-body code. The codes have been structured 
to allow users and developers to study and add different physical systems,
spanning from the simple pairing model to quantum dots and infinite
nuclear matter. Structuring the codes in for a example an abstract
system class and a solver class allows an eventual user to study
different physical systems and add new many-body solvers without
having to write a totally new program. As demonstrated in chapter 10, 
with few additions one can add 
the widely popular similarity renormalization group method. 

Till now we have limited our discussion to the construction of a
many-body code following the underlying mathematical expressions, 
including elements like how to structure a code in
terms of functions, how to modularize the code, how to develop classes
and how to verify and validate our computations (our checks provided
by the simple pairing model and many-body perturbation theory to
second order) against known results.  With the rewriting of our
equations in terms of efficient matrix and vector operations we have
also shown how to make our code more efficient.  Matrix and vector
operations can easily be parallelized, as demonstrated in our discussion of many-body perturbation theory to second and third order. Such algorithmic improvements
are necessary in order to be able to study complicated physical
systems.  Our codes can also be easily parallelized in order to run on
shared memory architectures using either OpenMP \cite{openmp} and/or
MPI/OpenMPI \cite{mpi,openmpi}. 

We conclude this section by summarizing and emphasizing some
topics we feel can help in structuring a large computational project.
Amongst these, the validation and verification of the correctness of
the employed algorithms and programs are central issues which can
save you a lot of time when developing a numerical project. In the
discussions above we used repeatedly the simple pairing model of
problem \ref{problem:pairingmodel}.  This model allows for benchmarks
against exact results. In addition, it provides analytical answers to
several approximations, from perturbation theory to specific terms in the solution of the coupled
cluster equations, the in-medium similarity renormalization group
approach of chapter 10 and the Green's function approach of chapter
11. 

It is also important to develop an understanding of how our
algorithms can go wrong and how they can be implemented in order to
run as efficiently as possible. When working on a numerical project it
is important to keep in mind that computing covers numerical as well
as symbolic computing and paper and pencil solutions. Furthermore,
version control is something we strongly recommend. It does not only
save you time in case you struggle with odd errors in a new version of
your code. It allows you also to make science reproducible.  Making
your codes available to a larger audience and providing proper
benchmarks allows fellow scientists to test what you have developed,
and perhaps come with considerable improvements and/or find flaws
or errors you were not aware of. 
Sharing your codes using for example modern version control
software makes thus your science reproducible and adds in a natural
way a sound ethical scientific element to what you develop.
In this chapter we have thus provided several code examples, hoping they can serve as good examples.

\section{Conclusions}
In this chapter we have presented many of the basic ingredients that
enter theoretical studies of infinite nuclear matter, with possible
extensions to the homogeneous electron gas in two and three dimensions
or other quantum mechanical systems.  We have focused on the
construction of a single-particle and many-body basis appropriate for
such systems, as well as introducing post Hartree-Fock many-body
methods like full configuration interaction theory, many-body
perturbation theory and coupled cluster theory. The results here,
albeit being obtained with a simpler model for the nuclear forces, can
easily be extended to more complicated and realistic models for
nuclear interactions and to include other many-body methods. We have
however, for pedagogical reasons, tried to keep the theoretical inputs
to the various many-body methods as simple as possible. The reader
should however, with the inputs from chapters 2-6, be able to have a
better of understanding of nuclear forces and how these can be derived
from the underlying theory for the strong force and effective field
theory.  The last exercise in this chapter replaces the simple Minnesota potential with
realistic interactions from effective field theory.

The subsequent chapters 9, 10 and 11 show how many of the theoretical
concepts and code elements discussed in this chapter can be used to
add other many-body methods, without having to develop a new numerical
project.  With a proper modularization and flexible classes, we can
add new physical systems as well as new many-body methods.  The codes
which have been developed in this chapter can be reused in the
development and analysis of the in-medium similarity renormalizaton
group approach of chapter 10 or the Green's function approach in chapter 11. Similarly, the theoretical concepts we
have developed in this chapter, such as the definition of a
single-particle basis using plane wave functions and correlations from
many-body perturbation theory or coupled cluster theory, can be used
in chapter 9, 10 and 11 as well. Chapter 9 for example, uses results
from coupled cluster theory in order to provide better ways to constrain
the Jastrow factor, which accounts for correlations beyond a mean-field picture, 
in Monte Carlo calculations.

  \section{Exercises}
  \begin{prob} \label{problem:prob8.1}
  Show that the one-body part of the Hamiltonian
      \begin{equation*}
          \hat{H}_0 = \sum_{pq} \element{p}{\hat{h}_0}{q} a^\dagger_p
          a_q
      \end{equation*}
  can be written, using standard annihilation and creation operators,
  in normal-ordered form as
      \[
          \hat{H}_0 = \sum_{pq} \element{p}{\hat{h}_0}{q}
          \left\{a^\dagger_p a_q\right\} + \sum_i
          \element{i}{\hat{h}_0}{i}
      \]
  Explain the meaning of the various symbols. Which reference vacuum
  has been used?
  \end{prob}

  \begin{prob} \label{problem:prob8.2}
  Show that the two-body part of the Hamiltonian
      \begin{equation*}
          \hat{H}_I = \frac{1}{4} \sum_{pqrs}
          \element{pq}{\hat{v}}{rs} a^\dagger_p a^\dagger_q a_s a_r
      \end{equation*}
  can be written, using standard annihilation and creation operators,
  in normal-ordered form as
      \begin{align*}
      \hat{H}_I &= \frac{1}{4} \sum_{pqrs} \element{pq}{\hat{v}}{rs}
      a^\dagger_p a^\dagger_q a_s a_r \nonumber \\ &= \frac{1}{4}
      \sum_{pqrs} \element{pq}{\hat{v}}{rs} \normord{a^\dagger_p
        a^\dagger_q a_s a_r} + \sum_{pqi} \element{pi}{\hat{v}}{qi}
      \normord{a^\dagger_p a_q} + \frac{1}{2} \sum_{ij}
      \element{ij}{\hat{v}}{ij}
      \end{align*}
  Explain again the meaning of the various symbols.
  \end{prob}

  \begin{prob}\label{problem:prob8.3}
  Derive the normal-ordered form of the threebody part of the
  Hamiltonian.
  \[
      \hat{H}_3 = \frac{1}{36} \sum_{\substack{ pqr \\ stu}}
      \element{pqr}{\hat{v}_3}{stu} a^\dagger_p a^\dagger_q
      a^\dagger_r a_u a_t a_s,
  \]
  and specify the contributions to the two-body, one-body and the
  constant part.
  \end{prob}

  \begin{prob}\label{problem:spbasissetup}
  Develop a program which sets up a single-particle basis for nuclear
  matter calculations with input a given number of nucleons and a user
  specificied density or Fermi momentum. Follow the setup discussed in
  Table \ref{tab:table1}.  You need to define the number of particles
  $A$ and the density of the system using
  \[
  \rho = g \frac{k_F^3}{6\pi^2}.
  \]
  Here you can either define the density itself or the Fermi momentum
  $k_F$.  With the density/Fermi momentum and a fixed number of
  nucleons $A$, we can define the length $L$ of the box used with our
  periodic boundary contributions via the relation
  \[
    V= L^3= \frac{A}{\rho}.
  \]
  We can then can use $L$ to define the spacing between various
  $k$-values, that is
  \[
    \Delta k = \frac{2\pi}{L}.
  \]
  \end{prob}

  \begin{prob}\label{problem:fourier}
  The interaction we will use for these calculations is a
  semirealistic nucleon-nucleon potential known as the Minnesota
  potential \cite{minnesota} which has the form, $V_{\alpha}\left(
  r\right)=V_{\alpha}\exp{-(\alpha r^{2})}$. The spin and isospin
  dependence of the Minnesota potential is given by,
  \[
  V\left( r\right)=\frac{1}{2}\left( V_{R}+\frac{1}{2}\left(
  1+P_{12}^{\sigma}\right) V_{T}+\frac{1}{2}\left(
  1-P_{12}^{\sigma}\right) V_{S}\right)\left(
  1-P_{12}^{\sigma}P_{12}^{\tau}\right),
  \]
  where $P_{12}^{\sigma}=\frac{1}{2}\left(
  1+\sigma_{1}\cdot\sigma_{2}\right)$ and
  $P_{12}^{\tau}=\frac{1}{2}\left( 1+\tau_{1}\cdot\tau_{2}\right)$ are
  the spin and isospin exchange operators, respectively.  Show that a
  Fourier transform to momentum space results in
  \[
  \langle \mathbf{k}_p \mathbf{k}_q \vert V_{\alpha}\vert
  \mathbf{k}_r\mathbf{k}_s\rangle=\frac{V_{\alpha}}{L^{3}}\left(\frac{\pi}{\alpha}\right)^{3/2}\exp{\frac{-q^{2}}{4\alpha}}\delta_{\vec{k}_{p}+\vec{k}_{q},\vec{k}_{r}+\vec{k}_{s}}.
  \]
  Write thereafter a function which sets up the full anti-symmetrized
  matrix elements for the Minnesota potential using the parameters
  listed in Table \ref{tab:minnesotatab}.
  \end{prob}

  \begin{prob}\label{problem:unitarity}
  Consider a Slater determinant built up of orthogonal single-particle
  orbitals $\psi_{\lambda}$, with $\lambda = 1,2,\dots,A$.

  The unitary transformation
  \[
  \psi_a = \sum_{\lambda} C_{a\lambda}\phi_{\lambda},
  \]
  brings us into the new basis.  The new basis has quantum numbers
  $a=1,2,\dots,A$.  Show that the new basis is orthogonal.
  \begin{enumerate}
  \item[a)] Show that the new Slater determinant constructed from the
    new single-particle wave functions can be written as the
    determinant based on the previous basis and the determinant of the
    matrix $C$.
  \item[b)] Show that the old and the new Slater determinants are
    equal up to a complex constant with absolute value unity.  Hint:
    $C$ is a unitary matrix.
  \end{enumerate}
  \end{prob}

  \begin{prob}\label{problem:referenceE}
  Use the ansatz for the ground state in second quantization
  \[
  |\Phi_0\rangle = \left(\prod_{i\le
    F}\hat{a}_{i}^{\dagger}\right)|0\rangle,
  \]
  where the index $i$ defines different single-particle states up to
  the Fermi level, to calculate using Wick's theorem (see the
  appendix) the expectation value
  \[
    E[\Phi_0]= E_{\mathrm{Ref}}= \sum_{i\le F}^A \langle i | \hat{h}_0
    | i \rangle + \frac{1}{2}\sum_{ij\le F}^A\langle
    ij|\hat{v}|ij\rangle.
  \]
  Insert thereafter the plane wave function basis for the various
  single-particle states and show that the above energy can be written
  as
  \[
    E[\Phi_0] = \sum_{i\le F}^A \langle k_i | \hat{t} | k_i \rangle +
    \frac{1}{2}\sum_{ij\le F}^A\langle
    k_ik_j|\hat{v}|k_ik_j\rangle,
  \]
  where we use the shorthand $\vert k_i\rangle = \vert
  \mathbf{k}_i,\sigma_i,\tau_{z_i}\rangle$ for the single-particle
  states in three dimensions.

  Replace then the discrete sums with integrals, that is
  \[
  \sum_i \rightarrow
  \sum_{\sigma_i}\sum_{\tau_{z_i}}\frac{L^3}{(2\pi)^3}\int_0^{\mathbf{k}_F}d\mathbf{k},
  \]
  and show that the energy per particle $A$ can be written as (for
  symmetric nuclear matter)
  \[
    \frac{E_{\mathrm{Ref}}}{A}=\frac{3\hbar^2k_F^2}{10M_N}+\frac{1}{2n}\frac{L^3}{(2\pi)^6}\sum_{\sigma_i\sigma_j}\sum_{\tau_{z_i}\tau_{z_j}}\int_0^{\mathbf{k}_F}d\mathbf{k}_i\int_0^{\mathbf{k}_F}d\mathbf{k}_j\langle
    k_ik_j|\hat{v}|k_ik_j\rangle,
  \]
  with the density $n=V/A=L^3/A$.

  Find the following expression for pure neutron matter. Use the
  Minnesota interaction and try to simplify the above six-dimensional
  integral for pure neutron matter (Hint: the interaction depends only
  the momentum transfer squared and fix one of the momentum
  integrations along the $z$-axis. Integrate out the dependence on the
  various angles).

  Finally, write a program which computes the above energy for pure
  neutron matter using the Minnesota potential.
  \end{prob}

  \begin{prob}\label{problem:hamiltoniansetup}
  We will assume that we can build various Slater determinants using
  an orthogonal single-particle basis $\psi_{\lambda}$, with $\lambda= 1,2,\dots,A$.

  The aim of this exercise is to set up specific matrix elements that
  will turn useful when we start our discussions of different
  many-body methods. In particular you will notice, depending on the
  character of the operator, that many matrix elements will actually
  be zero.

  Consider three $A$-particle Slater determinants $|\Phi_0$,
  $|\Phi_i^a\rangle$ and $|\Phi_{ij}^{ab}\rangle$, where the notation
  means that Slater determinant $|\Phi_i^a\rangle$ differs from
  $|\Phi_0\rangle$ by one single-particle state, that is a
  single-particle state $\psi_i$ is replaced by a single-particle
  state $\psi_a$.  It will later be interpreted as a so-called
  one-particle-one-hole excitation.  Similarly, the Slater determinant
  $|\Phi_{ij}^{ab}\rangle$ differs by two single-particle states from
  $|\Phi_0\rangle$ and is normally thought of as a
  two-particle-two-hole excitation.

  Define a general one-body operator $\hat{F} =
  \sum_{i}^A\hat{f}(x_{i})$ and a general two-body operator
  $\hat{G}=\sum_{i>j}^A\hat{g}(x_{i},x_{j})$ with $g$ being invariant
  under the interchange of the coordinates of particles $i$ and $j$.
  \begin{enumerate}
  \item[a)]
  \[
  \langle \Phi_0 \vert\hat{F}\vert\Phi_0\rangle,
  \]
  and
  \[
  \langle \Phi_0\vert\hat{G}|\Phi_0\rangle.
  \]
  \item[b)] Find thereafter
  \[
  \langle \Phi_0 |\hat{F}|\Phi_i^a\rangle,
  \]
  and
  \[
  \langle \Phi_0|\hat{G}|\Phi_i^a\rangle,
  \]
  \item[c)] Finally, find
  \[
  \langle \Phi_0 |\hat{F}|\Phi_{ij}^{ab}\rangle,
  \]
  and
  \[
  \langle \Phi_0|\hat{G}|\Phi_{ij}^{ab}\rangle.
  \]
  \item[d)] What happens with the two-body operator if we have a
    transition probability of the type
  \[
  \langle \Phi_0|\hat{G}|\Phi_{ijk}^{abc}\rangle,
  \]
  where the Slater determinant to the right of the operator differs by
  more than two single-particle states?
  \item[e)] With an orthogonal basis of Slater determinants
    $\Phi_{\lambda}$, we can now construct an exact many-body state as
    a linear expansion of Slater determinants, that is, a given exact
    state
  \[
  \Psi_i = \sum_{\lambda =0}^{\infty}C_{i\lambda}\Phi_{\lambda}.
  \]
  In all practical calculations the infinity is replaced by a given
  truncation in the sum.

  If you are to compute the expectation value of (at most) a two-body
  Hamiltonian for the above exact state
  \[
  \langle \Psi_i \vert \hat{H} \vert \Psi_i\rangle,
  \]
  based on the calculations above, which are the only elements which
  will contribute?  (there is no need to perform any calculation here,
  use your results from exercises a), b), and c)).

  These results simplify to a large extent shell-model calculations.
  \end{enumerate}
  \end{prob}

  \begin{prob}\label{problem:diagrams}
Write down the analytical expressions for diagrams (8) and (9) in Fig.~\ref{fig:goldstone} and discuss whether these
diagrams should be accounted for or not in the calculation of the energy per particle of infinite matter. 
If a Hartree-Fock basis is used, should these diagrams be included?  
Show also that diagrams (2), (6)-(7) and (10)-(16) are zero in infinite matter due to the lack of momentum conservation. 
\end{prob}

  \begin{prob}\label{problem:pairingmodel}

  We present a simplified Hamiltonian consisting of an unperturbed
  Hamiltonian and a so-called pairing interaction term. It is a model
  which to a large extent mimicks some central features of atomic
  nuclei, certain atoms and systems which exhibit superfluiditity or
  superconductivity.  To study this system, we will use a mix of
  many-body perturbation theory (MBPT), Hartree-Fock (HF) theory and
  full configuration interaction (FCI) theory. The latter will also
  provide us with the exact answer.  When setting up the Hamiltonian
  matrix you will need to solve an eigenvalue problem.

  We define first the Hamiltonian, with a definition of the model
  space and the single-particle basis. Thereafter, we present the
  various exercises (some of them are solved).

  The Hamiltonian acting in the complete Hilbert space (usually
  infinite dimensional) consists of an unperturbed one-body part,
  $\hat{H}_0$, and a perturbation $\hat{V}$.

  We limit ourselves to at most two-body interactions and our
  Hamiltonian is represented by the following operators
  \[
  \hat{H} = \sum_{\alpha\beta}\langle \alpha |h_0|\beta\rangle
  a_{\alpha}^{\dagger}a_{\beta}+\frac{1}{4}\sum_{\alpha\beta\gamma\delta}\langle
  \alpha\beta| V|\gamma\delta\rangle
  a_{\alpha}^{\dagger}a_{\beta}^{\dagger}a_{\delta}a_{\gamma},
  \]
  where $a_{\alpha}^{\dagger}$ and $a_{\alpha}$ etc.~are standard
  fermion creation and annihilation operators, respectively, and
  $\alpha\beta\gamma\delta$ represent all possible single-particle
  quantum numbers.  The full single-particle space is defined by the
  completeness relation
  \[
  \hat{{\bf 1}} = \sum_{\alpha=1}^{\infty}|\alpha \rangle \langle
  \alpha|.
  \]
  In our calculations we will let the single-particle states
  $|\alpha\rangle$ be eigenfunctions of the one-particle operator
  $\hat{h}_0$. Note that the two-body part of the Hamiltonian contains
  anti-symmetrized matrix elements.

  The above Hamiltonian acts in turn on various many-body Slater
  determinants constructed from the single-basis defined by the
  one-body operator $\hat{h}_0$.  As an example, the two-particle
  model space $\mathcal{P}$ is defined by an operator
  \[
  \hat{P} = \sum_{\alpha\beta =1}^{m}|\alpha\beta \rangle \langle
  \alpha\beta|,
  \]
  where we assume that $m=\dim(\mathcal{P})$ and the full space is
  defined by
  \[
  \hat{P}+\hat{Q}=\hat{{\bf 1}},
  \]
  with the projection operator
  \[
  \hat{Q} = \sum_{\alpha\beta =m+1}^{\infty}|\alpha\beta \rangle
  \langle \alpha\beta|,
  \]
  being the complement of $\hat{P}$.

  Our specific model consists of $N$ doubly-degenerate and equally
  spaced single-particle levels labelled by $p=1,2,\dots$ and spin
  $\sigma=\pm 1$.  
  We write the Hamiltonian as
  \[ \hat{H} = \hat{H}_0 + \hat{V} , \]
  where
  \[
  \hat{H}_0=\delta\sum_{p\sigma}(p-1)a_{p\sigma}^{\dagger}a_{p\sigma}
  \]
  and
  \[
  \hat{V}=-\frac{1}{2}g\sum_{pq}a^{\dagger}_{p+}
  a^{\dagger}_{p-}a_{q-}a_{q+}.
  \]
  Here, $H_0$ is the unperturbed Hamiltonian with a spacing between
  successive single-particle states given by $\delta$, which we will set
  to a constant value $\delta=1$ without loss of generality. The two-body
  operator $\hat{V}$ has one term only. It represents the pairing
  contribution and carries a constant strength $g$.

  The indices $\sigma=\pm$ represent the two possible spin values. The
  interaction can only couple pairs and excites therefore only two
  particles at the time.

  \begin{enumerate}
  \item[a)] Show that the unperturbed Hamiltonian $\hat{H}_0$ and
    $\hat{V}$ commute with both the spin projection $\hat{S}_z$ and
    the total spin $\hat{S}^2$, given by
  \[
    \hat{S}_z := \frac{1}{2}\sum_{p\sigma} \sigma
    a^{\dagger}_{p\sigma}a_{p\sigma}
  \]
  and
  \[
    \hat{S}^2 := \hat{S}_z^2 + \frac{1}{2}(\hat{S}_+\hat{S}_- +
    \hat{S}_-\hat{S}_+),
  \]
  where
  \[
    \hat{S}_\pm := \sum_{p} a^{\dagger}_{p\pm} a_{p\mp}.
  \]
  This is an important feature of our system that allows us to
  block-diagonalize the full Hamiltonian. We will focus on total spin
  $S=0$.  In this case, it is convenient to define the so-called pair
  creation and pair annihilation operators
  \[
  \hat{P}^{+}_p = a^{\dagger}_{p+}a^{\dagger}_{p-},
  \]
  and
  \[
  \hat{P}^{-}_p = a_{p-}a_{p+},
  \]
  respectively.
  \item[b)] Show that you can rewrite the Hamiltonian (with $\delta=1$)
    as
  \[
  \hat{H}=\sum_{p\sigma}(p-1)a_{p\sigma}^{\dagger}a_{p\sigma}
  -\frac{1}{2}g\sum_{pq}\hat{P}^{+}_p\hat{P}^{-}_q.
  \]
  \item[c)] Show also that Hamiltonian commutes with the product of
    the pair creation and annihilation operators.  This model
    corresponds to a system with no broken pairs. This means that the
    Hamiltonian can only link two-particle states in so-called
    spin-reversed states.

  \item[d)] Construct thereafter the Hamiltonian matrix for a system
    with no broken pairs and total spin $S=0$ for the case of the four
    lowest single-particle levels. Our system consists of four particles
    only that can occupy four doubly degenerate single-particle states.  
Our single-particle space consists of only the four lowest
    levels $p=1,2,3,4$.  You need to set up all possible Slater
    determinants.  Find all eigenvalues by diagonalizing the
    Hamiltonian matrix.  Vary your results for values of $g\in
    [-1,1]$.  We refer to this as the exact calculation. Comment the
    behavior of the ground state as function of $g$.
  \end{enumerate}
  \end{prob}

  \begin{prob}\label{problem:prob8.5}
  \begin{enumerate}
  \item[a)] We will now set up the Hartree-Fock equations by varying
    the coefficients of the single-particle functions. The
    single-particle basis functions are defined as
  \[
  \psi_p = \sum_{\lambda} C_{p\lambda}\psi_{\lambda}.
  \]
  where in our case $p=1,2,3,4$ and $\lambda=1,2,3,4$, that is the
  first four lowest single-particle orbits.  Set up the Hartree-Fock equations for
  this system by varying the coefficients $C_{p\lambda}$ and solve
  them for values of $g\in [-1,1]$.  Comment your results and compare
  with the exact solution. Discuss also which diagrams in
  Fig.~\ref{fig:goldstone} that can be affected by a Hartree-Fock
  basis. Compute the total binding energy using a Hartree-Fock basis
  and comment your results.

  \item[b)] We will now study the system using non-degenerate
    Rayleigh-Schr\"odinger perturbation theory to third order in the
    interaction.  If we exclude the first order contribution, all
    possible diagrams (so-called anti-symmetric Goldstone diagrams)
    are shown in Fig.~\ref{fig:goldstone}.

  Based on the form of the interaction, which diagrams contribute to
  the binding energy of the ground state?  Write down the expressions
  for the diagrams that contribute and find the contribution to the
  ground state energy as function $g\in [-1,1]$. Comment your results.
  Compare these results with those you obtained from the exact
  diagonalization with and without the $4p-4h$ state.  Discuss your
  results for a canonical Hartree-Fock basis and a non-canonical
  Hartree-Fock basis.

  Diagram 1 in Fig.~\ref{fig:goldstone} represents a second-order
  contribution to the energy and a so-called $2p-2h$ contribution to
  the intermediate states. Write down the expression for the wave
  operator in this case and compare the possible contributions with
  the configuration interaction calculations without the $4p-4h$
  Slater determinant. Comment your results for various values of $g\in
  [-1,1]$.

  We limit now the discussion to the canonical Hartree-Fock case
  only. To third order in perturbation theory we can produce diagrams
  with $1p-1h$, $2p-2h$ and $3p-3h$ intermediate excitations as shown in

  Define first linked and unlinked diagrams and explain briefly
  Goldstone's linked diagram theorem.  Based on the linked diagram
  theorem and the form of the pairing Hamiltonian, which diagrams will
  contribute to third order?

  Calculate the energy to third order with a canonical Hartree-Fock
  basis for $g\in [-1,1]$ and compare with the full diagonalization
  case in exercise b). Discuss the results.
  \end{enumerate}

  \end{prob}

  \begin{prob}\label{problem:prob8.6}
  This project serves as a continuation of the pairing model with the
  aim being to solve the same problem but now by developing a program
  that implements the coupled cluster method with double excitations
  only. In doing so you will find it convenient to write classes which
  define the single-particle basis and the Hamiltonian. Your functions
  that solve the coupled cluster equations will then just need to set
  up variables which point to interaction elements and single-particle
  states with their pertinent quantum numbers. 

  \begin{enumerate}

  \item[a)] Explain why no single excitations are involved in this
    model.

  \item[b)] Set up the coupled cluster equations for doubles
    excitations and convince yourself about their meaning and
    correctness.

  \item[c)] Write a class which holds single-particle data like
    specific quantum numbers, single-particle Hamiltonian etc. Write
    also a class which sets up and stores two-body matrix elements
    defined by the single-particle states.  Write thereafter
    functions/classes which implement the coupled cluster method with
    doubles only.

  \item[d)] Compare your results with those from the exact
    diagonalization with and without the $4p4h$ excitation. Compare
    also your results to perturbation theory at different orders, in
    particular to second order. Discuss your results.  
  \end{enumerate}
  \end{prob}

\begin{prob} \label{problem:amplitudes}
Derive the amplitude equations of Eq.~(\ref{eq:ccd}) starting with 
 \[
          0 = \langle\Phi_{i_1 \ldots i_n}^{a_1 \ldots a_n}\vert
          \overline{H}\vert \Phi_0\rangle.
 \]
\end{prob}
  \begin{prob}\label{problem:realisticforces}
Replace the Minnesota interaction model with realistic models for nuclear forces based on effective field theory. In particular 
replace the Minnesota interaction with the low-order (LO) contribution which includes a contact term and a one-pion exchange
term only. The expressions are discussed in section  \ref{subsec:forcemodels} and Eq.~(\ref{eq:eq_1PEci}). Reference \cite{ekstromPRX} contains a
detailed compilation of all terms up to order NNLO, with tabulated values for all constants. 
When adding realistic interaction models we recommend that you use the many-body perturbation theory codes to second order in the interaction, see the code link at \url{https://github.com/ManyBodyPhysics/LectureNotesPhysics/tree/master/Programs/Chapter8-programs/cpp/MBPT2/src/}. 
\end{prob}

  \section{Solutions to selected exercises}
  \begin{sol}{problem:prob8.1}
  To solve this problem, we start by introducing the shorthand label
  for single-particle states below the Fermi level $F$ as $i,j,\ldots
  \leq F$. For single-particle states above the Fermi level we reserve
  the labels $a,b,\ldots > F$, while the labels $p,q, \ldots$
  represent any possible single particle state.  Using the ansatz for
  the ground state $\vert \Phi_0$ as new reference vacuum state, the
  anticommutation relations are
  \[
  \left\{a_p^\dagger, a_q \right\}= \delta_{pq}, p, q \leq F,
  \]
  and
  \[
  \left\{a_p, a_q^\dagger \right\} = \delta_{pq}, \hspace{0.1cm} p, q
  > F.
  \]
  It is easy to see then that
  \[
          a_i|\Phi_0\rangle = |\Phi_i\rangle\ne 0, \hspace{0.5cm}
          a_a^\dagger|\Phi_0\rangle = |\Phi^a\rangle\ne 0,
  \]
  and
  \[
  a_i^\dagger|\Phi_0\rangle = 0 \hspace{0.5cm} a_a|\Phi_0\rangle = 0.
  \]
  We can then rewrite the one-body Hamiltonian as
   \begin{align*}
          \hat{H}_0 &= \sum_{pq} \element{p}{\hat{h}_0}{q} a^\dagger_p
          a_q \\ &= \sum_{pq} \element{p}{\hat{h}_0}{q}
          \left\{a^\dagger_p a_q\right\} + \delta_{pq\in i} \sum_{pq}
          \element{p}{\hat{h}_0}{q} \\ &= \sum_{pq}
          \element{p}{\hat{h}_0}{q} \left\{a^\dagger_p a_q\right\} +
          \sum_i \element{i}{\hat{h}_0}{i},
   \end{align*}
  where the curly brackets represent normal-ordering with respect to
  the new vacuum state. Withe respect to the new vacuum reference
  state, the
  \end{sol}

  \begin{sol}{problem:prob8.2}
  Using our anti-commutation rules, Wick's theorem discussed in the
  appendix and the definition of the creation and annihilation
  operators from the previous problem, we can rewrite the set of
  creation and annihilation operators of
      \begin{equation*}
          \hat{H}_I = \frac{1}{4} \sum_{pqrs}
          \element{pq}{\hat{v}}{rs} a^\dagger_p a^\dagger_q a_s a_r
      \end{equation*}
  as
  \begin{align*}
      a^\dagger_p a^\dagger_q a_s a_r &=\normord{a^\dagger_p
        a^\dagger_q a_s a_r} \\ & + \normord{
        \contraction{a^\dagger_p}{a}{{}^\dagger_q}{a}a^\dagger_p
        a^\dagger_q a_s a_r}+
      \normord{\contraction{a^\dagger_p}{a}{{}^\dagger_q a_s}{a}
        a^\dagger_p a^\dagger_q a_s a_r}+
      \normord{\contraction{}{a}{{}^\dagger_p
          a^\dagger_q}{a}a^\dagger_p a^\dagger_q a_s a_r} \\ & +
      \normord{\contraction{}{a}{{}^\dagger_p a^\dagger_q
          a_s}{a}a^\dagger_p a^\dagger_q a_s a_r}+
      \normord{\contraction[1.5ex]{}{a}{{}^\dagger_p a_q^\dagger
          a_s}{a}
        \contraction{a^\dagger_p}{a}{{}^\dagger_q}{a}a^\dagger_p
        a^\dagger_q a_s a_r}+ \normord{\contraction{}{a}{{}^\dagger_p
          a_q^\dagger}{a}\contraction[1.5ex]{a^\dagger_p}{a}{{}^\dagger_q
          a_s}{a}a^\dagger_p a^\dagger_q a_s a_r} \\ &=
      \normord{a^\dagger_p a^\dagger_q a_s a_r}+ \delta_{qs\in i}
      \normord{ a^\dagger_p a_r}- \delta_{qr \in i}
      \normord{a^\dagger_p a_s} - \delta_{ps \in i}
      \normord{a^\dagger_q a_r}\\ & + \delta_{pr \in i}
      \normord{a^\dagger_q a_s} + \delta_{pr \in i} \delta_{qs \in i}
      - \delta_{ps \in i} \delta_{qr \in i}.
  \end{align*}
   Inserting the redefinition of the creation and annihilation
   operators with respect to the new vacuum state, we have
      \begin{align*}
      \hat{H}_I &= \frac{1}{4} \sum_{pqrs} \element{pq}{\hat{v}}{rs}
      a^\dagger_p a^\dagger_q a_s a_r \\ &= \frac{1}{4} \sum_{pqrs}
      \element{pq}{\hat{v}}{rs} \normord{a^\dagger_p a^\dagger_q a_s
        a_r} + \frac{1}{4} \sum_{pqrs} \Bigl( \delta_{qs\in i}
      \element{pq}{\hat{v}}{rs} \normord{ a^\dagger_p a_r}\\ & -
      \delta_{qr \in i} \element{pq}{\hat{v}}{rs} \normord{a^\dagger_p
        a_s} - \delta_{ps \in i}\element{pq}{\hat{v}}{rs}
      \normord{a^\dagger_q a_r}\\ & + \delta_{pr \in i}
      \element{pq}{\hat{v}}{rs} \normord{a^\dagger_q a_s}+ \delta_{pr
        \in i} \delta_{qs \in i}- \delta_{ps \in i} \delta_{qr \in i}
      \Bigr) \\ &= \frac{1}{4} \sum_{pqrs} \element{pq}{\hat{v}}{rs}
      \normord{a^\dagger_p a^\dagger_q a_s a_r}\\ & + \frac{1}{4}
      \sum_{pqi} \Bigl(\element{pi}{\hat{v}}{qi} -
      \element{pi}{\hat{v}}{iq} - \element{ip}{\hat{v}}{qi} +
      \element{ip}{\hat{v}}{iq}\Bigr) \normord{a^\dagger_p a_q}\\ & +
      \frac{1}{4} \sum_{ij} \Bigl(\element{ij}{\hat{v}}{ij}-
      \element{ij}{\hat{v}}{ji}\Bigr)\\ &= \frac{1}{4} \sum_{pqrs}
      \element{pq}{\hat{v}}{rs} \normord{a^\dagger_p a^\dagger_q a_s
        a_r} + \sum_{pqi} \element{pi}{\hat{v}}{qi}
      \normord{a^\dagger_p a_q} + \frac{1}{2} \sum_{ij}
      \element{ij}{\hat{v}}{ij}.
      \end{align*}
  Summing up, we obtain a two-body part defined as
      \begin{equation*}
              \hat{V}_N = \frac{1}{4} \sum_{pqrs}
              \element{pq}{\hat{v}}{rs} \normord{a^\dagger_p
                a^\dagger_q a_s a_r},
      \end{equation*}
  a one-body part given by
      \begin{equation*}
              \hat{F}_N = \sum_{pqi} \element{pi}{\hat{v}}{qi}
              \normord{a^\dagger_p a_q},
      \end{equation*}
      and finally the so-called reference energy
      \begin{equation*}
                  E_{\mathrm{ref}}= \frac{1}{2} \sum_{ij}
                  \element{ij}{\hat{v}}{ij}.
      \end{equation*}
  which is the energy expectation value for the reference state.
  Thus, our normal-ordered Hamiltonian with at most a two-body
  nucleon-nucleon interaction is defined as
  \[
  \hat{H}_N =\frac{1}{4} \sum_{pqrs} \bra{pq}\hat{v}\ket{rs}
  \normord{a^\dagger_p a^\dagger_q a_s a_r} + \sum_{pq} f_q^p
  \normord{a^\dagger_p a_q}= \hat{V}_N + \hat{F}_N,
  \]
  with
  \[
  \hat{F}_N = \sum_{pq} f_q^p \normord{a^\dagger_p a_q},
  \]
  and
  \[
  \hat{V}_N = \frac{1}{4} \sum_{pqrs} \bra{pq}\hat{v}\ket{rs}
  \normord{a^\dagger_p a^\dagger_q a_s a_r},
  \]
  where
  \[
    f_q^p = \element{p}{\hat{h}_0}{q} + \sum_i
    \element{pi}{\hat{v}}{qi}
  \]
  \end{sol}

  \begin{sol}{problem:spbasissetup}
  The following python code sets up the quantum numbers for both
  infinite nuclear matter and neutron matter employing a cutoff in
  the value of $n$. The full code can be found at  \url{https://github.com/ManyBodyPhysics/LectureNotesPhysics/tree/master/Programs/Chapter8-programs/python/spstatescc.py}.
  \begin{lstlisting}
from numpy import *

nmax =2
nshell = 3*nmax*nmax
count = 1
tzmin = 1

print "Symmetric nuclear matter:"  
print "a, nx,   ny,   nz,   sz,   tz,   nx^2 + ny^2 + nz^2"
for n in range(nshell): 
    for nx in range(-nmax,nmax+1):
         for ny in range(-nmax,nmax+1):
            for nz in range(-nmax, nmax+1):  
                for sz in range(-1,1+1):
                    tz = 1
                    for tz in range(-tzmin,tzmin+1):
                        e = nx*nx + ny*ny + nz*nz
                        if e == n:
                            if sz != 0: 
                                if tz != 0: 
                                    print count, "  ",nx,"  ",ny, "  ",nz,"  ",sz,"  ",tz,"         ",e
                                    count += 1
                                    
                                    
nmax =1
nshell = 3*nmax*nmax
count = 1
tzmin = 1
print "------------------------------------"
print "Neutron matter:"                                    
print "a, nx,   ny,   nz,   sz,    nx^2 + ny^2 + nz^2"
for n in range(nshell): 
    for nx in range(-nmax,nmax+1):
         for ny in range(-nmax,nmax+1):
            for nz in range(-nmax, nmax+1):  
                for sz in range(-1,1+1):
                    e = nx*nx + ny*ny + nz*nz
                    if e == n:
                        if sz != 0: 
                            print count, "  ",nx,"  ",ny, "  ",sz,"  ",tz,"         ",e
                            count += 1     
  \end{lstlisting}                               
  \end{sol}

  \section*{Appendix, Wick's theorem}
  \addcontentsline{toc}{section}{Appendix} Wick's theorem is based on
  two fundamental concepts, namely $\textit{normal ordering}$ and
  $\textit{contraction}$. The normal-ordered form of
  $\hat{A}\hat{B}..\hat{X}\hat{Y}$, where the individual terms are
  either a creation or annihilation operator, is defined as
  \begin{align}
  \label{def: Normal ordering}
  \kpr{\hat{A}\hat{B}..\hat{X}\hat{Y}} \equiv
  (-1)^p\fpr{\text{creation operators}}\cdot\fpr{\text{annihilation
      operators}}.
  \end{align}
  The $p$ subscript denotes the number of permutations that is needed
  to transform the original string into the normal-ordered form. A
  contraction between to arbitrary operators $\hat{X}$ and $\hat{Y}$
  is defined as
  \begin{align}
  \contraction[0.5ex]{}{\hat{X}}{}{\hat{Y}}{} \hat{X}\hat{Y} \equiv
  \for{0}{\hat{X}\hat{Y}}{0}.
  \end{align}
  It is also possible to contract operators inside a normal ordered
  products. We define the original relative position between two
  operators in a normal ordered product as $p$, the so-called
  permutation number. This is the number of permutations needed to
  bring one of the two operators next to the other one. A contraction
  between two operators with $p \neq 0$ inside a normal ordered is
  defined as
  \begin{align}
  \kpr{\contraction[0.5ex]{}{\hat{A}}{\hat{B}..}{\hat{X}}\hat{A}\hat{B}..\hat{X}\hat{Y}}
  = (-1)^p
  \kpr{\contraction[0.5ex]{}{\hat{A}}{}{\hat{B}}\hat{A}\hat{B}..\hat{X}\hat{Y}}.
  \end{align}
  In the general case with $m$ contractions, the procedure is similar,
  and the prefactor changes to
  \begin{align}
  (-1)^{p_1 + p_2 + .. + p_m}.
  \end{align} 

  Wick's theorem states that every string of creation and annihilation
  operators can be written as a sum of normalordered products with all
  possible ways of contractions,
  \begin{align}
  \label{def: Wick's theorem}
  \hat{A}\hat{B}\hat{C}\hat{D}..\hat{R}\hat{X}\hat{Y}\hat{Z} &=
  \kpr{\hat{A}\hat{B}\hat{C}\hat{D}..\hat{R}\hat{X}\hat{Y}\hat{Z}}\\ &+
  \sum_{[1]} \kpr{ \contraction[0.5ex]{}{\hat{A}}{}{\hat{B}}
    \hat{A}\hat{B}\hat{C}\hat{D}..\hat{R}\hat{X}\hat{Y}\hat{Z}}\\ &+
  \sum_{[2]}
  \kpr{\contraction[0.5ex]{}{\hat{A}}{\hat{B}}{\hat{C}}\contraction[1.0ex]{\hat{A}}{\hat{B}}{\hat{C}}{\hat{D}}\hat{A}\hat{B}\hat{C}\hat{D}..\hat{R}\hat{X}\hat{Y}\hat{Z}}\\ &+
  ...\\ &+
  \sum_{[\frac{N}{2}]}\kpr{\contraction[0.5ex]{}{\hat{A}}{\hat{B}}{\hat{C}}\contraction[1.0ex]{\hat{A}}{\hat{B}}{\hat{C}}{\hat{D}}
    \hat{A}\hat{B}\hat{C}\hat{D}..\contraction[0.5ex]{}{\hat{R}}{\hat{X}}{\hat{Y}}\contraction[1.0ex]{\hat{R}}{\hat{X}}{\hat{Y}}{\hat{Z}}\ \hat{R}\hat{X}\hat{Y}\hat{Z}}.
  \end{align}

  The $\sum_{[m]}$ means the sum over all terms with $m$ contractions,
  while $\fpr{\frac{N}{2}}$ means the largest integer that not do not
  exceeds $\frac{N}{2}$ where $N$ is the number of creation and
  annihilation operators. When $N$ is even,
  \begin{align}
  \label{exp: Wick condition}
  \fpr{\frac{N}{2}} = \frac{N}{2},
  \end{align}
  and the last sum in Eq. (\ref{def: Wick's theorem}) is over fully
  contracted terms. When $N$ is odd,
  \begin{align}
  \fpr{\frac{N}{2}} \neq \frac{N}{2},
  \end{align}
  and none of the terms in Eq. (\ref{def: Wick's theorem}) are fully
  contracted.

  An important extension of Wick's theorem allow us to define
  contractions between normal-ordered strings of operators. This is
  the so-called generalized Wick's theorem,
  \begin{align}
  \label{def: Generalized Wick's theorem}
  \kpr{\hat{A}\hat{B}\hat{C}\hat{D}..}\kpr{\hat{R}\hat{X}\hat{Y}\hat{Z}..}
  &=
  \kpr{\hat{A}\hat{B}\hat{C}\hat{D}..\hat{R}\hat{X}\hat{Y}\hat{Z}}\\ &+
  \sum_{[1]} \kpr{
    \contraction[0.5ex]{}{\hat{A}}{\hat{B}\hat{C}\hat{D}..}{\hat{R}}
    \hat{A}\hat{B}\hat{C}\hat{D}..\hat{R}\hat{X}\hat{Y}\hat{Z}}\\ &+
  \sum_{[2]}
  \kpr{\contraction[0.5ex]{}{\hat{A}}{\hat{B}\hat{C}\hat{D}..}{\hat{R}}\contraction[1.0ex]{\hat{A}}{\hat{B}}{\hat{C}\hat{D}..\hat{R}}{\hat{X}}\hat{A}\hat{B}\hat{C}\hat{D}..\hat{R}\hat{X}\hat{Y}\hat{Z}}\\ &+
  ...
  \end{align}

  Turning back to the many-body problem, the vacuum expectation value
  of products of creation and annihilation operators can be written,
  according to Wick's theoren in Eq. (\ref{def: Wick's theorem}), as a
  sum over normal ordered products with all possible numbers and
  combinations of contractions,
  \begin{align}
  \for{0}{\hat{A}\hat{B}\hat{C}\hat{D}..\hat{R}\hat{X}\hat{Y}\hat{Z}}{0}
  &=
  \for{0}{\kpr{\hat{A}\hat{B}\hat{C}\hat{D}..\hat{R}\hat{X}\hat{Y}\hat{Z}}}{0}\\ &+
  \sum_{[1]} \for{0}{\kpr{\contraction[0.5ex]{}{\hat{A}}{}{\hat{B}}
      \hat{A}\hat{B}\hat{C}\hat{D}..\hat{R}\hat{X}\hat{Y}\hat{Z}}}{0}\\ &+
  \sum_{[2]}\for{0}{\kpr{\contraction[0.5ex]{}{\hat{A}}{\hat{B}}{\hat{C}}\contraction[1.0ex]{\hat{A}}{\hat{B}}{\hat{C}}{\hat{D}}\hat{A}\hat{B}\hat{C}\hat{D}..\hat{R}\hat{X}\hat{Y}\hat{Z}}}{0}\\ &+
  ... \\ &+ \sum_{[\frac{N}{2}]}
  \for{0}{\kpr{\contraction[0.5ex]{}{\hat{A}}{\hat{B}}{\hat{C}}\contraction[1.0ex]{\hat{A}}{\hat{B}}{\hat{C}}{\hat{D}}
      \hat{A}\hat{B}\hat{C}\hat{D}..\contraction[0.5ex]{}{\hat{R}}{\hat{X}}{\hat{Y}}\contraction[1.0ex]{\hat{R}}{\hat{X}}{\hat{Y}}{\hat{Z}}\ \hat{R}\hat{X}\hat{Y}\hat{Z}}}{0}.
  \end{align}

  All vacuum expectation values of normal ordered products without
  fully contracted terms are zero. Hence, the only contributions to
  the expectation value are those terms that $\textit{is}$ fully
  contracted,
  \begin{align}
  \for{0}{\hat{A}\hat{B}\hat{C}\hat{D}..\hat{R}\hat{X}\hat{Y}\hat{Z}}{0}
  &= \sum_{[all]}
  \for{0}{\kpr{\contraction[0.5ex]{}{\hat{A}}{\hat{B}}{\hat{C}}\contraction[1.0ex]{\hat{A}}{\hat{B}}{\hat{C}}{\hat{D}}
      \hat{A}\hat{B}\hat{C}\hat{D}..\contraction[0.5ex]{}{\hat{R}}{\hat{X}}{\hat{Y}}\contraction[1.0ex]{\hat{R}}{\hat{X}}{\hat{Y}}{\hat{Z}}\ \hat{R}\hat{X}\hat{Y}\hat{Z}}}{0}\\ &=
  \sum_{[all]}
  \contraction[0.5ex]{}{\hat{A}}{\hat{B}}{\hat{C}}\contraction[1.0ex]{\hat{A}}{\hat{B}}{\hat{C}}{\hat{D}}
  \hat{A}\hat{B}\hat{C}\hat{D}..\contraction[0.5ex]{}{\hat{R}}{\hat{X}}{\hat{Y}}\contraction[1.0ex]{\hat{R}}{\hat{X}}{\hat{Y}}{\hat{Z}}\ \hat{R}\hat{X}\hat{Y}\hat{Z}.
  \end{align}

  To obtain fully contracted terms, Eq. (\ref{exp: Wick condition})
  must hold. When the number of creation and annihilation operators is
  odd, the vacuum expectation value can be set to zero at once. When
  the number is even, the expectation value is simply the sum of terms
  with all possible combinations of fully contracted terms. Observing
  that the only contractions that give nonzero contributions are
  \begin{align}
  \contraction{}{a_{\alpha}}{}{a^{\dagger}_{\beta}}
  a_{\alpha}a^{\dagger}_{\beta} = \delta_{\alpha \beta},
  \end{align}
  the terms that contribute are reduced even more.

  Wick's theorem provides us with an algebraic method for easy
  determination of the terms that contribute to the matrix element.

  \begin{acknowledgement}
  We are much indebted to Thomas Papenbrock for many discussions on
  many-body theory.  Computational resources were provided by
  Michigan State University and the Research Council of Norway via the
  Notur project (Supercomputing grant NN2977K).  This work was
  supported by NSF Grant No.~PHY-1404159 (Michigan State University).
  \end{acknowledgement}

\end{document}